\documentclass[11pt]{article}
\usepackage{amsmath,amsfonts,amssymb,epsfig,color}
\usepackage{rotating}
\begin{document}
\title{Geometric Complexity Theory VII: 
Nonstandard quantum group for the plethysm problem
\\ {\small (extended abstract)}}
\author{Dedicated to Sri Ramakrishna \\ \\
Ketan D. Mulmuley \footnote{Part of this work was done while
the author was visiting I.I.T. Mumbai}
 \\
The University of Chicago \\ \\
(Technical Report TR-2007-14\\
Computer Science Department \\
The University of Chicago \\
September 2007) \\ Revised version \\
http://ramakrishnadas.cs.uchicago.edu 
}

\maketitle
\newtheorem{prop}{Proposition}[section]
\newtheorem{claim}[prop]{Claim}
\newtheorem{goal}[prop]{Goal}
\newtheorem{theorem}[prop]{Theorem}
\newtheorem{hypo}[prop]{Hypothesis}
\newtheorem{guess}[prop]{Guess}
\newtheorem{problem}[prop]{Problem}
\newtheorem{axiom}[prop]{Axiom}
\newtheorem{question}[prop]{Question}
\newtheorem{remark}[prop]{Remark}
\newtheorem{lemma}[prop]{Lemma}
\newtheorem{claimedlemma}[prop]{Claimed Lemma}
\newtheorem{claimedtheorem}[prop]{Claimed Theorem}
\newtheorem{cor}[prop]{Corollary}
\newtheorem{defn}[prop]{Definition}
\newtheorem{ex}[prop]{Example}
\newtheorem{conj}[prop]{Conjecture}
\newtheorem{obs}[prop]{Observation}
\newtheorem{phyp}[prop]{Positivity Hypothesis}
\newcommand{\bitlength}[1]{\langle #1 \rangle}
\newcommand{\ca}[1]{{\cal #1}}
\newcommand{\floor}[1]{{\lfloor #1 \rfloor}}
\newcommand{\ceil}[1]{{\lceil #1 \rceil}}
\newcommand{\gt}[1]{{\langle  #1 |}}
\newcommand{\C}{\mathbb{C}}
\newcommand{\N}{\mathbb{N}}
\newcommand{\R}{\mathbb{R}}
\newcommand{\Z}{\mathbb{Z}}
\newcommand{\frcgc}[5]{\left(\begin{array}{ll} #1 &  \\ #2 & | #4 \\ #3 & | #5
\end{array}\right)}

\newcommand{\cgc}[6]{\left(\begin{array}{ll} #1 ;& \quad #3\\ #2 ; & \quad #4
\end{array}\right| \left. \begin{array}{l} #5 \\ #6 \end{array} \right)}

\newcommand{\wigner}[6]
{\left(\begin{array}{ll} #1 ;& \quad #3\\ #2 ; & \quad #4
\end{array}\right| \left. \begin{array}{l} #5 \\ #6 \end{array} \right)}

\newcommand{\rcgc}[9]{\left(\begin{array}{ll} #1 & \quad #4\\ #2  & \quad #5
\\ #3 &\quad #6
\end{array}\right| \left. \begin{array}{l} #7 \\ #8 \\#9 \end{array} \right)}

\newcommand{\srcgc}[4]{\left(\begin{array}{ll} #1 & \\ #2 & | #4  \\ #3 & |
\end{array}\right)}

\newcommand{\arr}[2]{\left(\begin{array}{l} #1 \\ #2   \end{array} \right)}
\newcommand{\unshuffle}[1]{\langle #1 \rangle}
\newcommand{\ignore}[1]{}
\newcommand{\f}[2]{{\frac {#1} {#2}}}
\newcommand{\tableau}[5]{
\begin{array}{ccc} 
#1 & #2  &#3 \\
#4 & #5 
\end{array}}
\newcommand{\embed}[1]{{#1}^\phi}
\newcommand{\stab}{{\mbox {stab}}}
\newcommand{\limit}{{\mbox {lim}}}
\newcommand{\perm}{{\mbox {perm}}}
\newcommand{\trace}{{\mbox {trace}}}
\newcommand{\polylog}{{\mbox {polylog}}}
\newcommand{\sign}{{\mbox {sign}}}
\newcommand{\proj}{{\mbox {Proj}}}
\newcommand{\poly}{{\mbox {poly}}}
\newcommand{\std}{{\mbox {std}}}
\newcommand{\m}{\mbox}
\newcommand{\formula}{{\mbox {Formula}}}
\newcommand{\circuit}{{\mbox {Circuit}}}
\newcommand{\core}{{\mbox {core}}}
\newcommand{\orbit}{{\mbox {orbit}}}
\newcommand{\cycle}{{\mbox {cycle}}}
\newcommand{\ideal}{{\mbox {ideal}}}
\newcommand{\qed}{{\mbox {Q.E.D.}}}
\newcommand{\proof}{\noindent {\em Proof: }}
\newcommand{\weight}{{\mbox {wt}}}
\newcommand{\tab}{{\mbox {Tab}}}
\newcommand{\level}{{\mbox {level}}}
\newcommand{\vol}{{\mbox {vol}}}
\newcommand{\vect}{{\mbox {Vect}}}
\newcommand{\val}{{\mbox {wt}}}
\newcommand{\sym}{{\mbox {Sym}}}
\newcommand{\convex}{{\mbox {convex}}}
\newcommand{\spec}{{\mbox {spec}}}
\newcommand{\strong}{{\mbox {strong}}}
\newcommand{\adm}{{\mbox {Adm}}}
\newcommand{\eval}{{\mbox {eval}}}
\newcommand{\for}{{\quad \mbox {for}\ }}
\newcommand{\Q}{Q}
\newcommand{\mand}{{\quad \mbox {and}\ }}
\newcommand{\invlim}{{\mbox {lim}_\leftarrow}}
\newcommand{\directlim}{{\mbox {lim}_\rightarrow}}
\newcommand{\sformal}{{\cal S}^{\mbox f}}
\newcommand{\vformal}{{\cal V}^{\mbox f}}
\newcommand{\crystal}{\mbox{crystal}}
\newcommand{\conje}{\mbox{\bf Conj}}
\newcommand{\graph}{\mbox{graph}}
\newcommand{\ind}{\mbox{index}}

\newcommand{\rank}{\mbox{rank}}
\newcommand{\id}{\mbox{id}}
\newcommand{\str}{\mbox{string}}
\newcommand{\RSK}{\mbox{RSK}}
\newcommand{\wt}{\mbox{wt}}
\setlength{\unitlength}{.75in}

\begin{abstract} 
This article describes a  {\em nonstandard} quantum group 
that may be used to derive  a positive formula
for the plethysm problem,
just as the standard (Drinfeld-Jimbo) quantum group 
can be used to derive  the positive
Littlewood-Richardson rule for arbitrary complex semisimple Lie groups.
The sequel \cite{GCT8} gives conjecturally correct algorithms to 
construct canonical bases of the coordinate rings of these nonstandard
quantum groups and canonical bases of the dually paired
nonstandard deformations of the symmetric group algebra. 
A positive $\#P$-formula for the plethysm constant follows from the
conjectural properties of these canonical bases and the
duality and reciprocity conjectures herein.
\end{abstract}

\section{Introduction} \label{sintro}
The following is a fundamental problem in representation theory 
\cite{GCT6,macdonald,stanley}: 

\begin{problem} {\bf (Plethysm problem)} \label{pplethysm}

Find an explicit positive ($\#P$-) formula in the spirit of the Littlewood-Richardson rule  for the plethysm constant $a_{\lambda,\mu}^\pi$. 
For given partitions $\lambda,\mu$ and $\pi$,
this is the multiplicity  of 
the irreducible representation 
$V_\pi(H)$ of $H=GL_n(\C)$ 
in the irreducible representation 
$V_\lambda(G)$ of $G=GL(X)$,
where $X=V_\mu=V_\mu(H)$ is an irreducible representation of $H$.
Here $V_\lambda(G)$ is considered an $H$-module via 
the representation map $\rho:H\rightarrow G$.

\noindent {\bf (Generalized plethysm problem):} 

The same as above, letting  $H$  be any complex, 
semisimple (or, more generally, reductive)
classical Lie group,  $\lambda$ a dominant weight
of $G$, $\pi$ and $\mu$ dominant weights of $H$. 
\end{problem}

This article describes a  quantum group 
that may be used to derive such a positive formula, just as the 
standard (Drinfeld-Jimbo) quantum group \cite{drinfeld,jimbo,rtf} 
can be used to derive  the positive
Littlewood-Richardson rule for arbitrary complex semisimple Lie groups
\cite{kashiwaracrystal,littelmann,lusztigbook}; the results 
here were announced in \cite{GCT4} (most of the results here also hold
for nonclassical $H$, though we shall only worry about classical $H$ here).
 For the significance of 
a positive formula 
in the context of geometric complexity theory, see \cite{GCTflip1}.
The approach that we  wish to follow is: 
\begin{enumerate} 
\item Find a quantization of the homomorphism 
\begin{equation} \label{eqhomo1} 
 H \rightarrow G
\end{equation}
of the form
\begin{equation} \label{eqhomo2} 
 H_q \rightarrow G^H_q,
\end{equation}
where $H_q$ is the standard Drinfeld-Jimbo quantization of 
$H$, and $G^H_q$ is the new  nonstandard quantization of $G$ that we seek.
\item Develop a theory of canonical (local/global crystal) bases for the
representations of $G^H_q$ in the spirit of the
canonical bases \cite{kashiwaracrystal,lusztigcanonical} 
for the representations of the standard quantum group.
\item Derive the required explicit positive formula for the plethysm constant 
from the properties of the canonical  bases.
\end{enumerate}

The following addresses the first step.

\begin{theorem} (cf. Section~\ref{snonstdgroup}) \label{tnonstd}
There exists a possibly singular quantum group 
$G^H_q$ such that the homomorphism (\ref{eqhomo1}) can be quantized in
the form (\ref{eqhomo2}). 

Furthermore, all finite dimensional polynomial representations 
of $G^H_q$ are completely reducible, and a quantum analogue of
the Peter-Weyl theorem holds for the matrix coordinate ring of $G^H_q$.
\end{theorem}
For the 
precise meaning of the various terms here, see Section~\ref{snonstdgroup}.
Here and in what follows, we assume 
that the base field is $\C=\C(q)$, $q$ complex. But a suitable  algebraic 
extension of $\Q(q)$ will also suffice for our purposes;
see Section~\ref{srefined} for a discussion on the base field.

When $H=G$, $G^H_q$ specializes to the standard quantum group $H_q$.
When $H=GL(V)\times GL(W)$, $G=GL(X)$, $X=V\otimes W$ with natural $H$-action,
it reduces to the  quantum group in \cite{GCT4} for the Kronecker problem.

We call $G^H_q$ the {\em nonstandard quantum group} associated with
the embedding (\ref{eqhomo1}).
It can be singular in general. That is,
its determinant may vanish, and hence, the antipode need not exist.
Strictly speaking, it should hence be called a nonstandard quantum 
semi-group. We still use the term group, because 
this object has  characteristic features of the standard 
quantum group, such as semisimplicity of polynomial representations,
Peter-Weyl theorem, and most importantly, conjectural existence of
canonical bases for its representations and the matrix coordinate ring.

We also construct (Section~\ref{sbr})
a {\em nonstandard quantization}
${\cal B}^H_r={\cal B}^H_r(q)$ of the group algebra $\C[S_r]$ of the
symmetric group $S_r$ whose relationship with $G^H_q$ is 
conjecturally akin to that
 of the Hecke algebra  with the standard quantum group.
Specifically, let $X_q$ denote the irreducible representation $V_{q,\mu}$
of $H_q$ with highest weight $\mu$; it is the usual  quantization of 
$X=V_\mu$. 
Then:
\begin{conj} {\bf (Nonstandard duality)} \label{cduality}

\noindent (1)  The left  action of $G_q^H$ on $X_q^{\otimes r}$ 
and the right action  ${\cal B}_r^H (q)$ on  $X_q^{\otimes r}$ 
determine each other.

\noindent  (2) There is a one-to-one correspondence between the irreducible polynomial representations of
$G_q^H$ of degree $r$ and the irreducible representations of ${\cal B}_r^H$
so that, 
as a bimodule, 
\begin{equation} \label{enduality}
 X_q^{\otimes r} = \bigoplus_{\alpha} W_{q,\alpha}
 \otimes T_{q,\alpha},
\end{equation}
where $W_{q,\alpha}$ runs over the irreducible polynomial representations of 
$G_q^H$ of degree $r$, 
and $T_{q,\alpha}$ denotes the irreducible representation of 
${\cal B}_r^H (q)$ in correspondence with $W_{q,\alpha}$.  
\end{conj}

The irreducible representations
$W_{q,\alpha}$   here need not be
 $q$-deformations of the irreducible representations  
of  $G$, because $G_q^H$ is, in general, a nonflat deformation of
 $G$. This means  the Poincare series of $G_q^H$ need  not coincide 
with that of $G$.
Our first goal is to associate with 
each Weyl module $V_\lambda$ of $G$
a possibly reducible 
representation $V_{q,\lambda}^H$ of $G_q^H$, called 
the {\em $q$-analogue} of $V_\lambda$,  so that 
\[\limit_{q\rightarrow 1} V_{q,\lambda}^H \cong  V_\lambda\] as 
an $H$-module. In this context:

\begin{conj} {\bf (Nonstandard reciprocity)} \label{creciprocity}
Let $\lambda$ be a partition of size $r$.
Let \[
V_{q,\lambda}^H=\bigoplus_{\alpha} m^\alpha_\lambda W_{q,\alpha},\]
where $m^\alpha_\lambda$ denotes the multiplicity of the Specht module 
$S_\lambda$ of the symmetric group $S_r$ in 
$T_{q,\alpha}(1)=\limit_{q\rightarrow 1} T_{q,\alpha}$,
as defined in Section~\ref{srefined}. 
Then $V_{q,\lambda}^H$ is a $q$-analogue of $V_\lambda$ in the
sense defined above. 

Thus the multiplicity of the $G_q^H$-module 
$W_{q,\alpha}$ in $V_{q,\lambda}^H$ is equal to the multiplicity of 
the Specht module $S_{\lambda}$ in the specialization of $T_{q,\alpha}$ 
at $q=1$.
\end{conj} 
A more refined form of this conjecture is given in Section~\ref{srefined}.
Both duality and reciprocity  are supported by experimental evidence;
cf. Section~\ref{sexpi}.

By the conjectural reciprocity, 
\[ a_{\lambda,\mu}^\pi=\sum_\alpha m^\alpha_\lambda n^\alpha_{\pi},
\] 
where  $n_{\pi}^\alpha$ is the multiplicity 
of the irreducible $H_q$-module
$V_{q,\pi}$ in $W_{q,\alpha}$.
Hence  Problem~\ref{pplethysm}
can be  decomposed into the following two subproblems:

\noindent  {\bf (P1)}: Find  a positive ($\#P$-) formula for the multiplicity
 $n_{\pi}^\alpha$.

\noindent  {\bf (P2)}:  Find a positive ($\#P$-) formula for the multiplicity 
$m^\alpha_\lambda$.

The article \cite{GCT8} gives  conjecturally correct algorithms
to construct a canonical basis
of the matrix coordinate ring of $G_q^H$ whose conjectural properties 
would imply a positive formula as needed in the first problem, and 
a  canonical basis of ${\cal B}_r^H$ whose conjectural properties
would imply a positive formula as needed in the second problem.

At present, we cannot prove correctness of these algorithms
nor the required conjectural properties, 
because we are unable to deal with the high complexity of 
the nonstandard quantum group.  Specifically, as we shall see in
 Section~\ref{sminors}, the formulae for the 
minors of the nonstandard group 
turn out to be highly nonelementary in contrast to the 
elementary formulae for the minors of the standard quantum group.
 The coefficients of these formulae depend on the
multiplicative structural constants 
of   canonical bases akin to the canonical basis 
of the  coordinate ring of the 
standard quantum group constructed by  Kashiwara and Lusztig 
\cite{kashiwaraglobal,lusztigbook}.
To get  explicit formulae for these structural constants,
one  needs interpretations 
for them akin to the interpretations for 
the Kazhdan-Lusztig polynomials and multiplicative structural constants
of the canonical basis of the coordinate ring of the standard quantum group
in terms of perverse sheaves \cite{kazhdan1,lusztigcanonical,beilinson}.
Thus, the linear algebra for the nonstandard quantum group--i.e. the
theory of its minors--is already highly nonelementary 
in contrast to the linear algebra for the standard quantum group.
This is  why its representation theory may turn out  to  be far more complex.
In particular, we cannot explicitly construct nor classify its irreducible polynomial
representations. Of course, all this and much more would follow
if   correctness of the algorithms in \cite{GCT8} for
constructing canonical bases and their conjectural properties  can be proved.

\noindent {\bf Acknowledgement:} The author is grateful to David 
Kazhdan for helpful discussions  and comments, 
and to Milind Sohoni for helpful discussions, especially for bringing
the reference \cite{rossi} to our attention,
and for the help in explicit computations  in Section~\ref{sex2} in MATLAB.

\section{Nonstandard quantum group} \label{snonstdgroup}
We describe in this section the construction of the nonstandard 
quantum group $G_q^H$ in Theorem~\ref{tnonstd}. The reader may 
refer to \cite{GCT4} for the full details  in a nontrivial special case
of the plethysm problem, called  the Kronecker problem.
For the sake of simplicity, we assume here that $H=GL(V)$ (type $A$).
Let $X=V_\mu(H)$ be its irreducible polynomial representation.
The goal is to quantize the homomorphism 
\[ H=GL(V) \rightarrow G=GL(X).\] 
Let  ${\cal H}$ and ${\cal G}$ be the
Lie algebras of $H$ and $G$.  We
follow the  terminology  in \cite{klimyk}, which will be our
standard reference on quantum groups.

The standard quantum group $H_q=GL_q(V)$
associated with $GL(V)$ can be defined 
by first constructing 
the coordinate algebra ${\cal O}(M_q(V))$ of the standard quantum matrix space
$M_q(V)$ as a suitable FRT-algebra \cite{rtf}.
The coordinate ring
${\cal O}(GL_q(V))$ of $GL_q(V)$ is 
obtained by localizing ${\cal O}(M_q(V))$ at the suitably defined
quantum determinant. The
Drinfeld-Jimbo universal enveloping algebra $U_q({\cal G})$ 
\cite{drinfeld,jimbo} of $GL_q(V)$ can then be defined dually. Specifically, let 
$J$ be the maximal ideal of the elements in ${\cal O}(M_q(V))$ which vanish
at the identity--i.e. on which $\epsilon$, the counit, vanishes. Then 
$U_q({\cal G})$ can be identified with the space of linear functions
on ${\cal O}(M_q(V))$ which vanish on $J^r$ for some integer $r>0$
depending on the linear function.

Analogously, we  first construct the {nonstandard matrix 
coordinate ring}
${\cal O}(M_q^H(X))$ of the (virtual)
{nonstandard  matrix space}  $M_q^H(X)$, 
and then define the {nonstandard quantized universal enveloping algebra} 
$U_q^H({\cal G})$  by dualization. We define the {nonstandard 
quantum group} $G_q^H$ as the virtual object whose universal enveloping
algebra is $U_q^H({\cal G})$. 
The construction would yield natural bialgebra homomorphisms 
from $U_q({\cal H})$ to $U_q^H({\cal G})$ and from ${\cal O}(M_q^H(X))$ 
to ${\cal O}(M_q(V)$, thereby giving the desired quantizations of
the homomorphisms $U({\cal H}) \rightarrow U({\cal G})$ and
${\cal O}(M(X)) \rightarrow {\cal O}(M(V))$. This is what is meant 
by the quantization (\ref{eqhomo2}) of the map (\ref{eqhomo1}).
The determinant of $G_q^H$ may vanish, and hence, we cannot, in general,
define its coordinate ring ${\cal O}(G_q^H)$ by localizing
${\cal O}(M_q^H(X))$. Fortunately, this will not matter since
the coordinate ring ${\cal O}(M_q^H(X))$ and 
the nonstandard quantized algebra ${\cal B}_r^H(q)$ (Section~\ref{sbr}) 
together   contain conjecturally all the information
that we need (cf. Conjecture~\ref{creciprocity}),
and  have properties similar to that  of
the standard matrix coordinate ring ${\cal O}(M_q(V))$ and the Hecke algebra;
cf. Theorem~\ref{tnonstdformal} below.

The {\em nonstandard matrix coordinate ring} ${\cal O}(M_q^H(X))$ is
constructed as follows. 
Let $\hat R_{X,X}^H$ be the $\hat R$ matrix of $X_q=V_{q,\mu}$ considered
as an $H_q$-module \cite{klimyk}. Here and in what follows, 
we sometimes denote $X_q$ by $X$; 
the meaning should be clear from the context.
It is well known that $\hat R_{X,X}^H$ is diagonalizable
and that its each  eigenvalue 
is of the form $+$ or $-q^{a/2}$ for some  integer $a$ \cite{klimyk}. 
Let 
\begin{equation} \label{eqide}
I = P^{+,H}_{X,X} + P^{-,H}_{X,X},
\end{equation}
be the associated spectral decomposition of the identity, where 
$P^{+,H}_{X,X}$ and $P^{-,H}_{X,X}$ denote the projections of
$X_q\otimes X_q$ on the eigenspaces
of $\hat R_{X,X}^H$ 
for eigenvalues with $+$ and $-$ sign, respectively. Let ${\mathbf u}$ 
be a variable  matrix specifying a generic transformation from $X$ to $X$.
Let $u^i_j$ denote its variable entries.
Then ${\cal O}(M_q^H(X))$ is defined to be the FRT bialgebra 
\cite{rtf} associated 
with the transformation $P^{+,H}_{X,X}$, or equivalently, 
$P^{-,H}_{X,X}$. That is, it is the quotient of $\C\langle u^i_j\rangle$ 
modulo the relations 
\begin{equation}  \label{eqnonstd1}
P_{X,X}^{+,H} (\mathbf{u \otimes u}) = 
       (\mathbf{u \otimes u}) P_{X,X}^{+,H}, 
\end{equation}
or equivalently,
\begin{equation} \label{eqnonstd2}
P_{X,X}^{-,H} (\mathbf{u \otimes u}) = 
       (\mathbf{u \otimes u}) P_{X,X}^{-,H}.
\end{equation}

An alternative definition of ${\cal O}(M_q^H(X))$ is as follows.
Let $S_q^H(X\otimes X)$, the {\em symmetric subspace} of $X\otimes X$,
be the image of $P^{+,H}_{X,X}$, and 
$A_q^H(X\otimes X)$, the {\em antisymmetric subspace} of $X\otimes X$,
the image of $P^{-,H}_{X,X}$ \cite{klimyk}. 
In other words, $S_q^H(X\otimes  X)$ is defined by the equation
\begin{equation}\label{eqp-}
P^{-,H}_{X, X} \mathbf{x_1}\mathbf{x_2} =0,
\end{equation}
where $\mathbf{x_1}=\mathbf{x\otimes I}$ and
$\mathbf{x_2}=\mathbf{I\otimes x}$,
and $A_q^H(X\otimes X)$ is defined by the equation
\begin{equation}\label{eqp+}
P^{+,H}_{X,X} \mathbf{x_1}\mathbf{x_2} =0.
\end{equation}

The {\em braided
symmetric algebra} \cite{berenstein,rossi} $\C_q^H[X]$ of $X$ is defined to be 
the algebra over the entries $x_i$'s of $\mathbf{x}$ subject to 
the relation (\ref{eqp-}).
It will  be called the coordinate ring of the virtual {\em  quantum space} 
 $X^H_{sym}$.
Similarly, the {\em braided exterior algebra} $\wedge_q^H[X]$ of $X$
is defined to  be the algebra over the entries $x_i$'s of $\mathbf{x}$
subject to the relation (\ref{eqp+}).
It will called the  coordinate ring of the virtual {\em quantum  space}
$X^H_\wedge$.
Let $\C_q^{H,r}[X]$ and $\wedge_q^{H,r}[X]$ be the degree $r$ components  of 
$\C_q^H[X]$ and $\wedge_q^H[X]$, respectively.
It is known \cite{berenstein} that the dimensions of 
$\C_q^{H,r}[X]$ and $\wedge_q^{H,R}$ are bounded by the dimensions 
of the classical   $\C^r[X]$ and $\wedge^r[X]$,
respectively. But unlike in the standard setting, the dimensions can 
be strictly less \cite{berenstein,rossi}. That is, $\C_q^H[X]$ and
$\wedge_q^H[X]$  are, in general,
nonflat deformations of the classical symmetric and exterior 
algebras $\C[X]$ and $\wedge[X]$.
For example,  
$\wedge_q^{H,3}[X]=0$ when  ${\cal H}=sl_2(\C)$
 and $X$ is the four dimensional irreducible
representation of $sl_2(\C)$ \cite{berenstein}.

The equation  (\ref{eqnonstd1}) or (\ref{eqnonstd2}) 
after reformulation just says that 
the defining relation (\ref{eqp-}) of $X^H_{sym}$--or equivalently,
the defining relation (\ref{eqp+}) of $X^H_\wedge$--is preserved under the left 
and right actions of ${\mathbf u}$ on ${\mathbf x}$ given by 
${\mathbf x} \rightarrow {\mathbf u}{\mathbf x}$ and 
${\mathbf x}^t \rightarrow {\mathbf x}^t {\mathbf u}$.

This means $\C_q^{H}[X]$ and $\wedge_q^{H}[X]$ have  left and right
coactions of ${\cal O}(M_q^H(X))$. We  define the left and right 
{\em nonstandard minors}
of $G_q^H$ to be the matrix coefficients (in a suitable basis specified later) 
of the left and right coactions on $\wedge_q^H[X]$. 
If $\wedge_q^{H,\dim(X)}[X] \not  = 0$, then we  define the 
determinant of $G_q^H$ to be the matrix coefficient of the action of
${\cal O}(M_q^H(X))$ on $\wedge_q^{H,\dim(X)}[X]$. But it can vanish, 
as it does for ${\cal H}=sl_2(\C)$, $\dim(X)=4$.
The nonstandard minors   will be 
discussed  in more detail in Section~\ref{sminors}.

Let $J$ be the  ideal of elements in ${\cal O}(M_q^H(X))$ 
on which the counit  $\epsilon$ vanishes. Then the {\em nonstandard 
universal enveloping algebra} $U_q^H({\cal G})$ is defined to be 
the space of linear functions of ${\cal O}(M_q^H(X))$ which vanish 
on $J^r$ for some $r>0$ depending on the linear function.

The following is a precise form of Theorem~\ref{tnonstd}.
\begin{theorem} \label{tnonstdformal}
\noindent (1) 
There is a natural bialgebra 
homomorphism from ${\cal O}(M_q^H(X))$ to ${\cal O}(M_q(V))$. This
gives the desired quantization of the 
homomorphism  ${\cal O}(M(X)) \rightarrow {\cal O}(M(V))$.

\noindent (2) The matrix coordinate ring  ${\cal O}(M_q^H(X))$ of $G_q^H$ 
is cosemisimple. Hence,
its every finite dimensional corepresentation is completely reducible
as a direct sum of irreducible corepresentations.

\noindent (3) 
The $q$-analogue of the Peter-Weyl theorem holds: i.e.,
\[ 
{\cal O}(M_q^H(X))= \bigoplus_\alpha W_{q,\alpha}^* \otimes W_{q,\alpha},
\]
where $W_{q,\alpha}$ runs over all irreducible corepresentations of
${\cal O}(M_q^H(X))$. 

\noindent (4) 
The nonstandard enveloping algebra $U_q^H({\cal G})$ is 
a bialgebra with a compact real form (a $*$-structure) 
such that $X_q^{\otimes r}$ is its unitary representation with respect to 
the  Hermitian form on $X_q^{\otimes r}$ induced by the
standard Hermitian form on $X_q$.
There is a bialgebra homomorphism form
$U_q({\cal H})$ to $U_q^H({\cal G})$. This gives a desired quantization
of the homomorphism $U({\cal H}) \rightarrow U({\cal G})$. 
\end{theorem}

Here the standard Hermitian form on $X_q$ is the one that is
$U_q$-invariant, where $U_q \subseteq H_q$ is the compact real form
(the unitary subgroup) of $H_q$  in the sense of Woronowicz \cite{wor1}.
The special case of this theorem in the context of the Kronecker 
problem was proved in \cite{GCT4} on the basis of Woronowicz's work 
\cite{wor1}. The latter is no longer applicable in the general context here,
since the determinant of $G_q^H$ may vanish, and hence, we cannot, in general,
convert  ${\cal O}(M_q^H(X))$ into a Hopf algebra by localization at the
determinant. Fortunately, this does not matter since $U_q^H({\cal G})$
still has a compact real form, whose existence can be proved 
using the spectral properties of $\hat R_{X,X}^H$.

We also call $W_{q,\alpha}$ here a {\em polynomial
representation} of $G_q^H$.
By a polynomial representation of $U_q^H({\cal G})$ we mean 
a representation that is induced by a (finite dimensional)
corepresentation  of ${\cal O}(M_q^H(X))$.
It is completely reducible 
by cosemsimplicity of ${\cal O}(M_q^H(X))$. It may be conjectured
that every finite dimensional representation
of $U_q^H({\cal G})$ is completely reducible (as in the standard case),
though we shall not need this more general fact.

The standard Drinfeld-Jimbo enveloping algebra has an explicit 
presentation in the form of explicit generators ($e_i,f_i,K_i$) and 
explicit relations among them. It will be interesting to find an analogous
explicit presentation for $U_q^H({\cal G})$; cf. Section~\ref{sminors} for the
problems that arise in this context.

\ignore{
One can also  define nonstandard quantum Borel, parabolic and
Levi subgroups of
$G^H_q$, with respect to  the  Gelfand-Tsetlin basis of $X_q$,
just as in   the standard case. 
But there are  substantial
differences from the standard setting. For example, they can be singular--i.e.,
their determinants can vanish--and their
dimensions can be less than the classical dimensions.
It can be shown that  
the nonstandard quantized universal
enveloping algebra of each quantum Levi subgroup of  $G^H_q$
admits a compact real form, and 
that its matrix coordinate ring is cosemisimple with nonstandard 
Peter-Weyl decomposition,  as for  $G^H_q$.
}

\section{Nonstandard $q$-Schur algebra} 
In the standard setting, the $q$-Schur algebra ${\cal A}_r={\cal A}_r(q)$ is
defined to be the dual ${\cal O}(M_q(V))^*r$ of the degree $r$ 
component ${\cal O}(M_q(V))_r$ of the standard matrix coordinate
algebra ${\cal O}(M_q(V))$. Thus ${\cal A}_r(q)$ 
acts on $V^{\otimes r}$ from the left. It is known 
 \cite{klimyk} that it is the
centralizer in $\mbox{End}(V^{\otimes r})$ of the right action
of the Hecke algebra ${\cal H}_r(q)$ on $V^{\otimes r}$. 

Analogously, we define 
the {\em nonstandard $q$-Schur algebra} ${\cal A}^H_r={\cal A}^H_r(q)$ to
be the dual ${\cal O}(M^H_q(X))^*r$ of the degree $r$ 
component ${\cal O}(M^H_q(X))_r$ of the nonstandard matrix coordinate
algebra  ${\cal O}(M^H_q(X))$. Thus ${\cal A}^H_r(q)$ 
acts on $X^{\otimes r}$ from the left. As per the nonstadard duality
conjecture (Conjecture~\ref{cduality}),  it is the
centralizer in $\mbox{End}(X^{\otimes r})$ of the right action
of the nonstandard quantized  algebra ${\cal B}^H_r(q)$ (cf.
Section~\ref{sbr})  on $X^{\otimes r}$. 

Every irreducible corepresentation $W_{q,\alpha}$ of ${\cal O}(M_q^H(X))$
of degree $r$ can also be considered as a representation of ${\cal A}_r^H(q)$,
and conversely, every irreducible representation of ${\cal A}_r^H(q)$
arises in this way.
Theorem~\ref{tnonstdformal} now immediately implies:

\begin{theorem} \label{tnonstdformalschur}

\noindent (1) The nonstandard $q$-Schur algebra ${\cal A}^H_r(q)$
is semisimple. Hence,
its every finite dimensional representation is completely reducible
as a direct sum of irreducible representations.

\noindent (2) 
The $q$-analogue of the Peter-Weyl theorem in this case is the
Wederburn structure theorem for ${\cal A}_r^H(q)$:
\[ 
{\cal A}_r^H(q)= \bigoplus_\alpha W_{q,\alpha}^* \otimes W_{q,\alpha},
\]
where $W_{q,\alpha}$ runs over all irreducible representations of
${\cal A}_r^H(q)$.

\noindent (3) 
The nonstandard $q$-Schur algebra ${\cal A}^H_r(q)$
has  a compact real form (a $*$-structure) 
such that $X_q^{\otimes r}$ is its unitary representation with respect to 
the  Hermitian form on $X_q^{\otimes r}$ induced by the
standard Hermitian form on $X_q$.
\end{theorem}

\section{Nonstandard minors} \label{sminors}
In this section, we give a conjectural formula for  the Laplace expansion 
of the minors of $G^H_q$. 
The Laplace expansion for the standard quantum group $GL_q(V)$
is based on the  simple relation defining  the standard exterior  algebra 
$\wedge_q[V]$, namely
\[ v_i^2 =0 \quad \mbox{and} \quad 
 v_i v_j= -q ^{-1} v_j v_i, \quad \mbox{for} \quad i<j.\] 
This  explains why the Laplace expansion in the standard setting
is obtained from  the classical Laplace expansion by simply substituting 
$-q$ for $-1$. 
We need a similar  explicit formula for multiplication in
$\C_q^H[X]$ to get an explicit formula for Laplace expansion in
the nonstandard setting.

\subsection{Kronecker problem} \label{skronecker}
We begin with a special case that arises in the context of the
Kronecker problem \cite{GCT4} when $H=GL(V)\times GL(W)$ and
$X=V\otimes W$, with the natural $H$-action. 
The article \cite{GCT4} gives 
a formula for the column or row expansion of the
 minor of $GL_q^H(X)$ in this special case  in terms of fundamental 
Clebsch-Gordon coefficients for the standard quantum groups $GL_q(V)$ and
$GL_q(W)$. But this formula cannot be extended for the general 
Laplace expansion since Clebsch-Gordon coefficients are not
well defined when the underlying tensor products do not have multiplicity-free
decompositions as in the fundamental case. Here we give 
a formula for general Laplace expansion of the 
minors of $GL_q^H(X)$ in this case.

We begin by recalling  that when $V=W^*$
the braided symmetric algebra $\C^H_q[X]=\C^H[W^*\otimes W]$ is isomorphic to
the matrix coordinate ring  ${\cal O}(M_q(W))$ of the standard matrix space 
$M_q(W)$ \cite{GCT4}. For this, we have:

\begin{theorem} (Kashiwara and Lusztig \cite{kashiwaraglobal,lusztigbook}) 
 The coordinate ring ${\cal O}(M_q(W))$ has an (upper) canonical basis.
\end{theorem} 

This can be  naturally and easily extended to:

\begin{theorem} 
The braided symmetric coordinate algebra $\C_q^H[X]=\C_q^H[V\otimes W]$,
$H=GL(V)\times GL(W)$,
has an (upper)  canonical basis.
\end{theorem} 

The exterior form of this result is:

\begin{theorem} 
The exterior  coordinate algebra $\wedge_q^H[V\otimes W]$,
$H=GL(V)\times GL(W)$,
also has an (upper)  canonical basis.
\end{theorem} 

Lusztig \cite{lusztigbook} has conjectured that the multiplicative and comultiplicative
structural constants of the canonical basis of ${\cal O}(M_q(W))$ 
are polynomials in $q$ and $q^{-1}$ with nonnegative integer coefficients;
i.e., belong to $\N[q,q^{-1}]$. Analogous conjecture can be 
made for $\wedge_q^H[V\otimes W]$. Specifically, it can be conjectured 
that for any canonical basis elements $b$ and $b'$ in $\wedge_q^H[V\otimes W]$: 
\begin{equation} \label{eqmult}
b b'= \sum_{b''} \epsilon(b,b',b'') c_{b,b'}^{b''} b'', 
\end{equation} 
where the sign $\epsilon(b,b',b'')$ is $1$ or $-1$ and the coefficient
$c_{b,b'}^{b''} \in \N[q,q^{-1}]$. 
And conversely, any $b''\in \wedge_q^{H,r''}[V\otimes W]$ can be expressed as:

\begin{equation} \label{eqcomult}
b''= \sum_{b,b'} \epsilon'(b,b',b'') d^{b,b'}_{b''} b b', 
\end{equation}
where $b$ and $b'$ run over elements of 
$\wedge_q^{H,r}[V\times W]$ and $\wedge_q^{H,r'}[V\times W]$ respectively
with  $r''=r+r'$,
the sign $\epsilon'(b,b',b'')$ is $1$ or $-1$,
and $d^{b,b'}_{b''} \in \N[q,q^{-1}]$. 
To prove nonnegativity of the coefficients of $c_{b,b'}^{b''}$ and 
$d^{b,b'}_{b''}$, one needs interpretations for them
in terms of perverse sheaves \cite{beilinson} 
 in the spirit of Kazhdan-Lusztig 
\cite{kazhdan1} and Lusztig \cite{lusztigcanonical}. 

We  now define the (left  or right) 
minors of $G_q^H$ with respect to the canonical basis
$\wedge_q^{H,r}[V\otimes W]$
 to be the matrix coefficients of the (left or right)
coaction of ${\cal O}(M_q^H(V\otimes W))$. We shall call them 
(left or right) {\em canonical minors}. Then:

\begin{theorem} \label{tlaplace}
A canonical minor of degree $r''$ of $GL_q^H(X)$, $H=GL(V)\times GL(W)$,
admits a Laplace expansion
in terms of canonical minors of degree $r$ and $r'$ with $r''=r+r'$.
The coefficients of this Laplace expansion are  quadratic forms in
the structural constants $c_{b,b'}^{b''}$ and $d^{b,b'}_{b''}$ above.
\end{theorem}

An explicit  formula for Laplace expansion here (omitted)
is similar to the one 
in Proposition 6.1 of  \cite{GCT4} with 
these structural constants in place of the Clebsch-Gordon coefficients
there (which are not well defined for general Laplace expansion). 

\subsection{General nonstandard setting} 
Now let us turn to the general case.
The conjecturally correct 
algorithm in \cite{GCT8} for constructing a canonical basis of 
${\cal O}(M_q^H(X))$ also yields, as a byproduct,  conjectural canonical
bases of $\wedge_q^H[X]$ and  $\C_q^H[X]$ as implicitly sought in
\cite{berenstein}.
We  define the (left or right) minors of 
$G_q^H$ in general to be the 
matrix coefficients of the (left or right) coaction
of ${\cal O}(M_q^H(X))$ in this canonical basis of $\wedge_q^H[X]$.
We call these {\em nonstandard canonical minors}, or simply {\em nonstandard
minors}.

One can define 
structural constants $c_{b,b'}^{b''}$ and $d^{b,b'}_{b''}$ 
analogous to the ones in  (\ref{eqmult}) 
and (\ref{eqcomult})  in this case. With this:

\begin{theorem} \label{tlaplace2}
Analogue of Theorem~\ref{tlaplace}  holds in general.
\end{theorem}

Laplace expansion in the standard setting is 
used as a straightening relation to construct standard monomial bases 
of the coordinate ring and irreducible representations 
of $GL_q(X)$. In this sense, Laplace expansion is a mother relation
that governs the representation theory of the standard quantum group.
Similarly, the nonstandard Laplace expansions in Theorems~\ref{tlaplace} and 
\ref{tlaplace2} 
are  expected 
to be  mother relations governing the representation theory of the 
nonstandard quantum group $G_q^H$. In particular, an explicit interpretation
for the structural coefficients $c_{b,b'}^{b''}$ and $d^{b,b'}_{b''}$ akin
to the ones based on perverse sheaves for the
Kazhdan-Lusztig polynomials \cite{kazhdan1} 
 and the multiplicative structural constants
of the canonical basis for the standard quantum group \cite{lusztigbook}
is necessary to get fully explicit formulae for the nonstandard 
minors, and hence, for
constructing explicit bases for the irreducible polynomial
representations and the matrix coordinate ring of $G_q^H$.
In particular, this seems necessary  for 
proving  correctness of the algorithms in \cite{GCT8} 
for constructing
nonstandard canonical bases for the polynomial representations and
the matrix coordinate ring of $G_q^H$. 
This also seems necessary for finding an explicit presentation of
the nonstandard universal enveloping algebra $U_q^H({\cal G})$ in
the spirit of the explicit presentation of the Drinfeld-Jimbo enveloping 
algebra. 
Specifically, we expect the coefficients occuring in such an explicit 
presentation to depend on the structural constants such as
$c_{b,b'}^{b''}$ and $d^{b,b'}_{b''}$ above.

\section{Nonstandard quantized algebra} \label{sbr}
We now construct a nonstandard quantization $B^H_r(q)$
of the symmetric group ring $\C[S_r]$ which conjecturally has the
same relationship with $G^H_q$ that the Hecke algebra ${\cal H}_r(q)$, the 
standard deformation of $\C[S_r]$,  has with the
standard quantum group. For the sake of simplicity, we assume that
$H=GL(V)$.

Choose a standard embedding of $X=V_\mu(H)$  in $V^{\otimes d}$, where 
$d$ is the size of the partition $\mu$. That is, choose a Young symmetrizer
$c_\mu \in \C[S_r]$ such that $V^{\otimes d} \cdot c_\mu$, the image of
$V^{\otimes d}$ under the right action of $c_\mu$, is isomorphic to 
$X=V_\mu(H)$. Let 
$z_\mu \in {\cal H}_d(q)$ be the quantization of $c_\mu$ such that
$V_q^{\otimes d} \cdot z_\mu \cong X_q=V_{q,\mu}$. Here $V_q$ denotes the
quantization of $V$ and $V_{q,\mu}$ the irreducible $H_q$ module with
highest weight $\mu$. An explicit expression of $z_\mu$ may be found in
\cite{dipper}.
Let   $Z_q=V_q^{\otimes d}$. 
Let $\hat R_{Z,Z}^H$ denote the $\hat R$-matrix 
of $Z_q$ as an $H_q$-module. 
Let $r_{Z} \in {\cal H}_{2 d}(q)$, $1\le i <r$, be the element whose right
action on $Z_q \otimes Z_q=V_q^{\otimes 2 d}$
coincides with the action of $\hat R_{Z,Z}^H$. One can 
easily write down an explicit expression for $r_{Z}$ in terms of the
generators of ${\cal H}_{2 d}(q)$.

Now consider the right
action of ${\cal H}_s(q)$, $s=d r$,  on $Z_q^{\otimes r}=
V_q^{\otimes s}$,  which commutes with the left action of 
$H_q=GL_q(V)$. 
Let $r_{Z,i} \in {\cal H}_s(q)$, $1\le i <r$, be the element whose right
action on $Z_q^{\otimes r}$
coincides with the action of  $\hat R_{Z,Z}^H$ on the product of the 
$i$-th and $(i+1)$-st factors of $Z_q^{\otimes r}$. 
Thus $r_{Z,i}$ is the image of $r_Z$ under the obvious  embedding
of ${\cal H}_{2 d} (q)$ in ${\cal H}_s(q)$ depending on $i$. One 
can thus write down an explicit expression for $r_{Z,i}$ in terms of the
generators of ${\cal H}_s(q)$. 
Let \[r_{X,i}^H=  z_{\lambda,i} \cdot  z_{\lambda,i+1} \cdot r_{Z,i},\]
 where 
$z_{\lambda,i} \in {\cal  H}_s(q)$ denotes an explicit 
element whose action on 
the $i$-th factor of $Z_q^{\otimes r}$ coincides with the action of 
$z_\lambda$ on that factor--it is the image of $z_\lambda$ under the obvious
embedding of ${\cal H}_d(q)$ in ${\cal H}_s(q)$ depending on $i$.
Then  the right action of
$r_{X,i}^H$ on $Z_q^{\otimes r}$ 
corresponds to the action of  $\hat R_{X,X}^H$ on the product of the 
$i$-th and $(i+1)$-st factors of $X_q^{\otimes d} \subseteq Z_q^{\otimes d}$.
Let $p_{X,i}^{+,H}, p_{X,i}^{-,H} \in H_s(q)$ be the 
polynomials in $r_{X,i}^H$ whose actions
 on $Z_q^{\otimes r}$ correspond to the 
actions  of the positive and negative projection operators $P^{+,H}_{X,X}$
and $P^{-,H}_{X,X}$ in eq. (\ref{eqide}) on the tensor product of the 
$i$-th and $(i+1)$-st factors of $X_q^{\otimes d} \subseteq Z_q^{\otimes d}$;
one can write down these polynomials explicitly, using the known explicit 
spectral form of $\hat r_{X,i}^H$.

We define the {\em nonstandard quantized algebra} ${\cal B}_r^H(q)$ to be
the subalgebra of ${\cal H}_s(q)$ generated by the explicit elements 
$p_{X,i}^{+,H}$, or equivalently, $p_{X,i}^{-,H}$. In general, it is 
a nonflat deformation of $\C[S_r]$. That is, its dimension can be larger 
than that of $\C[S_r]$.
It can be shown to be  semisimple. Its right action on 
$X_q^{\otimes r}$  commutes with the left action $G_q^H$ by
the defining equation (\ref{eqnonstd1}) of $G_q^H$. Conjecture~\ref{cduality}
says that
its relationship with $G_q^H$ is akin to that of ${\cal H}_r(q)$ with the 
standard quantum group $G_q=GL_q(X)$. 

The Hecke algebra has an explicit presentation in terms of explicit 
relations among its generators. It will be interesting to find an
analogous explicit presentation for ${\cal B}_r^H(q)$.
Its  complexity  would
be  much higher than that of the Hecke algebra as indicated by the
concrete computations in \cite{GCT4}. Specifically, we expect 
an explicit presentation for  ${\cal B}_r^H(q)$ with defining relations
whose    coefficients are
akin to the structural constants 
$c_{b,b'}^{b''}$ and $d^{b,b'}_{b''}$ in Section~\ref{sminors} and have 
a topological interpretation akin to the one for Kazhdan-Lusztig polynomials.
Such an explicit presentation is needed to prove correctness of the 
algorithm in \cite{GCT8} to construct a canonical basis of ${\cal B}^H_r$.

\vspace{.1in} 

\noindent {\em Remark:} We can also define a (possibly singular)
quantum group $\hat G^H_q$, instead of $G^H_q$, by substituting
$\hat R^H_{X,X}$ in place of $P^{+,H}_{X,X}$ in the defining equation
(\ref{eqnonstd1}). One can then define a deformation $\hat {\cal B}^H_q(r)$
of $\C[S_r]$ that is conjecturally paired with $\hat G^H_q$, as
$G^H_q$ is with ${\cal B}^H_q(r)$. The main results (semisimplicity, and 
$q$-analogue of the Peter-Weyl theorem) also hold for these objects.
Furthermore, variants of the algorithms in \cite{GCT8} can be conjectured
to provide canonical bases for these as well. 
However, the Poincare series of $\hat G^H_q$ is much smaller than
that of $G^H_q$, and for this and other reasons, it does not seem possible
to use these objects in the context of the plethysm problem. However,
these may be interesting intermediate quantum objects to study nevertheless.

\section{Refined reciprocity} \label{srefined}
We now describe a refinement of the reciprocity conjecture 
(Conjecture~\ref{creciprocity})
that specifies precisely how the decomposion (\ref{enduality})
 of $X_q^{\otimes r}$,
\begin{equation} \label{enduality2}
X_q^{\otimes r} = \bigoplus W_{q,\alpha} \otimes T_{q,\alpha},
\end{equation}
as a $G^H_q \times {\cal B}^H_r(q)$-bimodule, tends to the 
decomposition 
\begin{equation} \label{eduality2}
X^{\otimes r}=\bigoplus_\lambda V_\lambda \otimes S_\lambda
\end{equation}
of $X^{\otimes r}$ as a $G\times S_r$-bimodule, as $q\rightarrow 1$,
and gives an explicit realization within $X_q^{\otimes r}$ of the $q$-analogue 
$V_{q,\lambda}^H$ of $V_\lambda$ as in Conjecture~\ref{creciprocity}.
Here, as usual, $V_\lambda$ denotes the Weyl module of $G$, and
$S_\lambda$ the Specht module of $S_r$.

First, we have to define the multiplicity $m^\alpha_\lambda$
of a Specht module 
$S_\lambda$ in the specialization $T_{q,\alpha}(1)$ of $T_{q,\alpha}$ 
at $q=1$. In this context, it may be remarked that
though ${\cal B}={\cal B}^H_r(q)$ is semisimple, its specialization
${\cal B}(1)$ at $q=1$ need not  be semisimple; see 
Section~\ref{sex1} for an example. Clearly, every representation of
$S_r$ is also a representation of ${\cal B}(1)$, though not always conversely.
But it may be conjectured that every irreducible ${\cal B}(1)$-representation
is also an irreducible $S_r$-representation, i.e., a Specht module.
Fix any (maximal) composition series of $T_{q,\alpha}(1)$ as a 
${\cal B}(1)$-module. We define the multiplicity $m^\alpha_\lambda$ to be 
the number of factors in this (or any such) series that are isomorphic to 
the specht module $S_\lambda$. 

Since ${\cal B}$ is semisimple (cf. Section~\ref{sbr}), it
admits a 
Wederburn structure decomposition of the form 
\begin{equation} \label{eqwederreci}
 {\cal B} = \bigoplus U^\alpha, \quad U^\alpha= T_{q,\alpha,L} \otimes 
T_{q,\alpha,R}, 
\end{equation}
where $\alpha$ is as in (\ref{enduality2}), and
$T_{q,\alpha,L}$ and $T_{q,\alpha,R}$ 
denote the left and right irreducible ${\cal B}$-modules 
indexed by $\alpha$. We call this a {\em complete} 
Wederburn structure decomposition.
 Here we are assuming that the base field is
$\C=\C(q)$, $q$ complex. This complete 
decomposition would also hold if the base field 
is instead an appropriate algebraic extension $K$ of $\Q(q)$. 
In the standard setting of
Hecke algebras, $K=\Q(q)$ suffices.
This need not be so in the nonstandard setting. That is,
an algebraic extension of $\Q(q)$ may be actually necessary
for a complete  decomposition of the above form to hold;
see Section~\ref{sex1}
for an example. If the base field is $\Q(q)$, 
each $U^\alpha$ in the Wederburn structure decomposition 
need not be, in general, 
of the form $T_{q,\alpha,L} \otimes T_{q,\alpha,R}$ as above,
but rather it would be
isomorphic to the endomorphism ring of  $T_{q,\alpha}$ 
over the  division algebra $End_{{\cal B}}(T_{q,\alpha})$. 
One has to take  similar 
variations of the  nonstandard $q$-analogue of the
Peter-Weyl theorem (Theorem~\ref{tnonstdformal} (3))
 and the duality conjecture 
(Conjecture~\ref{cduality}) if the base field is $\Q(q)$. 
However, for the reciprocity conjecture, it is necessary to 
take the base field as  $\C(q)$, $q$ complex,
 or an algebraic extension $K$ of $\Q(q)$
as described above. We assume this in the rest of this section.
See Section~\ref{sex1} for an example wherein reciprocity fails over
$\Q(q)$.

Fix any right cell, i.e., an irreducible right ${\cal B}$-subrepresentation
within $U^\alpha$. Let us denote it by $T_{q,\alpha,R}$ again.
Fix  a maximal composition series as a ${\cal B}(1)$-module
of the specialization $T_{q,\alpha,R}(1)$
of $T_{q,\alpha,R}$ at $q=1$:
\[ \hat T_{\alpha,0} \subset \hat T_{\alpha,1} \subset \cdots \subset
\hat  T_{\alpha,l(\alpha)} = T_{q,\alpha,R}(1).\] 

Let $\{x_i\}$ denote the upper canonical basis of $X_q$ as an $H_q$-module.

\begin{conj} \label{crecirefined} {\bf (Nonstandard refined reciprocity)}

There exists a   basis $Z_\alpha$ of $T_{q,\alpha,R}$ for each $\alpha$
with a filtration
\[ Z_{\alpha,0} \subset Z_{\alpha,1} \subset \cdots \subset 
Z_{\alpha,l(\alpha)}
 = Z_\alpha, \] 
such that:

\begin{enumerate}
\item  
The   specialization  $Z_{\alpha,i}(1)$ of $Z_{\alpha,i}$ at $q=1$
is a basis of $\hat T_{\alpha,i}$.

\item Let $z_{\alpha,i}^j$ denote the basis elements in
 $Z_{\alpha,i} \setminus Z_{\alpha,i-1}$.
Let $\lambda_{\alpha,i}$ be the partition such that
$\hat T_{\alpha,i}/\hat T_{\alpha,i-1}\cong S_{\lambda_{\alpha,i}}$ as 
a ${\cal B}(1)$-module (or equivalently as an $S_r$-module).
For any $\alpha,i$, define the left 
$G^H_q$-module  $W_{q,\alpha,i}=\cup_j X_q^{\otimes r} \cdot z_{\alpha,i}^j$.
By the duality conjecture (Conjecture~\ref{cduality}), 
\begin{equation}
W_{q,\alpha,i}
\subseteq W_{q,\alpha} \otimes T_{q,\alpha} \subseteq X_q^{\otimes r}.
\end{equation} 
We define its specialization $W_{1,\alpha,i}$ at $q=1$, also
denoted by $W_{q,\alpha,i}(1)$,  as follows.
Let $a(\alpha,i)$ be the largest nonnegative integer such
that the limit vector
\[ \mbox{lim}_{q\rightarrow 1} x_{i_1}\otimes \cdots \otimes x_{i_r}.z_{\alpha,i}^j/(q-1)^{a(\alpha,i)},\]
is well defined  for any $i_1,\ldots, i_r$ and $j$. 
We define  $W_{1,\alpha,i}$ to be  the span of such limits at $q=1$. 
Then, $W_{1,\alpha,i}$ is  a left $G$-module   contained within 
the  component $V_{\lambda_{\alpha,i}} \otimes S_{\lambda_{\alpha,i}} 
\subseteq X^{\otimes r}$ in (\ref{eduality2}).

\item For any fixed partition $\lambda$, 
\begin{equation} 
 \bigoplus_\alpha \bigoplus_i
W_{1,\alpha,i} = V_\lambda \otimes S_\lambda \subseteq X^{\otimes r},
\end{equation} 
where, for a given $\alpha$, $i$ ranges over all indices such 
that $\lambda_{\alpha,i}=\lambda$. 
\end{enumerate}
\end{conj} 

Furthermore, it may be conjectured that the canonical basis of $T_{q,\alpha,R}$
in terms of the ${\cal P}$-monomials as defined in \cite{GCT8}
has this property--this would make everything in the conjecture above
explicit.

The refined reciprocity conjecture basically
says that there is no information loss in
the nonstandard setting despite the lack of flatness. In fact, it can
be thought of as a variant of flatness.

\section{Evidence for duality and reciprocity} \label{sexpi}
Here we describe some concrete computations carried out in MATLAB/Maple
that support duality and reciprocity conjectures.

\noindent {\em Notation:} We denote the $q$-Weyl module
of $G_q$ for a partition $\lambda$ 
by $V_{q,\lambda}(G_q)$. We denote $V_{q,\lambda}(GL_q(\C^n))$
by $V_{q,\lambda}(n)$.

\subsection{Example 1} \label{sex1}
Let $H=GL(\C^2)$, ${\cal H}=gl(\C^2)$,  $X=V_{(3)}(H)$ is its four dimensional 
irreducible representation, and $G=GL(X)=GL(\C^4)$. 
Then $H_q=GL_q(\C^2)$, $G_q=GL_q(\C^4)$, and ${\cal H}_q=gl_q(\C^2)$.
We shall verify duality and reciprocity in this case for $r=3$.
This example is interesting because, as shown
in \cite{berenstein}, the degree three component 
$\wedge^{H,3}_q[X]$ of the braided exterior algebra vanishes in this case.
We expect that the results in this section can be extended to
any irreducible representation $X$ of $H$. But we shall confine ourselves
to the case $\dim(X)=4$, since this seems to be  the gist.

Let 
$\hat R=\hat R_{X,X}^H$ be the $\hat R$-matrix associated with $X_q$.
Let $P=P^H_{X,X}$ and $Q=Q^H_{X,X}$ be the projections on the 
eigenspaces in $X_q\otimes X_q$ for the positive  and negative eigenvalues of 
$\hat R^H_{X,X}$, respectively.
 Let $x_i= f^i x_0$, where $f$ is the usual operator
in ${\cal H}_q$, and $x_0$ is the highest weight vector in $X_q$. 
Matrices of 
$\hat R, P$ and $Q$ in the basis ${x_i\otimes x_j}$ of $X_q\otimes X_q$ 
can be  calculated from the known  explicit formulae; cf.
\cite{kassel,klimyk}.
The eigenvalues of 
$\hat R$ turn out to be   $q^{9/2}, -q^{-3/2}, q^{-11/2}$ and 
$-q^{-15/2}$. 
Explicit matrix  of  $P$
 in the basis $x_i\otimes x_j$ of $X_q\otimes X_q$ is given by

\begin{equation}\label{eqscaling}
P=\f 1 f {\cal P},
\end{equation}
where 
\begin{equation} \label{eqscalingfac} 
f = (q^4+1)(q^4-q^2+1)(q^2+1)/q^5
\end{equation}  and 
the matrix of ${\cal P}$ is as
specified  in Figure~\ref{fcalp} with the following sparse representation:
the entry $(j,v)$ in the $i$-row in Figure~\ref{fcalp} means 
${\cal P}(i,j)=v$. Thus the entry $(5,(q^4+1)/q^2)$ in the second row there
means ${\cal P}(2,5)=(q^4+1)/q^2$. The entries of ${\cal P}$-matrix  not shown
in Figure~\ref{fcalp} are all zero.
The scaling factor $f$ here is chosen so that the entries of
${\cal P}$-matrix  are polynomials in $q$ and $q^{-1}$. 
Explicit matrix of 
\begin{equation} \label{eqscalingq} 
 {\cal Q}=f Q
\end{equation}  is similar.

\subsubsection{Explicit presentation of ${\cal B}$} \label{sexppre}
Let ${\cal P}_1$ and ${\cal P}_2$ denote the ${\cal P}$ operators 
on the first two and the last two factors $X^{\otimes 3}$, respectively;
${\cal Q}_1$ and ${\cal Q}_2$ are defined similarly.
We have the trivial relations:
\[ {\cal Q}_i^2 = f {\cal Q}_i, \quad \mbox{and } {\cal P}_i^2 = f {\cal P}_i.
\]
The first nontrivial basic  relation among  ${\cal Q}_i$'s, as determined
with the help of a computer,  is:
\begin{equation} \label{eqasigmaq}
 \sum_\sigma a_\sigma {\cal Q}_\sigma = 0,
\end{equation}
where $\sigma$ ranges over the various strings of $1$'s and $2$'s 
as shown in Figure~\ref{fqreln},
 $a_\sigma \in \Q[q,q^{-1}]$ are as specified there,
and, for a string $\sigma=i_1 i_2 \cdots$, ${\cal Q}_\sigma$ 
denotes the monomial ${\cal Q}_{i_1} {\cal Q}_{i_2} \cdots$. 
The second   relation is obtained from this
 by simply interchaning ${\cal Q}_1$ and ${\cal Q}_2$. 
Simialrly, the first nontrivial basic relation among ${\cal P}_i$'s is 
\begin{equation}\label{eqbsigmap}
 \sum_\sigma b_\sigma {\cal P}_\sigma =0,
\end{equation}
where $\sigma$ ranges over  strings of $1$'s and $2$'s as in 
Figures~\ref{fpreln1}-\ref{fpreln2},
$b_\sigma$'s are as shown there, and ${\cal P}_\sigma$ is defined 
similarly.
The second  relation is obtained from this by simply interchanging 
${\cal P}_1$ and ${\cal P}_2$. 
All  coefficients in Figures~\ref{fqreln}-\ref{fpreln2}
as well as other figures in
this section are shown in factored forms, i.e., as products of
irreducible polynomials. One may ask if these coefficients have
a nice interpretation; we shall turn to this question in Section~\ref{sposi}.

Let ${\cal B}={\cal B}^H_3(q)$ be the nonstandard algebra in this case,
as defined in Section~\ref{sbr}.
It is isomorphic to the algebra   generated by ${\cal P}_i$'s 
subject to the two basic nontrivial 
relations among ${\cal P}_i$'s described above and
the trivial relations ${\cal P}_i^2 = f {\cal P}_i$,  or equivalently, to
the algebra generated by ${\cal Q}_i$'s subject to the two basic nontrivial
relations among ${\cal Q}_i$'s described above, and the trivial  relations
${\cal Q}_i^2 = f {\cal Q}_i$.

It is clear from these basic defining relations that $\{{\cal P}_\sigma\}$ or
$\{{\cal Q}_\sigma \}$, where $\sigma$ 
ranges over all strings of $1$'s and $2$'s of length at most $10$
without consecutive $1$'s or $2$'s, is  a basis of ${\cal B}$. Its 
dimension is $21$.

\subsubsection{Wederburn structure  decomposition}
Unlike for the Hecke algebras, for 
the complete 
Wederburn structure decomposition as in (\ref{eqwederreci}) to hold for
 ${\cal B}$, the base field has to contain 
the algebraic extension $K$ of $\Q(q)$ defined  as follows.
Let 
\begin{equation}
 disc=\left( 5\,{q}^{16}+8\,{q}^{12}-4\,{q}^{10}+18\,{q}^{8}-4\,{q}^{6}+8\,{q}^{4}+5 \right)  \left( {q}^{8}+1 \right) ^{2}{q}^{24},
\end{equation}
and
\[ x=disc^{1/2}.\] 
Since $disc$ is not a square, $x$ does not belong to $\Q(q)$.
Let $K=\Q(q)[x]$ be the algebraic extension of $\Q(q)$ obtained
by adjoining $x$. We assume that ${\cal B}$ is defined over this base
field.
It was found by computer that
${\cal B}$ has one one-dimensional irreducible representation $T_0$, and
five two-dimensional irreducible representations $T_i$, $1 \le i \le 5$, 
with a complete
Wederburn structure decomposition 
\begin{equation} 
{\cal B} =\bigoplus_i U_i, \quad U_i = T_{i,L} \otimes T_{i,R},
\end{equation} 
where 
the basis elements  of the various ${\cal B} \otimes
{\cal B}$-bimodules $U_i$ and the explicit representation matrices  of the
irreducible ${\cal B}$-representations $T_i$ are as follows.

Let $U_0$ be the $K$-span of $u_0 \in {\cal B}$,
where $u_0$ is as specified in Figures~\ref{fu01}-\ref{fu02}.
The coefficients in these and the following figures are in 
the basis $\{{\cal Q}_\sigma\}$. Let 
$U_i$, $1 \le i \le 5$, be the
$K$-span of the entries $u_i^1,u_i^{12},u_i^{21},u_i^2 \in {\cal B}$ 
of the matrix 
\[ u_i= \left[ \begin{array}{cc} u_i^1 &  u_i^{1 2} \\ u_i^{2 1} & u_i^2 \end{array} \right], \] 
where $u_1^1$ is as specified in 
Figure~\ref{fu1},
$u_2^1$ the element obtained from $u_1^1$ by substituting $-x$ for $x$,
and $u_3^1,u_4^1,u_5^1$ as specified in Figures~\ref{fu3}-\ref{fu5}.
Let $u_i^2$, $1\le i \le 5$, be the element obtained from $u_i^1$ 
by interchanging ${\cal Q}_1$ and ${\cal Q}_2$. 
Let $u_i^{12}=u_i^1 {\cal Q}_2$, and $u_i^{2 1}={\cal Q}_2 u_i^1$, for 
$1 \le i \le 5$.

Then it can be shown  that 
each $U_i$ has a left and right action of ${\cal B}$, and 
as a ${\cal B}\otimes {\cal B}$-bimodule
\begin{equation} 
{\cal B} = \bigoplus_i U_i.
\end{equation}
The columns of $u_i$ correspond to the  left cells and
the rows to right cells; i.e., the span of each column (row) is 
a left (resp. right) ${\cal B}$-module, which we shall denote by 
$T_{i,L}$ (resp. $T_{i,R}$). Thus,

\begin{equation} 
{\cal B} = \bigoplus_i T_{i,L} \otimes T_{i,R}.
\end{equation}

Here $T_0$, the span of $u_0$,  is the trivial one dimensional representation 
of ${\cal B}$, since it can be verified that:
\[ {\cal Q}_j u_0 = 0, \quad \mbox{for} \ j=1,2.
\]
The representation matrices $M_i^1$ and $M_i^2$  of ${\cal Q}_1$ and
 ${\cal Q}_2$
in the basis $\{u_i^1,u_i^{21}\}$ of $T_{i,L}$, $1 \le i \le 5$,  are as 
follows: 
\[ 
M_i^1=\left[ \begin{array}{cc} 0&1 \\ 0& f   \end{array} \right],
\]
where $f$ is the scaling factor in (\ref{eqscaling}),
\[ 
M_i^2=\left[ \begin{array}{cc} f&g_i \\ 0& 0   \end{array} \right],
\]
where $g_i$ are as shown in Figure~\ref{fgi}; $g_2$ is obtained from
$g_1$ by substituting $-x$ for $x$.

Let $T_i(1)$ denote the specialization of $T_i$ at $q=1$. 
It is a representation of ${\cal B}(1)$, the specialization of
${\cal B}$ at $q=1$.
Then $T_0(1)$ corresponds to the trivial one-dimensional
representation of $S_3$.
There is no one dimensional representation of ${\cal B}$
that specializes to the alternating (signed) one dimensional
representation of $S_3$. This implies that the degree three component
$\wedge^{H,3}_q[X]$ of 
the braided exterior algebra $\wedge_q^H[X]$ in this case 
is zero--as was already observed and 
 proved by other means  in \cite{berenstein}.

At $q=1$, the values of $f=f(q)$ and $g_i=g_i(q)$ are as follows:
\[ f(1)=g_1(1)=g_3(1)=g_4(1)=g_5(1)=4, \mbox{\ and \ }  g_2(1)=16. \]
Hence the ${\cal B}(1)$-modules
 $T_1(1),T_3(1),T_4(2)$ and $T_5(2)$ are all isomorphic, and it can
be verified that they are isomorphic to the Specht module $S_{(2,1)}$
of the symmetric group $S_3$ for the partition $(2,1)$. 
The module  $T_2(1)$ is reducible. Because it can be verified that it contains 
an irreducible ${\cal B}(1)$-module $T_2^1(1)$ isomorphic 
to the trivial one dimensional Specht module $S_{(3)}$ of the symmetric group
$S_3$, and
the quotient $T_2^2(1)=T_2(1)/T_2^1(1)$ is isomorphic to the 
one dimensional signed representation $S_{(1, 1, 1)}$ of $S_3$. 
But $T_2(1)$ is not completely reducible as a ${\cal B}(1)$ module. That is,
$T_2(1) \not \cong T_2^1(1) \oplus T_2^2(1)$, since 
it does not contain a submodule isomorphic to $S_{(1,1,1)}$. 
Thus, though ${\cal B}$ is semisimple for generic $q$, its specialization
${\cal B}(1)$ is not semisimple.

\subsubsection{Duality} 
Pick an element $u_i$ from each $U_i$, $1\le i \le 5$; say,
$u_i=u_i^1$, and $u_0$ is as before.
For $0 \le i \le 5$, let $W_i=X_q^{\otimes 3} \cdot u_i$, which has a 
left action of the nonstandard quantum group $G^H_q$. 
These are nonisomorphic irreducible representations of $G^H_q$. 
Their explicit decompositions as $H_q$-modules,
$H_q=GL_q(\C^2)$, were determined with the help of computer. They
are as follows.

The module $W_0$ is isomorphic to the sixteen dimensional degree three
component $\C_q^{H,3}[X]$ of the 
braided symmetric algebra 
\cite{berenstein} with the following decomposition as an $H_q$-module:
\[ W_0=V_{q,(9)}(2) \oplus V_{q,(7,2)}(2);\] 
recall that
$V_{q,\lambda}(n)$ denotes the $q$-Weyl module of $GL_q(n)$ corresponding
to the partition $\lambda$. 
This decomposition of $\C_q^{H,3}[X]$ in this case 
agrees with the one obtained  in \cite{berenstein}  by other means.

The modules $W_i$, $i>0$, are distinct 
irreducible representations 
of $G^H_q$ with the following decompositions as $H_q$-modules:
\begin{equation} 
\begin{array}{lcl} 
W_1&\cong& V_{q,(6,3)}(2), \\
W_2&\cong& V_{q,(6,3)}(2), \\
W_3&\cong& V_{q,(8,1)}(2), \\
W_4&\cong& V_{q,(5,4)}(2), \\
W_5&\cong& V_{q,(7,2)}(2).
\end{array}
\end{equation}
Their dimensions are $4,4,8,2$ and $6$, respectively. Though 
$W_1$ and $W_2$ are isomorphic as $H_q$-modules,
they  are nonisomorphic as $G^H_q$-modules; the matrix coefficients of
$W_2$ are obtained from those for $W_1$ by substituting $-x$ for $x$.

It can be  verified  that, as a $G^H_q\times {\cal B}$-bimodule,
\begin{equation}\label{eqdualityex1}
X_q^{\otimes 3} \cong \oplus_i W_i \otimes T_i,
\end{equation}
as per the  duality conjecture (Conjecture~\ref{cduality}).

\subsubsection{Reciprocity} 
Let $m^i_\mu$ denote the multiplicity of the Specht module $S_\mu$ of
the symmetric group $S_3$ in the ${\cal B}(1)$-module $T_i$. 
Then, we see that
\[
\begin{array}{l}
m^0_{(3)}=1, \\
m^1_{(2,1)}=m^3_{(2,1)}=m^4_{(2,1)}=m^5_{(2,1)}=1,   \\
m^2_{(3)}=m^2_{(1,1,1)}=1.
\end{array}
\]

Furthermore, it can be verified that the various
$G_q$-modules, $G_q=GL_q(\C^4)$, decompose as follows  when considered
 as $H_q$-modules:
\[ 
\begin{array}{lcl}
V_{q,(3)}(4)&\cong&  m^0_{(3)} W_0 \oplus m^2_{(3)} W_2,\\ 
&\cong& 
V_{q,(9)}(2) \oplus V_{q,(7,2)}(2) \oplus V_{q,(6,3)}(2), \quad \mbox{and}
\\ \\ 
V_{q,(2,1)}(4)&\cong&  m^1_{(2,1)} W_1 \oplus m^3_{(2,1)} W_3 \oplus 
m^4_{(2,1)} W_4 \oplus m^5_{(2,1)} W_5 \\
&\cong& V_{q,(6,3)}(2) \oplus V_{q,(8,1)}(2) \oplus V_{q,(5,4)}(2)
\oplus V_{q,(7,2)}(2).
\end{array}
\]
This verifies the nonstandard reciprocity conjecture 
(Conjecture~\ref{creciprocity}) 
in this case.

\subsubsection{Refined reciprocity}
Fix a right cell within $U_2$ isomorphic to the representation 
$T_{2,R}$; say, the one spanned by $u_2^1$ and $u_2^{12}$. We shall denote 
it by $T_{2,R}$ again. Let $z_0 \in T_{2,R}$ be the element such that 
$z_0 {\cal Q}_2=0$. Its coefficients are shown in Figure~\ref{fz0} in the
basis $\{Q_\sigma\}$.  Let $z_1=u_2^1$. 
Then the basis $Z=\{ z_0,z_1\}$ of $T_{2,R}$ admits a filtration
\[ Z_0=\{ z_0 \} \subseteq Z_1=\{z_0,z_1\},\] 
that yields at $q=1$ a composition series of $T_{2,R}(1)$ as 
a ${\cal B}(1)$-module:
\[ \hat T_{2,0}  \subset \hat T_{2,1} = T_{2,R}(1),
 \] 
where $\hat T_{2,0}$, spanned by the specialization $z_0(1)$ of $z_0$, 
 is the one dimensional trivial representation of
$S_3$, and $\hat T_{2,1} / \hat T_{2,0}$ is the one-dimensional
signed representation of $S_3$.

Let $W_{2,1}=X_q^{\otimes 3}\cdot z_1$ and 
$W_{2,0}=X_q^{\otimes 3}\cdot z_0$ be the  $G^H_q$-submodules of 
$X_q^{\otimes 3}$, and $W_{2,1}(1), W_{2,0}(1)$ their specializations
at $q=1$.
It can be  verified  that at $q=1$ we get:
\begin{equation}\label{eqreciveri}
 W_{2,1}(1) = \wedge^3(X) \subseteq X^{\otimes 3},
\quad \mbox{and} \quad 
W_0(1) \oplus  W_{2,0}(1) =  \sym^3(X) \subseteq X^{\otimes 3},
\end{equation}  where 
$\wedge^3(X)$ and $\sym^3(X)$ are 
the Weyl modules of $G=GL(X)$ for the 
partitions $(1,1,1)$ and $(3)$, respectively, and $W_0(1)$ the 
specialization of $W_0$ at $q=1$.

For example, Figures~\ref{fa1}-\ref{fb2}  show the nonzero coefficients of the 
elements  $a=(x_1 \otimes x_2 \otimes x_0) \cdot z_1$
and $b=(x_1 \otimes x_2 \otimes x_0) \cdot z_0$ in the monomial basis 
$\{x_i \otimes x_j \otimes x_k\}$ of $X_q^{\otimes 3}$.
It can be verified that the  specialization $a(1)$ at $q=1$ of $a$ 
 indeed  belongs to 
the subspace $\wedge^3(X) \subseteq X^{\otimes 3}$.
The specialization $b(1)$ of $b$, as it is, just vanishes, since its
coefficients are divisible by $(q-1)^2$.
But instead we consider the basis element $b'=b/(q-1)^2$ of
$W_{2,0}$. Then its specialization $b'(1)$ at $q=1$ indeed belongs to the 
subspace  $\sym^3(X)$ of $X^{\otimes 3}$.
The equation  (\ref{eqreciveri}) can be verified similarly.

Similarly it can be  verified  that 
\[ 
\lim_{q\rightarrow 1} \bigoplus_{i=1,3,4,5} 
(X_q^{\otimes 3} \cdot u_i^1 \cup
X_q^{\otimes 3} \cdot u_i^{12}) = V_{(2,1)} \otimes S_{(2,1)} \subseteq 
X^{\otimes 3}.
\]

This verifies the refined reciprocity conjecture in this case.
In particular, it explains what happens to the exterior 
and symmetric algebra components  here.
Specifically, though the braided exterior algebra component
$\wedge^{H,3}_q[X]=0$, 
\[W_{2,1}(1) = \wedge^{H,3}[X].\] 
Thus the $q$-deformation of
 $\wedge^H_3[X]$ has simply relocated 
itself as $W_{2,1}$ in the decomposition 
\[ X_q^{\otimes 3} = \oplus W_i \otimes T_i. \] 
Similarly, the symmetric algebra component $\C^{H,3}[X]$ splits in two parts, 
and the $q$-deformations of these parts, namely $W_0$ and $W_{2,0}$, get 
distributed in this decomposition. The situation for $V_{2,1}$ is similar.
 Thus, overall,  there is no information loss; 
the information has only been redistributed.
As per the refined reciprocity conjecture, this is a general phenomenon.

\subsubsection{Base field $\Q(q)$} \label{sbasefield}
Let us now see what happens if the base field is $\Q(q)$ instead.
The ${\cal B}$-representations $T_0,T_3,T_4,T_5$
are already defined over $\Q(q)$.
But $T_1$ and $T_2$ merge into a four dimensional ${\cal B}$-representation 
$T_{12}$ defined over $\Q(q)$. Explicitly, it can be realized within ${\cal B}$
as the linear span of the elements
\[\begin{array} {lcl}
v1&=&(u_1^1+u_2^1)/2, \\
v2&=&(u_1^1-u_2^1)/(2 x), \\
v3&=&(u_1^{21}+u_2^{21})/2, \\
v3&=&(u_1^{21}-u_2^{21})/(2 x).
\end{array}
 \]

Representation matrices 
of left multiplication by ${\cal Q}_1$ 
and ${\cal Q}_2$ in the basis $\{v_i\}$  are, respectively,
\[M^1=\left[ \begin {array}{cccc} {\frac {{q}^{10}+{q}^{6}+{q}^{4}+1}{{q}^{5}}}&0&a&1/2\,{q}^{-20}\\\noalign{\medskip}0&{\frac {{q}^{10}+{q}^{6}+{q}^{4}+1}{{q}^{5}}}&b&a\\\noalign{\medskip}0&0&0&0\\\noalign{\medskip}0&0&0&0\end {array} \right], 
\]
with 
\[
\begin{array}{lcl}
a&=& 1/2\,{\frac {3\,{q}^{16}+4\,{q}^{12}-2\,{q}^{10}+10\,{q}^{8}-2\,{q}^{6}+4\,{q}^{4}+3}{{q}^{8}}}, \\ \\
b&=&1/2\,{q}^{4} ( 5\,{q}^{32}+8\,{q}^{28}-4\,{q}^{26}+28\,{q}^{24}-4\,{q}^{22}+24\,{q}^{20}-8\,{q}^{18}+46\,{q}^{16} \\ && \quad \quad \quad  -8\,{q}^{14}+24\,{q}^{12}-4\,{q}^{10}+28\,{q}^{8}-4\,{q}^{6}+8\,{q}^{4}+5),
\end{array}
\]
and
\[
M^2=\left[ \begin {array}{cccc} 0&0&0&0\\\noalign{\medskip}0&0&0&0\\\noalign{\medskip}1&0&{\frac {{q}^{10}+{q}^{6}+{q}^{4}+1}{{q}^{5}}}&0\\\noalign{\medskip}0&1&0&{\frac {{q}^{10}+{q}^{6}+{q}^{4}+1}{{q}^{5}}}\end {array} \right].
\]

Similarly, the $G^H_q$-modules $W_0,W_3,W_4,W_5$ are already defined
over $\Q(q)$. The modules $W_1$ and $W_2$ merge into an eight-dimensional
$G^H_q$-module 
$W_{12}\cong X_q^{\otimes 3} \cdot v_i$, for any $i$--this is defined
over $\Q(q)$.
As an $H_q$-module,
\[
W_{1,2}\cong 2 \cdot V_{q,(6,3)}(2).
\]
A variant of the duality also holds. Specifically, the components 
$W_1\otimes T_1$ and $W_2 \otimes T_2$ in the decomposition 
(\ref{eqdualityex1}) of $X_q^{\otimes 3}$ 
merge into one sixteen dimensional $G^H_q \times {\cal B}$-bimodule
defined over $\Q(q)$. As a $G^H_q$ module, it is a direct sum of two
copies of $W_{12}$, and as a ${\cal B}$-module  a direct sum of four
copies of $T_{12}$. But for the reciprocity to hold, the base 
field has to be $K=\Q(q)[x]$ as before or larger. Indeed, it can be seen here
that the reciprocity conjecture fails over the base field $\Q(q)$.
This illustrates the need for base extension in general.

It may be illuminating to compare the $r=3$ case here with the 
one for the Kronecker problem treated in \cite{GCT4}. The one here
is basically a more complex version of the one in \cite{GCT4}, because
the basic defining relations here (Figures~\ref{fqreln}-\ref{fpreln2}) are
more complex versions of  the ones in \cite{GCT4}.

\subsubsection{On $r>3$ and positivity} \label{sposi}
Similar symbolic computations for $r=4$  seem beyond the reach
of desktop MATLAB/Maple. Fortunately, this case for the Kronecker
problem is within the reach, and will be treated in the next section.
The $r=4$ case,  $H=GL_2(\C)$, $X$ four dimensional, is expected
to be its  more complex version just as for $r=3$.

But it does not seem possible to progress much beyond  $r=3$ 
using the brute force computer-based approach that we are following here.
What is neeeded is an explicit presentation for ${\cal B}^H_r$ 
akin to the explicit presentation for the Hecke algebra, or the
one for $r=3$ in Section~\ref{sexppre}. That is, we need an explicit set of
generating   relations among ${\cal Q}_i$ or ${\cal P}_i$'s, each of
the form 
\begin{equation} \label{eqasigmapos}
\sum a_\sigma {\cal Q}_\sigma =0,
\end{equation}
or 
\begin{equation} \label{eqbsigmapos}
\sum b_\sigma {\cal P}_\sigma =0,
\end{equation}
where ${\cal Q}_{\sigma}$ and ${\cal P}_\sigma$,
for a string $\sigma=i_1 i_2 \cdots $ of symbols in 
$\{1,\cdots,r-1\}$, denote the monomials ${\cal Q}_{i_1} {\cal Q}_{i_2} \cdots$
and ${\cal P}_{i_1} {\cal P}_{i_2} \cdots$, respectively, and
each $a_\sigma$ and $b_\sigma$  has an explicit interpretation (formula).

The coefficients $a_\sigma$ and $b_\sigma$ in 
Figures~\ref{fqreln}-\ref{fpreln2}
for the $r=3$ case do not seem to have any obvious elementary interpretation.
Hence, in general, one can only expect nonelementary interpretations
for the coefficients $a_\sigma$ and $b_\sigma$ in 
(\ref{eqasigmapos})-(\ref{eqbsigmapos}). The following  numerical
analysis of these 
coefficients for the $r=3$ case 
 suggests that ${\cal B}^H_r$, in general,
may plausibly have an explicit presentation, the coefficients $a_\sigma$ and
$b_\sigma$ of whose generating relations have 
nonelementary interpretations in the spirit of the one for
Kazhdan-Lusztig polynomials.
By this we mean that each $a_\sigma$ 
has an explicit formula of the
form of an alternating sum
\begin{equation} \label{eqasigmaexpli}
 a_\sigma = (-1)^{d(\sigma)}  (q^{1/2}-q^{-1/2})^{d'(\sigma)}
(\sum_{j=0}^{s(\sigma)} (-1)^j a_\sigma^j), 
\end{equation}
for some nonnegative integers $d(\sigma),d'(\sigma),s(\sigma)$,
where
\begin{enumerate} 
\item  $s(\sigma)$ is small, say bounded by a polynomial of a fixed degree
in $r$ and $\dim(X)$ in the present case when $H=GL_2(\C)$, and 
in $r$, the rank of $H$  and the size of $\mu$ in the general 
plethysm problem (Problem~\ref{pplethysm}),
\item each $a_\sigma^j$ is a $-$-invariant (note that $a_\sigma$ is
$-$-invariant),  positive and  unimodal polynomial in
$q$ and $q^{-1}$; positive means each coefficient of $a_\sigma^j$ is
nonnegative, and unimodal means, if $a_\sigma^j(-k), \ldots, a_\sigma^j(k)$
 are the coefficients 
of $a_\sigma^j$, then 
\[ a_\sigma^j(-k) \le a_\sigma^j(-k+1) \le \cdots \le a_\sigma^j(-1) \le 
 a_\sigma^j(0) \le a_\sigma^j(1)
\le \cdots a_\sigma^j(k),\]
\item each $a_\sigma^j(s)$ has a topological interpretation akin to that
for Kazhdan-Lusztig polynomials, i.e., as the rank of an appropriate 
cohomology group. Then the duality $a_\sigma^j(-s)=a_\sigma^j(s)$
as per the $-$-invariance of $a_\sigma^j$ should come out as a consequence
of some form of Poincare duality and the unimodality as a consequence
of some form of Hard Lefschetz,
\end{enumerate} 
and each $b_\sigma$ 
has a similar  explicit formula of the
form 
\begin{equation} \label{eqbsigmaexpli}
b_\sigma = (-1)^{\bar d(\sigma)}  (q^{1/2}-q^{-1/2})^{\bar d'(\sigma)}
(\sum_{j=0}^{\bar s(\sigma)} (-1)^j b_\sigma^j).
\end{equation}

We shall call such an interpretation for $a_\sigma$ or $b_\sigma$, if it
exists, a {\em positive, unimodal, and  topological} interpretation.

Ideally speaking,   one would  like 
each $s(\sigma)$ and $\bar s(\sigma)$ above to be zero, but this may 
not always be possible for the reasons given below.
It is plausible that there exists 
some notion of  cohomological depth
that measures the extent of nonflatness, and which provides an
upper bound on  $s(\sigma)$ and $\bar s(\sigma)$ in such a topological
interepretation, if it exists.  For example, 
in the Kronecker problem, the braided symmetric and exterior algebras
$\C_q^H[X]$ and $\wedge^H_q[X]$ 
are flat deformations of the classical algebras $\C[X]$ and $\wedge[X]$.
In this case, one can expect an explicit presentation for ${\cal B}^H_q$
whose coefficients $a_\sigma$ and $b_\sigma$ 
have positive topological interpretation 
with $s(\sigma),\bar s(\sigma)=0$ in (\ref{eqasigmaexpli}) and 
(\ref{eqbsigmaexpli}).
 This is because $a_\sigma$ and $b_\sigma$
here are akin to
the structural constants $c_{b,b'}^{b''}$,
$d_{b''}^{b,b'}$ in Theorem~\ref{tlaplace},  which occur in the
defining Laplace relations for $G^H_q$, and which, in the Kroncker problem,
are conjecturally polynomials 
in $q$ and $q^{-1}$ with nonnegative coefficients 
for the reasons indicated there. But in general when $\C_q^H[X]$ and
$\wedge_q^H[X]$ are nonflat deformations, such cohomological depth
would not vanish, and hence  $s(\sigma)$ and $\bar s(\sigma)$ 
 may be nonzero, but still small
as indicated above.

We now turn to the analysis of the coefficients in the $r=3$ case 
mentioned above which suggests that  such an interpretation may plausibly
exist.
First let us oberve that the scaling factor $f$ in (\ref{eqscalingfac}) used in
the analysis so far is formally not the correct scaling factor.
To get the latter, we have to look at the formal expressions for 
$P$ and $Q$ in terms of $\hat R$. Since the eigenvalues 
of $\hat R$ in the present case are 
\[ q_1=q^{9/2},\quad  q_2=-q^{-3/2}, \quad q_3=q^{-11/2},\quad \mbox{and} 
\quad 
q_4=-q^{-15/2},\] 
we have
\begin{equation} \label{eqPinR}
P=\f{(\hat R-q_2)(\hat R-q_3) (\hat R-q_4)}{(q_1-q_2)(q_1-q_3)(q_1-q_4)}+
\f{(\hat R-q_1)(\hat R-q_2) (\hat R-q_4)}{(q_3-q_1)(q_3-q_2)(q_3-q_4)},
\end{equation}
and

\begin{equation}\label{eqQinR}
Q=\f{(\hat R-q_1)(\hat R-q_3) (\hat R-q_4)}{(q_2-q_1)(q_2-q_3)(q_2-q_4)}+
\f{(\hat R-q_1)(\hat R-q_2) (\hat R-q_3)}{(q_4-q1)(q_4-q_2)(q_4-q_3)}.
\end{equation}
Hence,
formally we should have defined the rescaled versions ${\cal P}$ and
${\cal Q}$ of $P$ and $Q$ by the equations
\begin{equation} \label{eqscalingP2}
{\cal P}=f_p P,
 \quad f_p=(q_1-q_2)(q_1-q_3)(q_1-q_4)(q_3-q_1)(q_3-q_2)(q_3-q_4),
\end{equation}
and 
\begin{equation} \label{eqscalingQ2}
{\cal Q}= f_q Q, \quad 
f_q=(q_2-q_1)(q_2-q_3)(q_2-q_4)(q_4-q_1)(q_4-q_3)(q_4-q_2),
\end{equation}
instead of the equations (\ref{eqscaling}) and (\ref{eqscalingq}).
The  scaling factor $f$ in (\ref{eqscalingfac})
 was the smallest factor chosen so that
the matrix coefficients of $P$ and $Q$ after rescaling become 
polynomials in $q,q^{-1}$. But this choice was dependendent on the
accidental cancellations in the numerators and denominators in
(\ref{eqPinR}) and (\ref{eqQinR}).
The choice of scaling makes no essential difference in
 Sections~\ref{sexppre}-\ref{sbasefield}.  
But it does matter in the study of
 positivity  below.

Hence, let us redefine ${\cal P}$ and ${\cal Q}$ as per 
(\ref{eqscalingP2}) and (\ref{eqscalingQ2}).
 Let us denote the coefficients of the old 
defining relations (\ref{eqasigmaq}) and (\ref{eqbsigmap})
 among ${\cal Q}_i$'s and
${\cal P}_i$'s by $a_\sigma'$ and $b_\sigma'$, and the coefficients of
the defining relations among the new ${\cal Q}_i$'s and ${\cal P}_i$'s
by $a_\sigma''$ and $b_\sigma''$. Then we have 
\[ a_\sigma''=  (\f{-(q-1)^2}{q})^{11-l(\sigma)}   \bar a_\sigma, \quad 
{\mbox with} \quad 
  \bar a_\sigma= (\hat f_q)^{11-l(\sigma)} a_\sigma',
\] and
\[
b_\sigma''=  (\f{-(q-1)^2}{q})^{11-l(\sigma)} \bar b_\sigma,
\quad \mbox{with} \quad 
\quad \bar b_\sigma= (\hat f_p)^{11-l(\sigma)} b_\sigma',
\] 
where 
$l(\sigma)$ denotes the length of $\sigma$,
\[
\begin{array}{lcl}
\hat f_p&=& \f {-q}{(q-1)^2} \f{f_p}{f} \\ \\
&=& 10+8\,q+2\,{q}^{4}+12\,{q}^{-1}+18\,{q}^{-6}+6\,{q}^{2}+4\,{q}^{3}+{q}^{5}\\
&& +14\,{q}^{-2}+18\,{q}^{-4}+16\,{q}^{-12}+16\,{q}^{-8}+{q}^{-27}+16\,{q}^{-11}\\
&&  +16\,{q}^{-10}+17\,{q}^{-7}+6\,{q}^{-24}+18\,{q}^{-5}+2\,{q}^{-26}+16\,{q}^{-3} \\ && +16\,{q}^{-9}+10\,{q}^{-22}+14\,{q}^{-20}+4\,{q}^{-25}+8\,{q}^{-23}+12\,{q}^{-21} +18\,{q}^{-18} \\ &&  +16\,{q}^{-19}+18\,{q}^{-16}+18\,{q}^{-17}+17\,{q}^{-15}+16\,{q}^{-14}+16\,{q}^{-13},
\end{array}
\]
and
\[
\begin{array}{lcl}
\hat f_q&=& \f {-q}{(q-1)^2} \f{f_q}{f} \\
 \\ &=& 10\,{q}^{-12}+2\,{q}^{-8}+{q}^{-31}+8\,{q}^{-27}+8\,{q}^{-11}+6\,{q}^{-10}+{q}^{-7} \\ && +2\,{q}^{-30}+10\,{q}^{-24}+6\,{q}^{-28}+10\,{q}^{-26}+4\,{q}^{-9}+6\,{q}^{-22} \\&&  +2\,{q}^{-20}+10\,{q}^{-25}+8\,{q}^{-23}+4\,{q}^{-29}+4\,{q}^{-21}+2\,{q}^{-18} \\ && +2\,{q}^{-19}+6\,{q}^{-16}+4\,{q}^{-17}+8\,{q}^{-15}+10\,{q}^{-14}+10\,{q}^{-13}.
\end{array}
\]
Both $f_p$ and $f_q$ are positive polynomials.

Let us define 
\begin{equation} 
\begin{array}{lcll}
\hat a_\sigma &=& \hat f_q^2 a_\sigma', & \mbox{for} \quad \sigma=121212121, \\
&=& \hat f_q a_\sigma', & \mbox{otherwise},
\end{array}
\end{equation} 
and 
\begin{equation} 
\begin{array}{lcll}
\hat b_\sigma &=&  b_\sigma', & \mbox{for} \quad \sigma=\emptyset, \mbox{and}\ 2, \\
&=& \hat f_p b_\sigma', & \mbox{otherwise}.
\end{array}
\end{equation} 
Since $f_p$ and $f_q$ are positive, the positivity properties of
$\hat a_\sigma$ and $\bar a_\sigma$ (also $\hat b_\sigma$
and $\bar b_\sigma$) are similar;  it turns out that 
the unimodularity properties are also 
similar. Hence we shall focus on $\hat a_\sigma$ and $\hat b_\sigma$ in
what follows.
Since $\hat a_\sigma$ is $-$-invariant, it is of the form
$\hat a_\sigma(0) + \sum_{t>0}  \hat a_\sigma(t) (q^t+q^{-t})$. 
Let $\hat A_\sigma$ be the vector $[a_\sigma(0), a_\sigma(1), \ldots]$;
the vector $\hat B_\sigma$ is defined similarly. 
Figure~\ref{fposq} shows $\hat A_\sigma$ for the various $\sigma$ in 
Figure~\ref{fqreln};
the vector for each $\sigma$ is obtained by concatenating the rows in
front of that $\sigma$. Figures~\ref{fposp1}-\ref{fposp3}
similarly show $\hat B_\sigma$
for the various $\sigma$ in Figures~\ref{fpreln1}-\ref{fpreln2};
 only the distinct 
$\hat B_\sigma$'s are shown. It may be seen 
the $\hat A_\sigma$'s are positive and nonincreasing.
Thus all $a_\sigma$ are positive and unimodal,
and hence, of the form (\ref{eqasigmaexpli}) with   $s(\sigma)=0$.
All $\hat B_\sigma$'s are positive and nonincreasing, except for 
$\sigma=121,1212$ and $21212$, for which each
$\hat B_\sigma$ is  positive and unimodal except at the tail.
Thus all $b_\sigma$, for $\sigma \not =
121,1212,21212$, are positive and unimodal,
and hence of the  form (\ref{eqbsigmaexpli}) with $\bar s(\sigma)=0$. For 
$\sigma=121,1212,21212$,
$b_\sigma$ seems to be of the form (\ref{eqbsigmaexpli})
 with $\bar s(\sigma)=1$,  both $b_\sigma^0$ 
and $b_\sigma^1$ being positive and unimodal,  $b_\sigma^0$ being the
dominant polynomial that  accounts for $b_\sigma$'s mostly 
positive and unimodal behaviour, and $b_\sigma^1$  the error
polynomial that accounts for the deviation at the tail.

The (co)multiplicative structural constants $c_{b,b'}^{b''}$
and $d_{b''}^{b,b'}$ for the canonical basis 
of the braided exterior algebra $\wedge_q^{H,r}[X]$, which occur in the
Laplace relations for the general nonstandard quantum group $G^H_q$ (cf.
Theorem~\ref{tlaplace2}), are akin  to the structure constants
$a_\sigma$ and $b_\sigma$ in (\ref{eqasigmapos}) and (\ref{eqbsigmapos}). 
Hence, we can
expect a similar positive topological interpretation for
$c_{b,b'}^{b''}$ and  $d_{b''}^{b,b'}$ (but not necessarily 
 unimodality since $c_{b,b'}^{b''}$ and $d_{b''}^{b,b'}$ need not be
 $-$-invariant).
The experimental evidence in \cite{GCT8} suggests that
the structure constants associated with the canonical bases of 
 the matrix coordinate ring of $G^H_q$ and the ring ${\cal B}^H_q$ defined
there may also have similar positive topological interpretations
(additionally unimodal for ${\cal B}^H_q$).

\begin{sidewaysfigure}[p!]
\[
\left[ \begin {array}{cccc} 1,{\frac { \left( {q}^{4}+1 \right)  \left( {q}^{4}-{q}^{2}+1 \right)  \left( {q}^{2}+1 \right) }{{q}^{5}}}&&&\\\noalign{\medskip}2,q \left( {q}^{4}+1 \right) &5,{\frac {{q}^{4}+1}{{q}^{2}}}&&\\\noalign{\medskip}3,{\frac { \left( {q}^{2}+1 \right)  \left( {q}^{8}-{q}^{6}+{q}^{4}-{q}^{2}+1 \right) }{{q}^{5}}}&6,{\frac { \left( q-1 \right)  \left( q+1 \right)  \left( {q}^{2}+q+1 \right)  \left( {q}^{2}-q+1 \right) }{{q}^{4}}}&9,{\frac {{q}^{2}+1}{q}}&\\\noalign{\medskip}4,{\frac {{q}^{8}-{q}^{6}+{q}^{4}+1}{{q}^{3}}}&7,{\frac { \left( q-1 \right) ^{2} \left( q+1 \right) ^{2} \left( {q}^{2}+q+1 \right)  \left( {q}^{2}-q+1 \right) }{{q}^{6}}}&10,{\frac { \left( q-1 \right)  \left( q+1 \right)  \left( {q}^{2}+q+1 \right)  \left( {q}^{2}-q+1 \right) }{{q}^{5}}}&13,{\frac {{q}^{4}+1}{{q}^{2}}}\\\noalign{\medskip}2,{\frac {{q}^{4}+1}{{q}^{2}}}&5,{\frac {{q}^{4}+1}{{q}^{5}}}&&\\\noalign{\medskip}3,{\frac { \left( {q}^{2}+1 \right) ^{2} \left( q-1 \right)  \left( q+1 \right) }{{q}^{2}}}&6,2\,{\frac {{q}^{2}+1}{q}}&9,-{\frac { \left( {q}^{2}+1 \right) ^{2} \left( q-1 \right)  \left( q+1 \right) }{{q}^{2}}}&\\\noalign{\medskip}4,{\frac { \left( q-1 \right) ^{2} \left( q+1 \right) ^{2} \left( {q}^{2}+q+1 \right)  \left( {q}^{2}-q+1 \right) }{{q}^{2}}}&7,{\frac {2\,{q}^{8}+{q}^{6}-2\,{q}^{2}+1}{{q}^{5}}}&10,-{\frac {{q}^{8}-{q}^{6}-2\,{q}^{4}-{q}^{2}+1}{{q}^{4}}}&13,-{\frac { \left( q-1 \right)  \left( q+1 \right)  \left( {q}^{2}+q+1 \right)  \left( {q}^{2}-q+1 \right) }{q}}\\\noalign{\medskip}8,{\frac { \left( {q}^{2}+1 \right)  \left( {q}^{8}-{q}^{6}+{q}^{4}-{q}^{2}+1 \right) }{{q}^{5}}}&11,{\frac { \left( {q}^{2}+1 \right) ^{2} \left( q-1 \right)  \left( q+1 \right) }{{q}^{4}}}&14,{\frac {{q}^{2}+1}{q}}&\\\noalign{\medskip}3,{\frac {{q}^{2}+1}{q}}&6,-{\frac { \left( q-1 \right)  \left( q+1 \right)  \left( {q}^{2}+q+1 \right)  \left( {q}^{2}-q+1 \right) }{{q}^{4}}}&9,{\frac { \left( {q}^{2}+1 \right)  \left( {q}^{8}-{q}^{6}+{q}^{4}-{q}^{2}+1 \right) }{{q}^{5}}}&\\\noalign{\medskip}4,{\frac { \left( q-1 \right)  \left( q+1 \right)  \left( {q}^{2}+q+1 \right)  \left( {q}^{2}-q+1 \right) }{q}}&7,-{\frac {{q}^{8}-{q}^{6}-2\,{q}^{4}-{q}^{2}+1}{{q}^{4}}}&10,{\frac {{q}^{8}-2\,{q}^{6}+{q}^{2}+2}{{q}^{3}}}&13,{\frac { \left( q-1 \right) ^{2} \left( q+1 \right) ^{2} \left( {q}^{2}+q+1 \right)  \left( {q}^{2}-q+1 \right) }{{q}^{2}}}\\\noalign{\medskip}{8,\frac { \left( q-1 \right)  \left( q+1 \right)  \left( {q}^{2}+q+1 \right)  \left( {q}^{2}-q+1 \right) }{{q}^{2}}}&11,2\,{\frac {{q}^{2}+1}{q}}&14,-{\frac { \left( q-1 \right)  \left( q+1 \right)  \left( {q}^{2}+q+1 \right)  \left( {q}^{2}-q+1 \right) }{{q}^{2}}}&\\\noalign{\medskip}12,q \left( {q}^{4}+1 \right) &15,{\frac {{q}^{4}+1}{{q}^{2}}}&&\\\noalign{\medskip}4,{\frac {{q}^{4}+1}{{q}^{2}}}&7,-{\frac { \left( q-1 \right)  \left( q+1 \right)  \left( {q}^{2}+q+1 \right)  \left( {q}^{2}-q+1 \right) }{{q}^{5}}}&10,{\frac { \left( q-1 \right) ^{2} \left( q+1 \right) ^{2} \left( {q}^{2}+q+1 \right)  \left( {q}^{2}-q+1 \right) }{{q}^{6}}}&13,{\frac {{q}^{8}+{q}^{4}-{q}^{2}+1}{{q}^{5}}}\\\noalign{\medskip}8,{\frac {{q}^{2}+1}{q}}&11,-{\frac { \left( {q}^{2}+1 \right) ^{2} \left( q-1 \right)  \left( q+1 \right) }{{q}^{4}}}&14,{\frac { \left( {q}^{2}+1 \right)  \left( {q}^{8}-{q}^{6}+{q}^{4}-{q}^{2}+1 \right) }{{q}^{5}}}&\\\noalign{\medskip}12,{\frac {{q}^{4}+1}{{q}^{2}}}&15,{\frac {{q}^{4}+1}{{q}^{5}}}&&\\\noalign{\medskip}16,{\frac { \left( {q}^{4}+1 \right)  \left( {q}^{4}-{q}^{2}+1 \right)  \left( {q}^{2}+1 \right) }{{q}^{5}}}&&&\end {array} \right] 
\]
\caption{${\cal P}$-matrix}
\label{fcalp}
\end{sidewaysfigure}

\begin{sidewaysfigure}[p!]
\[\begin{array}{|c|c|}\hline
\sigma & a_\sigma \\ \hline 
1&-({q}^{12}+{q}^{10}+2\,{q}^{8}+2\,{q}^{4}+{q}^{2}+1 ) ^{2} ( {q}^{8}-{q}^{6}+{q}^{4}-{q}^{2}+1 ) ^{2} ( {q}^{2}+1 ) ^{4} ( {q}^{4}-{q}^{2}+1 ) ^{4} ( {q}^{4}+1 ) ^{2}/q^{36}\\ \hline\noalign{\medskip}{121}& ( 1+7\,{q}^{4}+{q}^{2}+5\,{q}^{6}+18\,{q}^{8}+21\,{q}^{42}-107\,{q}^{20}+{q}^{50}-107\,{q}^{32}+73\,{q}^{14}+187\,{q}^{18}-14\,{q}^{16}+402\,{q}^{26} \\ & \quad -197\,{q}^{28}+20\,{q}^{40}+187\,{q}^{34}+ 73\,{q}^{38}-197\,{q}^{24}  +328\,{q}^{30}+{q}^{52}+328\,{q}^{22}+7\,{q}^{48}+5\,{q}^{46}+18\,{q}^{44}  \\ & \quad  +20\,{q}^{12}+21\,{q}^{10}-14\,{q}^{36} )  ( {q}^{2}+1 ) ^{2} ( {q}^{4}-{q}^{2}+1 ) ^{2}/q^{32}\\ \hline\noalign{\medskip}{12121}&-( 1+8\,{q}^{4}+3\,{q}^{2}+4\,{q}^{6}+33\,{q}^{8}+12\,{q}^{42}+80\,{q}^{20}+3\,{q}^{50}+80\,{q}^{32}+27\,{q}^{14}+113\,{q}^{18}+115\,{q}^{16}+360\,{q}^{26}  \\ & \quad  -9\,{q}^{28}+76\,{q}^{40}+113\,{q}^{34}  +27\,{q}^{38}-9\,{q}^{24} +253\,{q}^{30}+{q}^{52}+253\,{q}^{22}+8\,{q}^{48}+4\,{q}^{46}+33\,{q}^{44}+76\,{q}^{12}  \\ & \quad  +12\,{q}^{10}+115\,{q}^{36})/q^{26}\\ \hline\noalign{\medskip}{1212121}&( 3\,{q}^{36}+2\,{q}^{34}+8\,{q}^{32}+5\,{q}^{30}+17\,{q}^{28}+30\,{q}^{26}+11\,{q}^{24} \\ & \quad +61\,{q}^{22}-15\,{q}^{20}+108\,{q}^{18}-15\,{q}^{16}+61\,{q}^{14}  +11\,{q}^{12}+30\,{q}^{10}+17\,{q}^{8}+5\,{q}^{6}+8\,{q}^{4}+2\,{q}^{2}+3 )/q^{18}\\ \hline\noalign{\medskip}{121212121}&-( {q}^{20}+3\,{q}^{18}+{q}^{16}+5\,{q}^{14}-2\,{q}^{12}+16\,{q}^{10}-2\,{q}^{8}+5\,{q}^{6}+{q}^{4}+3\,{q}^{2}+1 )/q^{10}\\ \hline 
\noalign{\medskip}{12121212121}&1 \\ \hline \end {array} 
\]
\caption{Coefficients of the basic generating relation among ${\cal Q}_i$'s}
\label{fqreln}
\end{sidewaysfigure} 

\begin{sidewaysfigure}
\[\begin{array}{|c|c|}\hline
\sigma & b_\sigma \\ \hline 
\emptyset&-( {q}^{2}+1 ) ^{5} ( {q}^{4}+1 ) ^{3} ( {q}^{4}-{q}^{2}+1 ) ^{6} ( {q}^{4}+{q}^{3}+{q}^{2}+q+1 )  ( {q}^{4}-{q}^{3}+{q}^{2}-q+1 )  ( {q}^{6}-{q}^{3}+1 ) \\ & \quad \times    ( {q}^{6}+{q}^{3}+1 )   ( {q}^{6}-{q}^{5}+{q}^{4}-{q}^{3}+{q}^{2}-q+1 ) ( {q}^{6}+{q}^{5}+{q}^{4}+{q}^{3}+{q}^{2}+q+1 )  ( 2\,{q}^{8}-2\,{q}^{6}+3\,{q}^{4}-2\,{q}^{2}+2 ) 
\\ & \quad \times   ( {q}^{2}+q+1 ) ^{2} ( {q}^{2}-q+1 ) ^{2} ( q-1 ) ^{4}
 ( q+1 ) ^{4}/{{q}^{51}}\\ \hline\noalign{\medskip}2&( {q}^{2}+1 ) ^{4} ( {q}^{4}-{q}^{2}+1 ) ^{5} ( {q}^{4}+1 ) ^{2} ( {q}^{4}+{q}^{3}+{q}^{2}+q+1 )  ( {q}^{4}-{q}^{3}+{q}^{2}-q+1 ) 
\\ & \quad \times   ( {q}^{6}-{q}^{3}+1 )  ( {q}^{6}+{q}^{3}+1 ) 
  ( {q}^{6}-{q}^{5}+{q}^{4}-{q}^{3}+{q}^{2}-q+1 )  ( {q}^{6}+{q}^{5}+{q}^{4}+{q}^{3}+{q}^{2}+q+1 ) \\ & \quad \times ( 2\,{q}^{8}-2\,{q}^{6}+3\,{q}^{4}-2\,{q}^{2}+2 )  ( {q}^{2}+q+1 ) ^{2} ( {q}^{2}-q+1 ) ^{2} ( q-1 ) ^{4} ( q+1 ) ^{4}/{{q}^{46}}\\ \hline\noalign{\medskip}1&( {q}^{2}+1 ) ^{4} ( {q}^{4}-{q}^{2}+1 ) ^{4} ( {q}^{4}+1 ) ^{2} ( 2+7\,{q}^{4}-4\,{q}^{2}-9\,{q}^{6}+11\,{q}^{8}-10\,{q}^{42}-4\,{q}^{58}+7\,{q}^{56}-9\,{q}^{54} \\ & \quad  +12\,{q}^{20}-12\,{q}^{50}-11\,{q}^{32}-12\,{q}^{14}-10\,{q}^{18}+13\,{q}^{16}+16\,{q}^{26}  -11\,{q}^{28}+12\,{q}^{40}+16\,{q}^{34}+{q}^{38}+{q}^{24}+28\,{q}^{30} \\ & \quad  +11\,{q}^{52}+{q}^{22}+13\,{q}^{48}-12\,{q}^{46}+2\,{q}^{60}+13\,{q}^{44}+13\,{q}^{12}-12\,{q}^{10}+{q}^{36} )/{{q}^{46}}\\ \hline\noalign{\medskip}12&-( {q}^{2}+1 ) ^{3} ( {q}^{4}+1 )  ( {q}^{4}-{q}^{2}+1 ) ^{3} ( 2+7\,{q}^{4}-4\,{q}^{2}-9\,{q}^{6}+11\,{q}^{8}-10\,{q}^{42}-4\,{q}^{58}+7\,{q}^{56}-9\,{q}^{54}+12\,{q}^{20}
\\ & \quad   -12\,{q}^{50}-11\,{q}^{32}-12\,{q}^{14}  -10\,{q}^{18}+13\,{q}^{16}+16\,{q}^{26}-11\,{q}^{28}+12\,{q}^{40}+16\,{q}^{34}+{q}^{38}+{q}^{24}+28\,{q}^{30}+11\,{q}^{52}  \\ & \quad +{q}^{22}+13\,{q}^{48}-12\,{q}^{46}+2\,{q}^{60}+13\,{q}^{44}+13\,{q}^{12}-12\,{q}^{10}+{q}^{36} )/{{q}^{41}}\\ \hline\noalign{\medskip}21&-( {q}^{2}+1 ) ^{3} ( {q}^{4}+1 )  ( {q}^{4}-{q}^{2}+1 ) ^{3} ( 2+7\,{q}^{4}-4\,{q}^{2}-9\,{q}^{6}+11\,{q}^{8}-10\,{q}^{42}-4\,{q}^{58}+7\,{q}^{56}-9\,{q}^{54}  \\ & \quad +12\,{q}^{20}-12\,{q}^{50}-11\,{q}^{32}  -12\,{q}^{14}-10\,{q}^{18}+13\,{q}^{16}+16\,{q}^{26}-11\,{q}^{28}+12\,{q}^{40}+16\,{q}^{34}+{q}^{38}+{q}^{24}+28\,{q}^{30} \\ & \quad +11\,{q}^{52}+{q}^{22}+13\,{q}^{48}-12\,{q}^{46}+2\,{q}^{60}+13\,{q}^{44}+13\,{q}^{12}-12\,{q}^{10}+{q}^{36} )/{{q}^{41}}\\ \hline\noalign{\medskip}212&( {q}^{2}+1 ) ^{2} ( {q}^{4}-{q}^{2}+1 ) ^{2} ( 2+7\,{q}^{4}-4\,{q}^{2}-9\,{q}^{6}+11\,{q}^{8}-10\,{q}^{42}-4\,{q}^{58}+7\,{q}^{56}-9\,{q}^{54}+12\,{q}^{20} \\ & \quad -12\,{q}^{50}-11\,{q}^{32}-12\,{q}^{14}-10\,{q}^{18}+13\,{q}^{16}+16\,{q}^{26}-11\,{q}^{28}+12\,{q}^{40}+16\,{q}^{34}+{q}^{38}+{q}^{24}+28\,{q}^{30}+11\,{q}^{52}  \\ & \quad  +{q}^{22}+13\,{q}^{48}-12\,{q}^{46}+2\,{q}^{60}+13\,{q}^{44}+13\,{q}^{12}-12\,{q}^{10}+{q}^{36} )/{{q}^{36}}\\ \hline\noalign{\medskip}121&( {q}^{2}+1 ) ^{2} ( {q}^{4}+1 ) ^{2} ( {q}^{4}-{q}^{2}+1 ) ^{2} ( 2+2\,{q}^{4}-4\,{q}^{2}-2\,{q}^{6}-2\,{q}^{8}-9\,{q}^{42}+27\,{q}^{20}  \\ & \quad  -4\,{q}^{50}+27\,{q}^{32} -13\,{q}^{14}-48\,{q}^{18}
+{q}^{16}-110\,{q}^{26}+53\,{q}^{28}-3\,{q}^{40}-48\,{q}^{34}-13\,{q}^{38}+53\,{q}^{24}-77\,{q}^{30} \\ & \quad  +2\,{q}^{52}-77\,{q}^{22}+2\,{q}^{48}-2\,{q}^{46}-2\,{q}^{44}-3\,{q}^{12}-9\,{q}^{10}+{q}^{36} )/{{q}^{36}}\\ \hline\noalign{\medskip}1212&-( {q}^{2}+1 )  ( {q}^{4}+1 )  ( {q}^{4}-{q}^{2}+1 )  ( 2+2\,{q}^{4}-4\,{q}^{2}-2\,{q}^{6}-2\,{q}^{8}-9\,{q}^{42}+27\,{q}^{20}-4\,{q}^{50}+27\,{q}^{32} \\ & \quad  -13\,{q}^{14}-48\,{q}^{18}+{q}^{16}-110\,{q}^{26}+53\,{q}^{28}-3\,{q}^{40}-48\,{q}^{34}-13\,{q}^{38}+53\,{q}^{24}-77\,{q}^{30}+2\,{q}^{52}-77\,{q}^{22}  \\ & \quad  +2\,{q}^{48}-2\,{q}^{46}-2\,{q}^{44}-3\,{q}^{12}-9\,{q}^{10}+{q}^{36} )/{{q}^{31}}\\ \hline \end {array} 
\]
\caption{The first eight terms of the basic generating relation among ${\cal P}_i$'s}
\label{fpreln1}
\end{sidewaysfigure}

\begin{sidewaysfigure}
\[ \begin {array}{|c|c|}\hline
\sigma & b_\sigma \\ \hline 
 2121&-( {q}^{2}+1 )  ( {q}^{4}+1 )  ( {q}^{4}-{q}^{2}+1 )  ( 2+2\,{q}^{4}-4\,{q}^{2}-2\,{q}^{6}-2\,{q}^{8}-9\,{q}^{42}+27\,{q}^{20}-4\,{q}^{50}+27\,{q}^{32}
 \\ & \quad -13\,{q}^{14}  -48\,{q}^{18}+{q}^{16}-110\,{q}^{26}+53\,{q}^{28}  -3\,{q}^{40}-48\,{q}^{34}-13\,{q}^{38}+53\,{q}^{24}-77\,{q}^{30}+2\,{q}^{52}-77\,{q}^{22} \\ & \quad+2\,{q}^{48}-2\,{q}^{46}-2\,{q}^{44}-3\,{q}^{12}-9\,{q}^{10}+{q}^{36} )/{{q}^{31}}\\ \hline\noalign{\medskip}21212&(2+2\,{q}^{4}-4\,{q}^{2}-2\,{q}^{6}-2\,{q}^{8}-9\,{q}^{42}+27\,{q}^{20}-4\,{q}^{50}+27\,{q}^{32}-13\,{q}^{14}-48\,{q}^{18}+{q}^{16}-110\,{q}^{26} \\ & \quad +53\,{q}^{28}-3\,{q}^{40}-48\,{q}^{34}-13\,{q}^{38}+53\,{q}^{24}-77\,{q}^{30}+2\,{q}^{52}-77\,{q}^{22}+2\,{q}^{48}-2\,{q}^{46}-2\,{q}^{44}-3\,{q}^{12} \\ & \quad -9\,{q}^{10}+{q}^{36})/{{q}^{26}}\\ \hline\noalign{\medskip}12121&( {q}^{2}+1 ) ^{2} ( {q}^{4}+1 ) ^{2} ( {q}^{4}-{q}^{2}+1 ) ^{2} ( 3\,{q}^{16}+2\,{q}^{14}+14\,{q}^{8}+2\,{q}^{2}+3 )  ( {q}^{8}+{q}^{4}-{q}^{2}+1 )  ( {q}^{8}-{q}^{6}+{q}^{4}+1 )/{{q}^{26}}\\ \hline\noalign{\medskip}121212&-( {q}^{2}+1 )  ( {q}^{4}+1 )  ( {q}^{4}-{q}^{2}+1 )  ( 3\,{q}^{16}+2\,{q}^{14}+14\,{q}^{8}+2\,{q}^{2}+3 )  ( {q}^{8}+{q}^{4}-{q}^{2}+1 )  ( {q}^{8}-{q}^{6}+{q}^{4}+1 )/{{q}^{21}}\\ \hline\noalign{\medskip}212121&-( {q}^{2}+1 )  ( {q}^{4}+1 )  ( {q}^{4}-{q}^{2}+1 )  ( 3\,{q}^{16}+2\,{q}^{14}+14\,{q}^{8}+2\,{q}^{2}+3 )  ( {q}^{8}+{q}^{4}-{q}^{2}+1 )  ( {q}^{8}-{q}^{6}+{q}^{4}+1 )/{{q}^{21}}\\ \hline\noalign{\medskip}2121212&( 3\,{q}^{16}+2\,{q}^{14}+14\,{q}^{8}+2\,{q}^{2}+3 )  ( {q}^{8}+{q}^{4}-{q}^{2}+1 )  ( {q}^{8}-{q}^{6}+{q}^{4}+1 )/{{q}^{16}}\\ \hline\noalign{\medskip}1212121&-( {q}^{2}+1 ) ^{2} ( {q}^{4}+1 ) ^{2} ( {q}^{4}-{q}^{2}+1 ) ^{2} ( 3\,{q}^{16}-{q}^{14}+3\,{q}^{12}-3\,{q}^{10}+12\,{q}^{8}-3\,{q}^{6}+3\,{q}^{4}-{q}^{2}+3 )/{{q}^{18}} \\ \hline
 12121212&( {q}^{2}+1 )  ( {q}^{4}+1 )  ( {q}^{4}-{q}^{2}+1 )  ( 3\,{q}^{16}-{q}^{14}+3\,{q}^{12}-3\,{q}^{10}+12\,{q}^{8}-3\,{q}^{6}+3\,{q}^{4}-{q}^{2}+3 )/{{q}^{13}}\\ \hline\noalign{\medskip}21212121&( {q}^{2}+1 )  ( {q}^{4}+1 )  ( {q}^{4}-{q}^{2}+1 )  ( 3\,{q}^{16}-{q}^{14}+3\,{q}^{12}-3\,{q}^{10}+12\,{q}^{8}-3\,{q}^{6}+3\,{q}^{4}-{q}^{2}+3 )/{{q}^{13}}\\ \hline\noalign{\medskip}212121212&-(3\,{q}^{16}-{q}^{14}+3\,{q}^{12}-3\,{q}^{10}+12\,{q}^{8}-3\,{q}^{6}+3\,{q}^{4}-{q}^{2}+3)/{{q}^{8}}\\ \hline\noalign{\medskip}121212121&( {q}^{2}+1 ) ^{2} ( {q}^{4}+1 ) ^{2} ( {q}^{4}-{q}^{2}+1 ) ^{2}/{{q}^{10}}\\ \hline\noalign{\medskip}1212121212&-( {q}^{2}+1 )  ( {q}^{4}+1 )  ( {q}^{4}-{q}^{2}+1 )/{{q}^{5}}\\ \hline\noalign{\medskip}2121212121&-( {q}^{2}+1 )  ( {q}^{4}+1 )  ( {q}^{4}-{q}^{2}+1 )/{{q}^{5}}\\ \hline\noalign{\medskip}21212121212&1 \\ \hline \end {array}
\]
\caption{The last fourteen  terms of the basic generating relation among ${\cal P}_i$'s}
\label{fpreln2}
\end{sidewaysfigure}

\begin{sidewaysfigure}
\[
\begin {array}{|c|c|}\hline
\sigma & Coefficient \\ \hline 
\emptyset &( {q}^{4}+{q}^{3}+{q}^{2}+q+1 )  ( {q}^{4}-{q}^{3}+{q}^{2}-q+1 )  ( {q}^{6}-{q}^{3}+1 )  ( {q}^{6}+{q}^{3}+1 )  ( {q}^{6}-{q}^{5}+{q}^{4}-{q}^{3}+{q}^{2}-q+1 ) \\ & \quad \times  ( {q}^{6}+{q}^{5}+{q}^{4}+{q}^{3}+{q}^{2}+q+1 )  ( 2\,{q}^{8}-2\,{q}^{6}+3\,{q}^{4}-2\,{q}^{2}+2 )  ( {q}^{2}+q+1 ) ^{2} ( {q}^{2}-q+1 ) ^{2} \\ & \quad \times   ( {q}^{4}+1 ) ^{2} ( q-1 ) ^{4} ( q+1 ) ^{4} ( {q}^{2}+1 ) ^{4} ( {q}^{4}-{q}^{2}+1 ) ^{5}/{{q}^{46}}\\ \hline\noalign{\medskip}2&-( {q}^{2}+1 ) ^{3} ( {q}^{4}+1 )  ( {q}^{4}-{q}^{2}+1 ) ^{3} ( 2+13\,{q}^{12}-12\,{q}^{10}+11\,{q}^{8}+7\,{q}^{4}-4\,{q}^{2}-9\,{q}^{6}+{q}^{36}-10\,{q}^{42} \\ & \quad +12\,{q}^{20}-12\,{q}^{50}-11\,{q}^{32}-12\,{q}^{14}-10\,{q}^{18}+13\,{q}^{16}+16\,{q}^{26}-11\,{q}^{28}+12\,{q}^{40}+16\,{q}^{34}+{q}^{38}+{q}^{24}+28\,{q}^{30} \\ & \quad +11\,{q}^{52}+{q}^{22}+13\,{q}^{48}-12\,{q}^{46}+13\,{q}^{44}-4\,{q}^{58}+7\,{q}^{56}-9\,{q}^{54}+2\,{q}^{60} )/{{q}^{41}}\\ \hline\noalign{\medskip}1&-( {q}^{2}+1 ) ^{3} ( {q}^{4}+1 )  ( {q}^{4}-{q}^{2}+1 ) ^{3} ( 2+13\,{q}^{12}-12\,{q}^{10}+11\,{q}^{8}+7\,{q}^{4}-4\,{q}^{2}-9\,{q}^{6}+{q}^{36}-10\,{q}^{42} \\ & \quad +12\,{q}^{20}-12\,{q}^{50}-11\,{q}^{32}-12\,{q}^{14}-10\,{q}^{18}+13\,{q}^{16}+16\,{q}^{26}-11\,{q}^{28}+12\,{q}^{40}+16\,{q}^{34}+{q}^{38}+{q}^{24}+28\,{q}^{30} \\ & \quad +11\,{q}^{52}+{q}^{22}+13\,{q}^{48}-12\,{q}^{46}+13\,{q}^{44}-4\,{q}^{58}+7\,{q}^{56}-9\,{q}^{54}+2\,{q}^{60} )/{{q}^{41}}\\ \hline\noalign{\medskip}12&( 2+13\,{q}^{12}-12\,{q}^{10}+11\,{q}^{8}+7\,{q}^{4}-4\,{q}^{2}-9\,{q}^{6}+{q}^{36}-10\,{q}^{42}+12\,{q}^{20}-12\,{q}^{50}-11\,{q}^{32}-12\,{q}^{14}-10\,{q}^{18} \\ & \quad +13\,{q}^{16}+16\,{q}^{26}-11\,{q}^{28}+12\,{q}^{40}+16\,{q}^{34}+{q}^{38}+{q}^{24}+28\,{q}^{30}+11\,{q}^{52}+{q}^{22}+13\,{q}^{48}-12\,{q}^{46}+13\,{q}^{44} \\ & \quad -4\,{q}^{58}+7\,{q}^{56}-9\,{q}^{54}+2\,{q}^{60} )  ( {q}^{2}+1 ) ^{2} ( {q}^{4}-{q}^{2}+1 ) ^{2}/{{q}^{36}}\\ \hline\noalign{\medskip}21&( 2+13\,{q}^{12}-12\,{q}^{10}+11\,{q}^{8}+7\,{q}^{4}-4\,{q}^{2}-9\,{q}^{6}+{q}^{36}-10\,{q}^{42}+12\,{q}^{20}-12\,{q}^{50}-11\,{q}^{32}-12\,{q}^{14}-10\,{q}^{18} \\ & \quad +13\,{q}^{16}+16\,{q}^{26}-11\,{q}^{28}+12\,{q}^{40}+16\,{q}^{34}+{q}^{38}+{q}^{24}+28\,{q}^{30}+11\,{q}^{52}+{q}^{22}+13\,{q}^{48}-12\,{q}^{46}+13\,{q}^{44} \\ & \quad -4\,{q}^{58}+7\,{q}^{56}-9\,{q}^{54}+2\,{q}^{60} )  ( {q}^{2}+1 ) ^{2} ( {q}^{4}-{q}^{2}+1 ) ^{2}/{{q}^{36}}\\ \hline\noalign{\medskip}212&-( {q}^{2}+1 )  ( {q}^{4}+1 )  ( {q}^{4}-{q}^{2}+1 )  ( 2-3\,{q}^{12}-9\,{q}^{10}-2\,{q}^{8}+2\,{q}^{4}-4\,{q}^{2}-2\,{q}^{6}+{q}^{36}-9\,{q}^{42}+27\,{q}^{20} \\ & \quad -4\,{q}^{50}+27\,{q}^{32}-13\,{q}^{14}-48\,{q}^{18}+{q}^{16}-110\,{q}^{26}+53\,{q}^{28}-3\,{q}^{40}-48\,{q}^{34}-13\,{q}^{38}+53\,{q}^{24}-77\,{q}^{30} \\ & \quad +2\,{q}^{52}-77\,{q}^{22}+2\,{q}^{48}-2\,{q}^{46}-2\,{q}^{44} )/{{q}^{31}}\\ \hline\noalign{\medskip}121&-( {q}^{2}+1 )  ( {q}^{4}+1 )  ( {q}^{4}-{q}^{2}+1 )  ( 2-3\,{q}^{12}-9\,{q}^{10}-2\,{q}^{8}+2\,{q}^{4}-4\,{q}^{2}-2\,{q}^{6}+{q}^{36}-9\,{q}^{42}+27\,{q}^{20} \\ & \quad -4\,{q}^{50}+27\,{q}^{32}-13\,{q}^{14}-48\,{q}^{18}+{q}^{16}-110\,{q}^{26}+53\,{q}^{28}-3\,{q}^{40}-48\,{q}^{34}-13\,{q}^{38}+53\,{q}^{24}-77\,{q}^{30}+2\,{q}^{52} \\ & \quad -77\,{q}^{22}+2\,{q}^{48}-2\,{q}^{46}-2\,{q}^{44} )/{{q}^{31}}\\ \hline\noalign{\medskip}1212&(2-3\,{q}^{12}-9\,{q}^{10}-2\,{q}^{8}+2\,{q}^{4}-4\,{q}^{2}-2\,{q}^{6}+{q}^{36}-9\,{q}^{42}+27\,{q}^{20}-4\,{q}^{50}+27\,{q}^{32}-13\,{q}^{14}-48\,{q}^{18} \\ & \quad +{q}^{16}-110\,{q}^{26}+53\,{q}^{28}-3\,{q}^{40}-48\,{q}^{34}-13\,{q}^{38}+53\,{q}^{24}-77\,{q}^{30}+2\,{q}^{52}-77\,{q}^{22}+2\,{q}^{48}-2\,{q}^{46}-2\,{q}^{44})/{{q}^{26}}\\ \hline\noalign{\medskip}2121&(2-3\,{q}^{12}-9\,{q}^{10}-2\,{q}^{8}+2\,{q}^{4}-4\,{q}^{2}-2\,{q}^{6}+{q}^{36}-9\,{q}^{42}+27\,{q}^{20}-4\,{q}^{50}+27\,{q}^{32}-13\,{q}^{14}-48\,{q}^{18}+{q}^{16} \\ & \quad -110\,{q}^{26}+53\,{q}^{28}-3\,{q}^{40}-48\,{q}^{34}-13\,{q}^{38}+53\,{q}^{24}-77\,{q}^{30}+2\,{q}^{52}-77\,{q}^{22}+2\,{q}^{48}-2\,{q}^{46}-2\,{q}^{44})/{{q}^{26}}\\ \hline \end {array}  
\]
\caption{First nine coefficients of $u_0$}
\label{fu01}
\end{sidewaysfigure}

\begin{sidewaysfigure}
\[
\begin {array}{|c|c|}\hline
\sigma & Coefficient \\ \hline 
21212&-( {q}^{2}+1 )  ( {q}^{4}+1 )  ( {q}^{4}-{q}^{2}+1 )  ( 3\,{q}^{16}+2\,{q}^{14}+14\,{q}^{8}+2\,{q}^{2}+3 )  ( {q}^{8}-{q}^{6}+{q}^{4}+1 ) ( {q}^{8}+{q}^{4}-{q}^{2}+1 )/{{q}^{21}}\\ \hline \noalign{\medskip}
12121&-( {q}^{2}+1 )  ( {q}^{4}+1 )  ( {q}^{4}-{q}^{2}+1 )  ( 3\,{q}^{16}+2\,{q}^{14}+14\,{q}^{8}+2\,{q}^{2}+3 )  ( {q}^{8}-{q}^{6}+{q}^{4}+1 )  ( {q}^{8}+{q}^{4}-{q}^{2}+1 )/{{q}^{21}} \\ \hline 
\noalign{\medskip} 121212&( 3\,{q}^{16}+2\,{q}^{14}+14\,{q}^{8}+2\,{q}^{2}+3 )  ( {q}^{8}+{q}^{4}-{q}^{2}+1 )  ( {q}^{8}-{q}^{6}+{q}^{4}+1 )/{{q}^{16}}\\ \hline\noalign{\medskip}212121&( 3\,{q}^{16}+2\,{q}^{14}+14\,{q}^{8}+2\,{q}^{2}+3 )  ( {q}^{8}+{q}^{4}-{q}^{2}+1 )  ( {q}^{8}-{q}^{6}+{q}^{4}+1 )/{{q}^{16}}\\ \hline\noalign{\medskip}2121212&( {q}^{2}+1 )  ( {q}^{4}-{q}^{2}+1 )  ( {q}^{4}+1 )  ( 3\,{q}^{16}-{q}^{14}+3\,{q}^{12}-3\,{q}^{10}+12\,{q}^{8}-3\,{q}^{6}+3\,{q}^{4}-{q}^{2}+3 )/{{q}^{13}}\\ \hline\noalign{\medskip}1212121&( {q}^{2}+1 )  ( {q}^{4}-{q}^{2}+1 )  ( {q}^{4}+1 )  ( 3\,{q}^{16}-{q}^{14}+3\,{q}^{12}-3\,{q}^{10}+12\,{q}^{8}-3\,{q}^{6}+3\,{q}^{4}-{q}^{2}+3 )/{{q}^{13}}\\ \hline\noalign{\medskip}12121212&-(3\,{q}^{16}-{q}^{14}+3\,{q}^{12}-3\,{q}^{10}+12\,{q}^{8}-3\,{q}^{6}+3\,{q}^{4}-{q}^{2}+3)/{{q}^{8}}\\ \hline\noalign{\medskip}21212121&-(3\,{q}^{16}-{q}^{14}+3\,{q}^{12}-3\,{q}^{10}+12\,{q}^{8}-3\,{q}^{6}+3\,{q}^{4}-{q}^{2}+3)/{{q}^{8}}\\ \hline\noalign{\medskip}212121212&-( {q}^{2}+1 )  ( {q}^{4}-{q}^{2}+1 )  ( {q}^{4}+1 )/{{q}^{5}}\\ \hline\noalign{\medskip}121212121&-( {q}^{2}+1 )  ( {q}^{4}-{q}^{2}+1 )  ( {q}^{4}+1 )/{{q}^{5}}\\ \hline\noalign{\medskip}1212121212&1\\ \hline\noalign{\medskip}2121212121&1 \\ \hline \end  {array}  
\]
\caption{Last twelve coefficients of $u_0$}
\label{fu02}
\end{sidewaysfigure}

\begin{sidewaysfigure}[p!]
\[\begin {array}{|c|c|}\hline
\sigma & Coefficient \\ \hline 
 1&1/2\,( {q}^{4}-{q}^{2}+1 ) ^{2} ( {q}^{8}-{q}^{6}+{q}^{4}-{q}^{2}+1 ) ^{2} ( {q}^{4}+1 ) ^{2} ( {q}^{2}+1 ) ^{4} \\ & \quad \times  ( x+3\,{q}^{28}+4\,{q}^{24}-2\,{q}^{22}+10\,{q}^{20}-2\,{q}^{18}+4\,{q}^{16}+3\,{q}^{12} )/{{q}^{40}}\\ \hline\noalign{\medskip}121&-1/2\,( {q}^{2}+1 ) ^{2} ( 2\,{q}^{18}-295\,{q}^{28}-516\,{q}^{36}+x+210\,{q}^{26}+3\,{q}^{56}-3\,{q}^{54}+47\,{q}^{46}+9\,{q}^{52}-{q}^{48} \\ & \quad +2\,{q}^{50}-84\,{q}^{24}-295\,{q}^{40}+604\,{q}^{34}+462\,{q}^{30}-x{q}^{2}+47\,{q}^{22}-9\,{q}^{20}x+19\,{q}^{10}x-{q}^{26}x+{q}^{28}x-3\,{q}^{14}\\ & \quad +{q}^{24}x+4\,{q}^{22}x  +30\,{q}^{14}x+462\,{q}^{38}-516\,{q}^{32}+19\,{q}^{18}x+210\,{q}^{42}-{q}^{20}+9\,{q}^{16}-24\,{q}^{16}x-24\,{q}^{12}x\\ & \quad -9\,{q}^{8}x+4\,{q}^{6}x+{q}^{4}x-84\,{q}^{44}+3\,{q}^{12} )/{{q}^{36}}\\ \hline\noalign{\medskip}12121&1/2\,({q}^{18}-2\,{q}^{28}+22\,{q}^{36}+x+45\,{q}^{26}+2\,{q}^{46}+3\,{q}^{48}+22\,{q}^{24}+24\,{q}^{40}+45\,{q}^{34}+92\,{q}^{30}+18\,{q}^{22}+{q}^{20}x \\ & \quad +6\,{q}^{10}x+2\,{q}^{14}+{q}^{14}x+18\,{q}^{38}-2\,{q}^{32}+{q}^{42}+24\,{q}^{20}+9\,{q}^{16}+{q}^{16}x+{q}^{6}x+{q}^{4}x+9\,{q}^{44}+3\,{q}^{12})/{{q}^{30}}\\ \hline\noalign{\medskip}1212121&-1/2\,(22\,{q}^{20}+6\,{q}^{16}+6\,{q}^{24}+2\,{q}^{26}+2\,{q}^{14}+2\,{q}^{30}+2\,{q}^{10}+3\,{q}^{28}-2\,{q}^{22}-2\,{q}^{18}+3\,{q}^{12}+x)/{{q}^{20}}\\ \hline\noalign{\medskip}121212121&1
\\ \hline \end {array}
\]
\caption{Coefficients of $u_1^1$}
\label{fu1}
\end{sidewaysfigure}

\begin{sidewaysfigure}[p!]
\[
 \begin {array}{|c|c|}\hline
\sigma & Coefficient \\ \hline  1&( {q}^{8}-{q}^{6}+{q}^{4}-{q}^{2}+1 ) ^{2} ( {q}^{12}+{q}^{10}+2\,{q}^{8}+2\,{q}^{4}+{q}^{2}+1 ) ^{2} ( {q}^{2}+1 ) ^{4} ( {q}^{4}-{q}^{2}+1 ) ^{4}/{{q}^{32}}\\ \hline\noalign{\medskip}121&-( 1-4\,{q}^{10}+14\,{q}^{8}-30\,{q}^{14}+44\,{q}^{28}+73\,{q}^{16}+3\,{q}^{2}+14\,{q}^{32}-30\,{q}^{26}+73\,{q}^{24}+3\,{q}^{38} \\ & \quad + 102\,{q}^{20}-53\,{q}^{18}+{q}^{40}+44\,{q}^{12}-53\,{q}^{22}-4\,{q}^{30}+5\,{q}^{4}+5\,{q}^{36} )  ( {q}^{2}+1 ) ^{2} ( {q}^{4}-{q}^{2}+1 ) ^{2}/{{q}^{26}}\\ \hline\noalign{\medskip}12121&(3+72\,{q}^{18}+14\,{q}^{28}+3\,{q}^{36}+20\,{q}^{26}+10\,{q}^{24}+2\,{q}^{34}+2\,{q}^{30}+36\,{q}^{22}+14\,{q}^{8}+7\,{q}^{4} \\ & \quad +2\,{q}^{2}+7\,{q}^{32}+2\,{q}^{6}-10\,{q}^{20}-10\,{q}^{16}+10\,{q}^{12}+20\,{q}^{10}+36\,{q}^{14})/{{q}^{18}}\\ \hline\noalign{\medskip}1212121&-(1-2\,{q}^{12}+14\,{q}^{10}-2\,{q}^{8}+{q}^{4}+3\,{q}^{2}+4\,{q}^{6}+{q}^{20}+4\,{q}^{14}+3\,{q}^{18}+{q}^{16})/{{q}^{10}}\\ \hline\noalign{\medskip}121212121&1\\ \hline \end{array}  
\]
\caption{Coefficients of $u_3^1$}
\label{fu3}
\end{sidewaysfigure}

\begin{sidewaysfigure}[p!]
\[
 \begin {array}{|c|c|}\hline
\sigma & Coefficient \\ \hline  1&( {q}^{2}+1 ) ^{2} ( {q}^{4}-{q}^{2}+1 ) ^{2} ( {q}^{4}+1 ) ^{2} ( {q}^{8}-{q}^{6}+{q}^{4}-{q}^{2}+1 ) ^{2} ( {q}^{12}+{q}^{10}+2\,{q}^{8}+2\,{q}^{4}+{q}^{2}+1 ) ^{2}/{{q}^{30}}\\ \hline\noalign{\medskip}121&-(1+75\,{q}^{18}-49\,{q}^{28}+42\,{q}^{36}+206\,{q}^{26}-{q}^{46}+{q}^{52}+7\,{q}^{48}-49\,{q}^{24}+40\,{q}^{40}+75\,{q}^{34}+158\,{q}^{30}+158\,{q}^{22} \\ & \quad +22\,{q}^{8}+7\,{q}^{4}+17\,{q}^{14}+17\,{q}^{38}+{q}^{32}-{q}^{6}+{q}^{20}+42\,{q}^{16}+22\,{q}^{44}+40\,{q}^{12}){{q}^{26}}\\ \hline\noalign{\medskip}12121&(3+80\,{q}^{18}+10\,{q}^{28}+3\,{q}^{36}+26\,{q}^{26}-3\,{q}^{24}+{q}^{34}+5\,{q}^{30}+52\,{q}^{22}+10\,{q}^{8}+5\,{q}^{4}+52\,{q}^{14}+{q}^{2}+5\,{q}^{32} \\ & \quad +5\,{q}^{6}-19\,{q}^{20}-19\,{q}^{16}-3\,{q}^{12}+26\,{q}^{10})/{{q}^{18}}\\ \hline\noalign{\medskip}1212121&-(1-2\,{q}^{12}+14\,{q}^{10}-2\,{q}^{8}+3\,{q}^{2}+5\,{q}^{6}+{q}^{20}+5\,{q}^{14}+3\,{q}^{18})/{{q}^{10}}\\ \hline\noalign{\medskip}121212121&1\\ \hline \end{array}  
\]
\caption{Coefficients of $u_4^1$}
\label{fu4}
\end{sidewaysfigure}

\begin{sidewaysfigure}[p!]
\[
 \begin {array}{|c|c|}\hline
\sigma & Coefficient \\ \hline  1&( {q}^{2}+1 ) ^{2} ( {q}^{4}+1 ) ^{2} ( {q}^{12}+{q}^{10}+2\,{q}^{8}+2\,{q}^{4}+{q}^{2}+1 ) ^{2} ( {q}^{4}-{q}^{2}+1 ) ^{4}/{{q}^{26}}\\ \hline\noalign{\medskip}121&-( {q}^{36}+3\,{q}^{34}+10\,{q}^{32}+19\,{q}^{30}+33\,{q}^{28}+53\,{q}^{26}+64\,{q}^{24}+91\,{q}^{22}+84\,{q}^{20}+116\,{q}^{18}+84\,{q}^{16}+91\,{q}^{14} \\ & \quad +64\,{q}^{12}+53\,{q}^{10}+33\,{q}^{8}+19\,{q}^{6}+10\,{q}^{4}+3\,{q}^{2}+1 )  ( {q}^{4}-{q}^{2}+1 ) ^{2}/{{q}^{22}}\\ \hline\noalign{\medskip}12121&(80\,{q}^{16}+3\,{q}^{26}+26\,{q}^{24}+4\,{q}^{22}+{q}^{32}+7\,{q}^{28}+50\,{q}^{20}+3\,{q}^{6}+3\,{q}^{2}+50\,{q}^{12}+1+3\,{q}^{30}+7\,{q}^{4}-14\,{q}^{18} \\ & \quad -14\,{q}^{14}+4\,{q}^{10}+26\,{q}^{8})/{{q}^{16}}\\ \hline\noalign{\medskip}1212121&-(3+5\,{q}^{12}-2\,{q}^{10}+14\,{q}^{8}+5\,{q}^{4}+{q}^{2}-2\,{q}^{6}+{q}^{14}+3\,{q}^{16})/{{q}^{8}}\\ \hline\noalign{\medskip}121212121&1\\ \hline \end{array}  
\]
\caption{Coefficients of $u_5^1$}
\label{fu5}
\end{sidewaysfigure}

\begin{figure} [h!]
\[\begin {array}{|c|c|}\hline g_1&-1/2\,{\frac {-3\,{q}^{28}-4\,{q}^{24}+2\,{q}^{22}-10\,{q}^{20}+2\,{q}^{18}-4\,{q}^{16}-3\,{q}^{12}+x}{{q}^{20}}}\\ \hline\noalign{\medskip}g_3&{\frac { \left( {q}^{4}+1 \right) ^{2}}{{q}^{4}}}\\ \hline\noalign{\medskip}g_4&{\frac { \left( {q}^{2}+1 \right) ^{2} \left( {q}^{4}-{q}^{2}+1 \right) ^{2}}{{q}^{6}}}\\ \hline\noalign{\medskip}g_5&{\frac { \left( {q}^{2}+1 \right) ^{2} \left( {q}^{8}-{q}^{6}+{q}^{4}-{q}^{2}+1 \right) ^{2}}{{q}^{10}}}\\ \hline \end {array}
\]
\caption{The elements $g_i$} 
\label{fgi}
\end{figure}

\begin{sidewaysfigure} [p!]
\[\begin {array}{|c|c|}\hline
\sigma & Coefficient \\ \hline  1&1/2\,( {q}^{2}+1 ) ^{5} ( {q}^{4}-{q}^{2}+1 ) ^{3} ( {q}^{8}-{q}^{6}+{q}^{4}-{q}^{2}+1 ) ^{2} ( {q}^{4}+1 ) ^{3} ( -3\,{q}^{28}-4\,{q}^{24}+2\,{q}^{22} \\ & \quad -10\,{q}^{20}+2\,{q}^{18}-4\,{q}^{16}-3\,{q}^{12}+x ) /{{q}^{45}}\\ \hline\noalign{\medskip}12&-1/2\,( {q}^{2}+1 ) ^{4} ( {q}^{4}-{q}^{2}+1 ) ^{2} ( {q}^{8}-{q}^{6}+{q}^{4}-{q}^{2}+1 ) ^{2} ( {q}^{4}+1 ) ^{2} ( -3\,{q}^{28}-4\,{q}^{24}+2\,{q}^{22} \\ & \quad -10\,{q}^{20}+2\,{q}^{18}-4\,{q}^{16}-3\,{q}^{12}+x ) /{{q}^{40}}\\ \hline\noalign{\medskip}121&-1/2\,( {q}^{2}+1 ) ^{3} ( x+516\,{q}^{32}-462\,{q}^{38}-47\,{q}^{22}+{q}^{20}+84\,{q}^{24}+{q}^{48}-210\,{q}^{42}+84\,{q}^{44}-47\,{q}^{46}-210\,{q}^{26} \\ & \quad -2\,{q}^{18}-9\,{q}^{16}+3\,{q}^{14}+x{q}^{28}-x{q}^{26}+4\,x{q}^{22}+x{q}^{24}-24\,x{q}^{12}+19\,x{q}^{10}+30\,x{q}^{14}-9\,x{q}^{8}-24\,x{q}^{16} \\ & \quad +4\,x{q}^{6}-9\,x{q}^{20}+x{q}^{4}-x{q}^{2}+19\,x{q}^{18}+295\,{q}^{28}+295\,{q}^{40}-604\,{q}^{34}+516\,{q}^{36}-2\,{q}^{50}+3\,{q}^{54}-3\,{q}^{56} \\ & \quad -9\,{q}^{52}-462\,{q}^{30}-3\,{q}^{12} )  ( {q}^{4}+1 )  ( {q}^{4}-{q}^{2}+1 ) /{{q}^{41}}\\ \hline\noalign{\medskip}1212&1/2\,( {q}^{2}+1 ) ^{2} ( x+516\,{q}^{32}-462\,{q}^{38}-47\,{q}^{22}+{q}^{20}+84\,{q}^{24}+{q}^{48}-210\,{q}^{42}+84\,{q}^{44}-47\,{q}^{46}-210\,{q}^{26} \\ & \quad -2\,{q}^{18}-9\,{q}^{16}+3\,{q}^{14}+x{q}^{28}-x{q}^{26}+4\,x{q}^{22}+x{q}^{24}-24\,x{q}^{12}+19\,x{q}^{10}+30\,x{q}^{14}-9\,x{q}^{8}-24\,x{q}^{16} \\ & \quad +4\,x{q}^{6}-9\,x{q}^{20}+x{q}^{4}-x{q}^{2}+19\,x{q}^{18}+295\,{q}^{28}+295\,{q}^{40}-604\,{q}^{34}+516\,{q}^{36}-2\,{q}^{50}+3\,{q}^{54}-3\,{q}^{56} \\ & \quad -9\,{q}^{52}-462\,{q}^{30}-3\,{q}^{12} ) /{{q}^{36}}\\ \hline\noalign{\medskip}12121&1/2\, ( x+2\,{q}^{32}-18\,{q}^{38}-18\,{q}^{22}-24\,{q}^{20}-22\,{q}^{24}-3\,{q}^{48}-{q}^{42}-9\,{q}^{44}-2\,{q}^{46}-45\,{q}^{26}-{q}^{18}-9\,{q}^{16} \\ & \quad -2\,{q}^{14}+6\,x{q}^{10}+x{q}^{14}+x{q}^{16}+x{q}^{6}+x{q}^{20}+x{q}^{4}+2\,{q}^{28}-24\,{q}^{40}-45\,{q}^{34}-22\,{q}^{36}-92\,{q}^{30}-3\,{q}^{12} ) \\ & \quad \times  ( {q}^{2}+1 )  ( {q}^{4}+1 )  ( {q}^{4}-{q}^{2}+1 ) /{{q}^{35}}\\ \hline\noalign{\medskip}121212&-1/2\,(x+2\,{q}^{32}-18\,{q}^{38}-18\,{q}^{22}-24\,{q}^{20}-22\,{q}^{24}-3\,{q}^{48}-{q}^{42}-9\,{q}^{44}-2\,{q}^{46}-45\,{q}^{26}-{q}^{18}-9\,{q}^{16} \\ & \quad -2\,{q}^{14}+6\,x{q}^{10}+x{q}^{14}+x{q}^{16}+x{q}^{6}+x{q}^{20}+x{q}^{4}+2\,{q}^{28}-24\,{q}^{40}-45\,{q}^{34}-22\,{q}^{36}-92\,{q}^{30}-3\,{q}^{12})/{{q}^{30}}\\ \hline\noalign{\medskip}1212121&-1/2\,( -2\,{q}^{10}-22\,{q}^{20}+2\,{q}^{18}-2\,{q}^{14}-3\,{q}^{12}-6\,{q}^{16}-2\,{q}^{30}+2\,{q}^{22}-6\,{q}^{24}-3\,{q}^{28}-2\,{q}^{26}+x )  ( {q}^{2}+1 ) \\ & \quad \times  ( {q}^{4}+1 )  ( {q}^{4}-{q}^{2}+1 )/{{q}^{25}}\\ \hline\noalign{\medskip}12121212&1/2\,(-2\,{q}^{10}-22\,{q}^{20}+2\,{q}^{18}-2\,{q}^{14}-3\,{q}^{12}-6\,{q}^{16}-2\,{q}^{30}+2\,{q}^{22}-6\,{q}^{24}-3\,{q}^{28}-2\,{q}^{26}+x)/{{q}^{20}}\\ \hline\noalign{\medskip}121212121&-( {q}^{4}+1 )  ( {q}^{4}-{q}^{2}+1 )  ( {q}^{2}+1 )/{{q}^{5}}\\ \hline\noalign{\medskip}1212121212&1 \\ \hline \hline \end {array}  
\]
\caption{Nonzero coefficients of $z_0$}
\label{fz0}
\end{sidewaysfigure}

\begin{sidewaysfigure}[p]
\[
\begin {array}{|c|c|}\hline
Monomial & Coefficient \\ \hline  x_2\otimes x_1 \otimes x_0 &-1/2\,( {q}^{4}+1 ) ^{2} ( {q}^{2}-q+1 )  ( {q}^{2}+q+1 )  ( {q}^{8}+1 ) ^{2} ( 5\,x+173\,{q}^{32}-192\,{q}^{38}-192\,{q}^{22}+91\,{q}^{20}+117\,{q}^{24}+11\,{q}^{48} \\ & \quad -45\,{q}^{42}+24\,{q}^{44}-37\,{q}^{46}-131\,{q}^{26}-45\,{q}^{18}+24\,{q}^{16}-37\,{q}^{14}+19\,x{q}^{12}-40\,x{q}^{10}-7\,x{q}^{14}+19\,x{q}^{8}+8\,x{q}^{16} \\ & \quad -7\,x{q}^{6}+5\,x{q}^{20}+8\,x{q}^{4}-17\,x{q}^{2}-17\,x{q}^{18}+173\,{q}^{28}+91\,{q}^{40}-131\,{q}^{34}+117\,{q}^{36}-310\,{q}^{30}+11\,{q}^{12} )/{{q}^{51}}\\ \hline\noalign{\medskip} x_1\otimes x_2 \otimes x_0 &1/2\,( {q}^{2}+1 )  ( {q}^{2}+q+1 )  ( {q}^{2}-q+1 )  ( {q}^{4}+1 )  ( {q}^{8}+1 ) ^{2} ( 5\,x+173\,{q}^{32}-192\,{q}^{38}-192\,{q}^{22}+91\,{q}^{20}+117\,{q}^{24} \\ & \quad +11\,{q}^{48}-45\,{q}^{42}+24\,{q}^{44}-37\,{q}^{46}-131\,{q}^{26}-45\,{q}^{18}+24\,{q}^{16}-37\,{q}^{14}+19\,x{q}^{12}-40\,x{q}^{10}-7\,x{q}^{14}+19\,x{q}^{8} \\ & \quad +8\,x{q}^{16}-7\,x{q}^{6}+5\,x{q}^{20}+8\,x{q}^{4}-17\,x{q}^{2}-17\,x{q}^{18}+173\,{q}^{28}+91\,{q}^{40}-131\,{q}^{34}+117\,{q}^{36}-310\,{q}^{30}+11\,{q}^{12} )/{{q}^{46}}\\ \hline\noalign{\medskip} x_0\otimes x_3 \otimes x_0 &-1/2\,( q+1 )  ( q-1 )  ( {q}^{4}+1 )  ( {q}^{2}+q+1 ) ^{2} ( {q}^{2}-q+1 ) ^{2} ( {q}^{8}+1 ) ^{2} ( 3\,x+727\,{q}^{32}-500\,{q}^{38} \\ & \quad -330\,{q}^{22}+191\,{q}^{20}+460\,{q}^{24}+59\,{q}^{48}-359\,{q}^{42}+192\,{q}^{44}-110\,{q}^{46}-603\,{q}^{26}-138\,{q}^{18}+76\,{q}^{16}-35\,{q}^{14} \\ & \quad -20\,x{q}^{22}+5\,x{q}^{24}+70\,x{q}^{12}-58\,x{q}^{10}-77\,x{q}^{14}+45\,x{q}^{8}+48\,x{q}^{16}-46\,x{q}^{6}+23\,x{q}^{20}+30\,x{q}^{4}-15\,x{q}^{2} \\ & \quad -32\,x{q}^{18}+587\,{q}^{28}+402\,{q}^{40}-780\,{q}^{34}+584\,{q}^{36}-44\,{q}^{50}+11\,{q}^{52}-685\,{q}^{30}+7\,{q}^{12} )/{{q}^{53}}\\ \hline\noalign{\medskip} x_2\otimes x_0 \otimes x_1 &1/2\,( {q}^{4}+1 ) ^{2} ( {q}^{2}-q+1 )  ( {q}^{2}+q+1 )  ( {q}^{8}+1 ) ^{2} ( 5\,x+173\,{q}^{32}-192\,{q}^{38}-192\,{q}^{22}+91\,{q}^{20} \\ & \quad +117\,{q}^{24}+11\,{q}^{48}-45\,{q}^{42}+24\,{q}^{44}-37\,{q}^{46}-131\,{q}^{26}-45\,{q}^{18}+24\,{q}^{16}-37\,{q}^{14}+19\,x{q}^{12}-40\,x{q}^{10} \\ & \quad -7\,x{q}^{14}+19\,x{q}^{8}+8\,x{q}^{16}-7\,x{q}^{6}+5\,x{q}^{20}+8\,x{q}^{4}-17\,x{q}^{2}-17\,x{q}^{18}+173\,{q}^{28}+91\,{q}^{40}-131\,{q}^{34} \\ & \quad +117\,{q}^{36}-310\,{q}^{30}+11\,{q}^{12} ) /{{q}^{48}}\\ \hline\noalign{\medskip} x_1\otimes x_1 \otimes x_1 &-1/2\,( {q}^{2}+1 ) ^{2} ( {q}^{4}+1 )  ( q-1 )  ( q+1 )  ( {q}^{2}+q+1 )  ( {q}^{2}-q+1 )  ( {q}^{8}+1 ) ^{2} ( 5\,x+173\,{q}^{32}-192\,{q}^{38}-192\,{q}^{22} \\ & \quad +91\,{q}^{20}+117\,{q}^{24}+11\,{q}^{48}-45\,{q}^{42}+24\,{q}^{44}-37\,{q}^{46}-131\,{q}^{26}-45\,{q}^{18}+24\,{q}^{16}-37\,{q}^{14}+19\,x{q}^{12}-40\,x{q}^{10} \\ & \quad -7\,x{q}^{14}+19\,x{q}^{8}+8\,x{q}^{16}-7\,x{q}^{6}+5\,x{q}^{20}+8\,x{q}^{4}-17\,x{q}^{2}-17\,x{q}^{18}+173\,{q}^{28}+91\,{q}^{40}-131\,{q}^{34}+117\,{q}^{36} \\ & \quad -310\,{q}^{30}+11\,{q}^{12} ) /{{q}^{47}}\\ \hline\noalign{\medskip} x_0\otimes x_2 \otimes x_1 &1/2\,( {q}^{4}+1 )  ( {q}^{2}+q+1 )  ( {q}^{2}-q+1 )  ( {q}^{8}+1 ) ^{2} ( -3\,x-1316\,{q}^{32}+1364\,{q}^{38}+419\,{q}^{22}-200\,{q}^{20}-577\,{q}^{24} \\ & \quad -464\,{q}^{48}+957\,{q}^{42}-617\,{q}^{44}+613\,{q}^{46}+748\,{q}^{26}+5\,{q}^{30}x+134\,{q}^{18}-76\,{q}^{16}+35\,{q}^{14}-25\,x{q}^{28}+35\,x{q}^{26} \\ & \quad +67\,x{q}^{22}-28\,x{q}^{24}-117\,x{q}^{12}+97\,x{q}^{10}+110\,x{q}^{14}-48\,x{q}^{8}-85\,x{q}^{16}+44\,x{q}^{6}-112\,x{q}^{20}-30\,x{q}^{4}+15\,x{q}^{2} \\ & \quad +123\,x{q}^{18}-816\,{q}^{28}-1224\,{q}^{40}+1325\,{q}^{34}-1132\,{q}^{36}+257\,{q}^{50}+85\,{q}^{54}-55\,{q}^{56}+11\,{q}^{58}-108\,{q}^{52}\\ & \quad +1220\,{q}^{30}-7\,{q}^{12} )/{{q}^{50}}\\ \hline 
\end{array}
\] 
\caption{First five nonzero coefficients  of $a \in X_q^{\otimes 3}$}
\label{fa1}
\end{sidewaysfigure}

\begin{sidewaysfigure}[p!]
\[
\begin {array}{|c|c|}\hline
Monomial & Coefficient \\ \hline 
x_0\otimes x_0 \otimes x_2 &-1/2\,( {q}^{2}+1 )  ( {q}^{2}+q+1 )  ( {q}^{2}-q+1 )  ( {q}^{4}+1 )  ( {q}^{8}+1 ) ^{2} ( 5\,x+173\,{q}^{32}-192\,{q}^{38}-192\,{q}^{22}+91\,{q}^{20}+117\,{q}^{24} \\ & \quad +11\,{q}^{48}-45\,{q}^{42}+24\,{q}^{44}-37\,{q}^{46}-131\,{q}^{26}-45\,{q}^{18}+24\,{q}^{16}-37\,{q}^{14}+19\,x{q}^{12}-40\,x{q}^{10}-7\,x{q}^{14} \\ & \quad +19\,x{q}^{8}+8\,x{q}^{16}-7\,x{q}^{6}+5\,x{q}^{20}+8\,x{q}^{4}-17\,x{q}^{2}-17\,x{q}^{18}+173\,{q}^{28}+91\,{q}^{40}-131\,{q}^{34}+117\,{q}^{36} \\ & \quad -310\,{q}^{30}+11\,{q}^{12} )/{{q}^{46}}\\ \hline\noalign{\medskip} x_0\otimes x_1 \otimes x_2 &1/2\,( {q}^{4}+1 )  ( {q}^{2}+q+1 )  ( {q}^{2}-q+1 )  ( {q}^{8}+1 ) ^{2} ( 3\,x+951\,{q}^{32}-1060\,{q}^{38}-363\,{q}^{22}+176\,{q}^{20}+449\,{q}^{24}+248\,{q}^{48} \\ & \quad -592\,{q}^{42}+395\,{q}^{44}-451\,{q}^{46}-532\,{q}^{26}-97\,{q}^{18}+65\,{q}^{16}-35\,{q}^{14}+8\,x{q}^{28}-27\,x{q}^{26}-31\,x{q}^{22}+16\,x{q}^{24} \\ & \quad +81\,x{q}^{12}-85\,x{q}^{10}-62\,x{q}^{14}+40\,x{q}^{8}+59\,x{q}^{16}-27\,x{q}^{6}+64\,x{q}^{20}+25\,x{q}^{4}-15\,x{q}^{2}-97\,x{q}^{18}+654\,{q}^{28} \\ & \quad +797\,{q}^{40}-898\,{q}^{34}+828\,{q}^{36}-129\,{q}^{50}-61\,{q}^{54}+18\,{q}^{56}+52\,{q}^{52}-998\,{q}^{30}+7\,{q}^{12} ) /{{q}^{49}}\\ \hline\noalign{\medskip} x_0\otimes x_0 \otimes x_3 &-1/2\,( {q}^{4}+1 )  ( q-1 ) ^{2} ( q+1 ) ^{2} ( {q}^{8}+1 ) ^{2} ( {q}^{2}+q+1 ) ^{3} ( {q}^{2}-q+1 ) ^{3} ( 3\,x+275\,{q}^{32}-94\,{q}^{38}-220\,{q}^{22} \\ & \quad +132\,{q}^{20}+275\,{q}^{24}-35\,{q}^{42}+7\,{q}^{44}-279\,{q}^{26}-94\,{q}^{18}+65\,{q}^{16}-35\,{q}^{14}+25\,x{q}^{12}-26\,x{q}^{10}-15\,x{q}^{14} \\ & \quad +22\,x{q}^{8}+3\,x{q}^{16}-26\,x{q}^{6}+25\,x{q}^{4}-15\,x{q}^{2}+250\,{q}^{28}+65\,{q}^{40}-220\,{q}^{34}+132\,{q}^{36}-279\,{q}^{30}+7\,{q}^{12} ) /{{q}^{50}}\\ \hline\end {array}  
\]
\caption{Last four nonzero coefficients of $a$}
\label{fa2}
\end{sidewaysfigure}

\begin{sidewaysfigure}[p!]
\[
\begin {array}{|c|c|}\hline
Monomial & Coefficient \\ \hline x_3\otimes x_0 \otimes x_0 &1/2\,( {q}^{4}+1 )  ( q-1 ) ^{2} ( q+1 ) ^{2} ( {q}^{8}+1 ) ^{2} ( {q}^{2}+q+1 ) ^{3} ( {q}^{2}-q+1 ) ^{3} ( 5\,x+1214\,{q}^{32}-847\,{q}^{38}-525\,{q}^{22} \\ & \quad +289\,{q}^{20}+714\,{q}^{24}+107\,{q}^{48}-525\,{q}^{42}+289\,{q}^{44}-178\,{q}^{46}-847\,{q}^{26}-178\,{q}^{18}+107\,{q}^{16}-55\,{q}^{14}-25\,x{q}^{22} \\ & \quad +5\,x{q}^{24}+130\,x{q}^{12}-113\,x{q}^{10}-113\,x{q}^{14}+75\,x{q}^{8}+75\,x{q}^{16}-60\,x{q}^{6}+45\,x{q}^{20}+45\,x{q}^{4}-25\,x{q}^{2}-60\,x{q}^{18} \\ & \quad  +920\,{q}^{28} +714\,{q}^{40}-1139\,{q}^{34}+920\,{q}^{36}-55\,{q}^{50}+11\,{q}^{52}-1139\,{q}^{30}+11\,{q}^{12} )/{{q}^{57}}\\ \hline\noalign{\medskip} x_2\otimes x_1 \otimes x_0 &-1/2\,( {q}^{4}+1 )  ( q-1 ) ^{2} ( q+1 ) ^{2} ( {q}^{2}+q+1 ) ^{2} ( {q}^{2}-q+1 ) ^{2} ( {q}^{8}+1 ) ^{2} ( -5\,x-1170\,{q}^{32}+977\,{q}^{38}+425\,{q}^{22} \\ & \quad -207\,{q}^{20}-565\,{q}^{24}-371\,{q}^{48}+636\,{q}^{42}-451\,{q}^{44}+397\,{q}^{46}+523\,{q}^{26}+5\,{q}^{30}x+106\,{q}^{18}-89\,{q}^{16}+55\,{q}^{14}  \\ & \quad -20\,x{q}^{28}+20\,x{q}^{26}+43\,x{q}^{22}-20\,x{q}^{24}-119\,x{q}^{12}+97\,x{q}^{10}+71\,x{q}^{14}-45\,x{q}^{8}-59\,x{q}^{16}+28\,x{q}^{6}-95\,x{q}^{20}  \\ & \quad -37\,x{q}^{4}+25\,x{q}^{2}+87\,x{q}^{18}-676\,{q}^{28}-1021\,{q}^{40}+891\,{q}^{34}-849\,{q}^{36}+173\,{q}^{50}+52\,{q}^{54}-44\,{q}^{56}+11\,{q}^{58}  \\ & \quad -82\,{q}^{52}+1002\,{q}^{30}-11\,{q}^{12} ) /{{q}^{56}}\\ \hline\noalign{\medskip} x_1\otimes x_2 \otimes x_0 &1/2\,( {q}^{4}+1 )  ( q-1 ) ^{2} ( q+1 ) ^{2} ( {q}^{2}+q+1 ) ^{2} ( {q}^{2}-q+1 ) ^{2} ( {q}^{8}+1 ) ^{2} ( -8\,x-1447\,{q}^{32}+1598\,{q}^{38}+590\,{q}^{22}  \\ & \quad -304\,{q}^{20}-648\,{q}^{24}-496\,{q}^{48}+927\,{q}^{42}-742\,{q}^{44}+696\,{q}^{46}+778\,{q}^{26}+5\,{q}^{30}x+153\,{q}^{18}-89\,{q}^{16}+72\,{q}^{14}  \\ & \quad -25\,x{q}^{28}+40\,x{q}^{26}+58\,x{q}^{22}-40\,x{q}^{24}-124\,x{q}^{12}+138\,x{q}^{10}+96\,x{q}^{14}-76\,x{q}^{8}-114\,x{q}^{16}+39\,x{q}^{6}-116\,x{q}^{20} \\ & \quad -33\,x{q}^{4}   +32\,x{q}^{2}+152\,x{q}^{18}-1064\,{q}^{28}-1329\,{q}^{40}+1319\,{q}^{34}-1386\,{q}^{36}+244\,{q}^{50}+96\,{q}^{54}-55\,{q}^{56} \\ & \quad +11\,{q}^{58}-134\,{q}^{52}+1516\,{q}^{30}-18\,{q}^{12} ) /{{q}^{55}}\\ \hline\noalign{\medskip} x_0\otimes x_3 \otimes x_0 &1/2\,( {q}^{4}+1 )  ( q-1 ) ^{2} ( q+1 ) ^{2} ( {q}^{8}+1 ) ^{2} ( {q}^{2}-q+1 ) ^{3} ( {q}^{2}+q+1 ) ^{3} ( -3\,x-1693\,{q}^{32}+1146\,{q}^{38}+641\,{q}^{22}-404\,{q}^{20} \\ & \quad -1020\,{q}^{24}-178\,{q}^{48}+814\,{q}^{42}-567\,{q}^{44}+296\,{q}^{46}+1268\,{q}^{26}+266\,{q}^{18}-155\,{q}^{16}+53\,{q}^{14}+5\,x{q}^{26}+45\,x{q}^{22} \\ & \quad -25\,x{q}^{24}   -164\,x{q}^{12}+123\,x{q}^{10}+170\,x{q}^{14}-108\,x{q}^{8}-131\,x{q}^{16}+96\,x{q}^{6}-60\,x{q}^{20}-65\,x{q}^{4}+23\,x{q}^{2}  \\ & \quad +78\,x{q}^{18}-1376\,{q}^{28}  -1006\,{q}^{40} +1709\,{q}^{34}-1491\,{q}^{36}+107\,{q}^{50}+11\,{q}^{54}-55\,{q}^{52}+1449\,{q}^{30}-7\,{q}^{12} ) /{{q}^{58}}\\ \hline\noalign{\medskip}x_2\otimes x_0 \otimes x_1 &-1/2\,( {q}^{2}+1 )  ( {q}^{4}+1 )  ( q-1 ) ^{2} ( q+1 ) ^{2} ( {q}^{2}-q+1 ) ^{2} ( {q}^{2}+q+1 ) ^{2} ( {q}^{8}+1 ) ^{2} ( 5\,x+964\,{q}^{32}-627\,{q}^{38}  \\ & \quad -431\,{q}^{22}+224\,{q}^{20}+582\,{q}^{24}+100\,{q}^{48}-431\,{q}^{42}+224\,{q}^{44}-143\,{q}^{46}-627\,{q}^{26}-143\,{q}^{18}+100\,{q}^{16}-55\,{q}^{14}-25\,x{q}^{22}  \\ & \quad +5\,x{q}^{24}+108\,x{q}^{12}-87\,x{q}^{10}-87\,x{q}^{14}+50\,x{q}^{8}+50\,x{q}^{16}-45\,x{q}^{6}+42\,x{q}^{20}+42\,x{q}^{4}-25\,x{q}^{2}-45\,x{q}^{18}+645\,{q}^{28}  \\ & \quad +582\,{q}^{40}-860\,{q}^{34}+645\,{q}^{36}-55\,{q}^{50}+11\,{q}^{52}-860\,{q}^{30}+11\,{q}^{12} ) /{{q}^{53}}\\ \hline 
\end{array} 
\] 
\caption{First five nonzero coefficients of $b$}
\label{fb1}
\end{sidewaysfigure}

\begin{sidewaysfigure}[p!]
\[
\begin {array}{|c|c|}\hline
Monomial & Coefficient \\ \hline 
x_1\otimes x_1 \otimes x_1 &1/2\,( {q}^{2}+1 ) ^{2} ( {q}^{2}-q+1 )  ( {q}^{2}+q+1 )  ( {q}^{4}+1 )  ( q-1 ) ^{2} ( q+1 ) ^{2} ( {q}^{8}+1 ) ^{2} ( 3\,x+277\,{q}^{32}-621\,{q}^{38}-165\,{q}^{22}  \\ & \quad +97\,{q}^{20}+83\,{q}^{24}+125\,{q}^{48}-291\,{q}^{42}+291\,{q}^{44}-299\,{q}^{46}-255\,{q}^{26}-47\,{q}^{18}-17\,{q}^{14}+5\,x{q}^{28}-20\,x{q}^{26}-15\,x{q}^{22}  \\ & \quad +20\,x{q}^{24}+5\,x{q}^{12}-41\,x{q}^{10}-25\,x{q}^{14}+31\,x{q}^{8}+55\,x{q}^{16}-11\,x{q}^{6}+21\,x{q}^{20}-4\,x{q}^{4}-7\,x{q}^{2}-65\,x{q}^{18}+388\,{q}^{28}  \\ & \quad +308\,{q}^{40}-428\,{q}^{34}+537\,{q}^{36}-71\,{q}^{50}-44\,{q}^{54}+11\,{q}^{56}+52\,{q}^{52}-514\,{q}^{30}+7\,{q}^{12} )/{{q}^{52}}\\ \hline\noalign{\medskip} x_0\otimes x_2 \otimes x_1&-1/2\,( {q}^{2}+1 )  ( {q}^{4}+1 )  ( q-1 ) ^{2} ( q+1 ) ^{2} ( {q}^{2}-q+1 ) ^{2} ( {q}^{2}+q+1 ) ^{2} ( {q}^{8}+1 ) ^{2} ( -3\,x-1651\,{q}^{32}+1803\,{q}^{38}  \\ & \quad +486\,{q}^{22}-287\,{q}^{20}-679\,{q}^{24}-567\,{q}^{48}+1181\,{q}^{42}-1013\,{q}^{44}+814\,{q}^{46}+920\,{q}^{26}+5\,{q}^{30}x+147\,{q}^{18}-72\,{q}^{16} \\ & \quad +35\,{q}^{14}   -25\,x{q}^{28}+45\,x{q}^{26}+78\,x{q}^{22}-60\,x{q}^{24}-133\,x{q}^{12}+122\,x{q}^{10}+138\,x{q}^{14}-87\,x{q}^{8}-167\,x{q}^{16}  \\ & \quad  +49\,x{q}^{6}-131\,x{q}^{20}-28\,x{q}^{4} +15\,x{q}^{2}+170\,x{q}^{18}-1270\,{q}^{28}-1556\,{q}^{40}+1669\,{q}^{34}-1825\,{q}^{36}+296\,{q}^{50}  \\ & \quad  +107\,{q}^{54}-55\,{q}^{56}+11\,{q}^{58}-178\,{q}^{52}+1547\,{q}^{30}-7\,{q}^{12} ) /{{q}^{55}}\\ \hline\noalign{\medskip} x_1\otimes x_0 \otimes x_2 &-1/2\,( {q}^{4}+1 ) ^{2} ( q-1 ) ^{2} ( q+1 ) ^{2} ( {q}^{8}+1 ) ^{2} ( {q}^{2}-q+1 ) ^{3} ( {q}^{2}+q+1 ) ^{3} ( 3\,x+275\,{q}^{32}-94\,{q}^{38}-220\,{q}^{22}  \\ & \quad +132\,{q}^{20}+275\,{q}^{24}-35\,{q}^{42}+7\,{q}^{44}-279\,{q}^{26}-94\,{q}^{18}+65\,{q}^{16}-35\,{q}^{14}+25\,x{q}^{12}-26\,x{q}^{10} \\ & \quad -15\,x{q}^{14}+22\,x{q}^{8}   +3\,x{q}^{16}-26\,x{q}^{6}+25\,x{q}^{4}-15\,x{q}^{2}+250\,{q}^{28}+65\,{q}^{40}-220\,{q}^{34}+132\,{q}^{36}-279\,{q}^{30}+7\,{q}^{12} ) /{{q}^{51}}\\ \hline\noalign{\medskip} x_0\otimes x_1 \otimes x_2 &-1/2\,( {q}^{4}+1 ) ^{2} ( q-1 ) ^{2} ( q+1 ) ^{2} ( {q}^{8}+1 ) ^{2} ( {q}^{2}-q+1 ) ^{3} ( {q}^{2}+q+1 ) ^{3} ( 3\,x+275\,{q}^{32}-94\,{q}^{38}-220\,{q}^{22}  \\ & \quad +132\,{q}^{20}+275\,{q}^{24}-35\,{q}^{42}+7\,{q}^{44}-279\,{q}^{26}-94\,{q}^{18}+65\,{q}^{16}-35\,{q}^{14}+25\,x{q}^{12}-26\,x{q}^{10}-15\,x{q}^{14}+22\,x{q}^{8}  \\ & \quad +3\,x{q}^{16}-26\,x{q}^{6}+25\,x{q}^{4}-15\,x{q}^{2}+250\,{q}^{28}+65\,{q}^{40}-220\,{q}^{34}+132\,{q}^{36}-279\,{q}^{30}+7\,{q}^{12} )/{{q}^{54}}\\ \hline\noalign{\medskip} x_0\otimes x_0 \otimes x_3 &1/2\,( {q}^{2}+1 )  ( {q}^{4}+1 ) ^{2} ( {q}^{4}-{q}^{2}+1 )  ( q-1 ) ^{2} ( q+1 ) ^{2} ( {q}^{8}+1 ) ^{2} ( {q}^{2}-q+1 ) ^{3} ( {q}^{2}+q+1 ) ^{3} ( 3\,x+275\,{q}^{32}  \\ & \quad -94\,{q}^{38}-220\,{q}^{22}+132\,{q}^{20}+275\,{q}^{24}-35\,{q}^{42}+7\,{q}^{44}-279\,{q}^{26}-94\,{q}^{18}+65\,{q}^{16}-35\,{q}^{14}+25\,x{q}^{12}-26\,x{q}^{10}-15\,x{q}^{14}  \\ & \quad +22\,x{q}^{8}+3\,x{q}^{16}-26\,x{q}^{6}+25\,x{q}^{4}-15\,x{q}^{2}+250\,{q}^{28}+65\,{q}^{40}-220\,{q}^{34}+132\,{q}^{36}-279\,{q}^{30}+7\,{q}^{12} )/{{q}^{55}} \\ \hline \end {array} 
\]
\caption{Last five nonzero coefficients of $b$}
\label{fb2}
\end{sidewaysfigure}

\subsection{Example 2} \label{sex2}
Now we verify the duality and reciprocity conjectures for the special 
case of the Kronecker problem (Section~\ref{skronecker}), when $H=GL(V) \times GL(W)$,
$V=W=\C^2$ and $G=GL(X)$, $X=V\otimes W \cong \C^4$, and $r=4$.
Thus $G_q=GL_q(\C^4)$, and $H_q=GL_q(\C^2) \times GL_q(\C^2)$. 
Let ${\cal B}={\cal B}_r^H$ be the nonstandard algebra in this case 
 and  $P_i=p_{X,i}^{+,H}, Q_i=p_{X,i}^{-,X}$, $i<r$, 
the positive and negative projection operators as in Section~\ref{sbr}. 
Let ${\cal P}_i$ and ${\cal Q}_i$ be the rescaled versions of $P_i$ and
$Q_i$ as defined in \cite{GCT4}. 
Then ${\cal B}$ is generated by ${\cal P}_i$, or equivalently, 
${\cal Q}_i$. The explicit generating relations among ${\cal P}_i$'s 
and  ${\cal Q}_i$'s turn out to be very complicated.
 For example, Figures~\ref{fb41}-\ref{fb43} reproduced from  \cite{GCT4} 
shows a typical generating relation among ${\cal Q}_i$'s 
with 74 terms. There are several
dozen such relations. Because of the nature of these generating relations,
there is no good ``standard monomial basis'' for ${\cal B}$ as for the 
Hecke algebra or for the $r=3$ case in Section~\ref{sexppre}.
 Fortunately, this makes no difference as far
as duality and reciprocity is concerned, as we shall see here, and
also as far as existence of a canonical basis is concerned, as we
shall see in \cite{GCT8}. 

It was verified by computer that ${\cal B}$ is of dimension $114$ \cite{GCT4}. 
Since it is semisimple, it admits a Wederburn 
structure decomposition. It turns out that a complete 
Wederburn structure  decomposition of the form (\ref{eqwederreci}) works over
$\Q(q)$ itself; i.e., no algebraic extension of $\Q(q)$ is necessary here,
just as in the case of Hecke algebras. This may be conjectured to be
the case for the Kronecker problem in general, though it is not 
so for the plethysm problem in general as we already saw in Section~\ref{sex1}.

So let 
\begin{equation}  \label{eqwederex2}
 {\cal B}= \otimes_i T_{i,L} \otimes T_{i,R}, 
\end{equation}
be the complete Wederburn structure decomposition of ${\cal B}$, where 
$T_i=T_{i,L}$ ranges over all irreducible left ${\cal B}$-modules.

\subsubsection{Irreducible representations}
We describe  these  $T_i$ next.
There are two distinct irreducible representations of ${\cal B}$ 
of dimension $1,2,3$ and $5$ each, and one of dimension $6$.
Since 
\[ 114=1^2+1^2 + 2^2+2^2 + 3^2 + 3^2 + 5^2 + 5^2 + 6^2,\] 
this is consistent with the Wederburn 
structure decomposition in (\ref{eqwederex2}). 

Let $S_{q,\lambda}$ denote the $q$-Specht module of the Hecke algebra 
${\cal H}_r(q)$ for the partition $\lambda$, and 
$KL_\lambda$ its Kazhdan-Lusztig basis ordered appropriately.
Since, in this case,
${\cal B}={\cal B}_r^H(q)  \subseteq {\cal H}_r(q) \otimes {\cal H}_r(q)$,
the tensor product $S_{q,\lambda} \otimes S_{q,\mu}$ is 
a representation of ${\cal B}$. 
In particular, 
\[T_{q,\lambda}=S_{q,\lambda} \otimes S_{q,(r)} 
\cong S_{q,(r)} \otimes S_{q,\lambda}, \]
where $S_{q,(r)}$ is the trivial one dimensional $q$-Specht module, 
is an irreducible ${\cal B}$-module, which specializes at $q=1$ to 
the Specht module $S_\lambda$ of the symmetric group $S_r$. 

Let
\begin{equation} 
\begin{array} {lcl} 
T_0 &=& T_{q,(4)}, \\
T_1 &=& T_{q,(1,1,1,1)}, \\
T_2&=& T_{q,(2,2)}, \\
T_3&=& T_{q,(2,1,1)}, \\
T_4&=& T_{q,(3,1)},\\
T_5&=&S_{q,(3,1)} \otimes S_{q,(2,2)} \cong  S_{q,(2,1,1)} \otimes S_{q,(2,2)}.\\
\end{array}
\end{equation} 

These are irreducible ${\cal B}$-modules. Their dimensions are
$1,1,2,3,3$ and $6$ respectively.

To get the other two dimensional irreducible ${\cal B}$-module,
we analyze how the tensor product $S_{q,(2,2)} \otimes S_{q,(2,2)}$ 
decomposes as a ${\cal B}$-module. It decomposes as:
\[S_{q,(2,2)} \otimes S_{q,(2,2)} \cong 
T_{q,(4)} \oplus T_{q,(1,1,1,1)} \oplus T_6,\] 
where $T_6$ is the other two dimensional irreducible ${\cal B}$-module
that we were looking for. Explicitly, a basis of $T_6$ in terms of 
the Kazhdan-Lusztig basis $KL_{(2,2)} \otimes KL_{(2,2)}$ of 
$S_{q,(2,2)} \otimes S_{q,(2,2)}$ is given by the rows of 
the matrix
\[
\left[
\begin{array}{llll}
    1& \f{1+q}{2 q^{1/2}} & \f{1+q}{2 q^{1/2}}&                 0 \\
                 0& \f{1+q}{2 q^{1/2}}& \f{1+q}{2 q^{1/2}}&                 1
\end{array}
\right].
\]

Matrix representations of the right action of the
generators ${\cal Q}_i$'s of ${\cal B}$ on this 
basis are:
\[
{\cal Q}_1={\cal Q}_3=\left[
\begin{array} {ll}
 (1+q)^2/q&         0 \\
 (1+q^2)/q&         0
 \end{array}
\right]
\]

\[
{\cal Q}_2=\left[
\begin{array} {ll}
0& (1+q^2)/q \\
0&  (1+q)^2/q
 \end{array}
\right]
\]

The specialization of $T_6$ at $q=1$ is isomorphic to the Specht 
module $S_{(2,2)}$ of $S_4$. But $T_6$  is nonisomorphic to $T_2$, whose
specialization at $q=1$ is the same.

To get the five dimensional irreducible ${\cal B}$-modules, we
analyze how the tensor products $S_{q,(2,1,1)} \otimes S_{q,(2,1,1)}$
and $S_{q,(3,1)} \otimes S_{q,(2,1,1)}$ 
decompose as  ${\cal B}$-modules. 
We have
\[
S_{q,(2,1,1)} \otimes S_{q,(2,1,1)} \cong T_{q,(2,1,1)} \oplus T_{q,(4)} 
\oplus T_7,\] 
where $T_7$ is the first five dimensional irreducible ${\cal B}$-representation
that we were looking for. 
Explicitly, its basis in terms of the Kazhdan-Lusztig basis 
$KL_{(2,1,1)}\otimes KL_{(2,1,1)}$ is given by the rows of the
matrix:

\[
\begin{array} {lllllll}
w_1=[&0& -(1+q)/(2 q^{1/2})&     -(1+q)^2/(2 q)& 0&   0&                  0]\\
w_2=[&-1& -(1+q)/(2 q^{1/2})& 0&  0&  (1+q)/(2 q^{1/2})&                  1]\\
w_3=[&0& 0&     -(1+q)^2/(2 q)&   0& -(1+q)/(2 q^{1/2})&                  0]\\
v_1=[& 0& -(1+q)/(2 q^{1/2})& -(1+q)^2/(2 q) &-1& -(1+q)/(2 q^{1/2})&   0]\\
v_2=[&1&  (1+q)/(2 q^{1/2})& 1&  0&  (1+q)/(2 q^{1/2})&     1]
\end{array}
\]

Matrix representations of the right action
of  ${\cal Q}_i$'s in this basis are:

\[
{\cal Q}_1=
\left[
\begin{array} {llllll}
  (1+q)^2/q&          0&          0&          0&          0\\
  (1+q^2)/q&          0&          0& -(1+q^2)/q&          0\\
          0&          0&          0&  (1+q)^2/q&          0\\
          0&          0&          0&  (1+q)^2/q&          0\\
  (q-1)^2/q&          0&          0& -(1+q^2)/q&          0\\
\end{array}
\right]
\]

\[
{\cal Q}_2=
\left[
\begin{array} {llllll}
 0&  \f{(1+q)^2}{2 q}&              0&              0& -\f{(1+q)^2}{2 q}\\
0&     \f {(1+q)^2}{q}&              0&              0&              0\\
0& -\f{(1+q)^2}{2 q}&              0&              0& -\f{(1+q)^2}{2 q}\\
0&              0&              0&              0&     -\f{1+q^2}{q}\\
0&              0&              0&              0&      \f{(1+q)^2}{q}\\
 \end{array}
\right]
\]

\[
{\cal Q}_3=
\left[
\begin{array} {llllll}
          0&          0&          0&  (1+q)^2/q&          0\\
          0&          0& -(1+q^2)/q&  (1+q^2)/q&          0\\
          0&          0&  (1+q)^2/q&          0&          0\\
          0&          0&          0&  (1+q)^2/q&          0\\
          0&          0&  (q-1)^2/q& -(1+q^2)/q&          0\\
\end{array}
\right]
\]

Let $V$ denote the 
 span  of the vectors $v_1$ and $v_2$, and $V(1)$ its specialization
at $q=1$. It can be checked that $V(1)$ is isomorphic to 
 the Specht module $S_{(2,2)}$ of $S_4$, and 
the quotient $T_7(1)/V$, where $T_7(1)$ denotes the specialization
of $T_7$ at $q=1$, is
isomorphic to the Specht module $S_{(3,1)}$ of $S_4$.

Finally, 
\[
S_{q,(3,1)} \otimes S_{q,(2,1,1)} \cong T_{q,(3,1)} \oplus T_{q,(1,1,1,1)} 
\oplus T_8,\] 
where $T_8 \not \cong T_7$ is the second
five dimnsional irreducible ${\cal B}$-representation
that we were looking for. Its basis and representation matrices are similar.

This specifies all irreducible representations of ${\cal B}$.

\subsubsection{Duality} 
Using the explicit representations $T_i$ above, the Wederburn 
structure decomposition (\ref{eqwederex2}) of ${\cal B}$ was explicitly 
determined with the help of a computer. The explicit bases of
the structure components $U_i=T_{i,L}\otimes T_{i,R}$
in (\ref{eqwederex2}) are far too complex to be given
here. 

Fix any $u_i \in U_i$, $0 \le i \le 8$, and let $W_i= X_q^{\otimes r} \cdot 
 u_i$ be the
corresponding left representation of the nonstandard quantum group
$G^H_q$. Computer experiments indicate that these are 
nonisomorphic irreducible representations of $G^H_q$ with
the following decompositions as $H_q$-modules,
$H_q=GL_q(\C^2) \times GL_q(\C^2)$. (Recall that $V_{q,\lambda}(n)$
is the $q$-Weyl module of $GL_q(\C^n)$).

\[
\begin{array}{lcl} 
W_0&\cong& V_{q,(4)}(2) \otimes  V_{q,(4)}(2) \oplus 
V_{q,(3,1)}(2) \otimes  V_{q,(3,1)}(2) \oplus 
V_{q,(2,2)}(2) \otimes  V_{q,(2,2)}(2), \\
W_1&\cong& V_{q,(2,2)}(2) \otimes  V_{q,(2,2)}(2), \\
W_2&\cong& V_{q,(4)}(2) \otimes  V_{q,(2,2)}(2) \oplus 
V_{q,(2,2)}(2) \otimes  V_{q,(4)}(2), \\
W_3&\cong& V_{q,(3,1)}(2) \otimes  V_{q,(3,1)}(2), \\
W_4&\cong& V_{q,(3,1)}(2) \otimes  V_{q,(4)}(2) \oplus 
V_{q,(4)}(2) \otimes  V_{q,(3,1)}(2), \\
W_5&\cong& V_{q,(2,2)}(2) \otimes  V_{q,(3,1)}(2) \oplus 
V_{q,(3,1)}(2) \otimes  V_{q,(2,2)}(2), \\
W_6&\cong& V_{q,(2,2)}(2) \otimes  V_{q,(2,2)}(2),\\
W_7&\cong& V_{q,(3,1)}(2) \otimes  V_{q,(3,1)}(2).\\
\end{array}
\]

Their dimensions are $35,1,10,9,30,6,1$ and $9$, respectively.
The module $W_8$ turns out to be zero when  $\dim(V)=\dim(W)=2$, as here;
however, it would be nonzero for general $\dim(V)$ and $\dim(W)$.
Furthermore,
\[ X_q^{\otimes 4} = \bigoplus_i W_i \otimes T_i,\] 
 in accordance with the duality conjecture.

\noindent {\em Remark:} These computations are not  final. 
The main problem is that the symbolic computations needed here are too 
heavy for MATLAB/Maple to handle. Hence, in some of the computations
$q$ was set to a fixed real value (such as $.5$). This introduces 
floating point errors in various calculations.
As far as we can see, this does not affect the decomposition   above.
But this has to be double checked by other means.

\subsubsection{Reciprocity}
Let $m^i_\mu$ denote the multiplicity of the Specht module $S_\mu$ of
$S_4$ in $T_i$. Then it can be verified  that
\[ 
\begin{array}{l} 
m^0_{(4)}=1, \\
m^1_{(1,1,1,1)}=1, \\
m^2_{(2,2)}=1, \\
m^3_{(2,1,1)}=1, \\
m^4_{(3,1)}=1, \\
m^5_{(3,1)}=m^5_{(2,1,1)}=1,\\
m^6_{(2,2)}=1, \\
m^7_{(3,1)}=m^7_{(2,2)}=1, \\
m^8_{(2,1,1)}=m^8_{(2,2)}=1. \\
\end{array}
\]

All other $m^i_\mu$'s are zero. 
It can now be seen that, as $H_q$-modules, $H_q=GL_q(2) \times GL_q(2)$, 
we have
\[
\begin{array}{lcl}
V_{q,(4)}(4) &\cong& m^0_{(4)} W_0 \\ 
&\cong &
 V_{q,(4)}(2) \otimes  V_{q,(4)}(2) \oplus 
V_{q,(3,1)}(2) \otimes  V_{q,(3,1)}(2)   \\ 
&&  \quad \oplus
V_{q,(2,2)}(2) \otimes  V_{q,(2,2)}(2), \\ \\ 
V_{q,(3,1)}(4) &\cong& m^4_{(3,1)} W_4 \oplus m^5_{(3,1)} W_5 \oplus
m^7_{(3,1)} W_7 \\ &\cong& 
V_{q,(3,1)}(2) \otimes  V_{q,(4)}(2) \oplus 
V_{q,(4)}(2) \otimes  V_{q,(3,1)}(2) \\
&& \quad \oplus 
V_{q,(2,2)}(2) \otimes  V_{q,(3,1)}(2) \oplus 
V_{q,(3,1)}(2) \otimes  V_{q,(2,2)}(2)  \\
&&\quad \oplus
V_{q,(3,1)}(2) \otimes  V_{q,(3,1)}(2), \\ \\ 
V_{q,(2,2)}(4)&\cong & m^2_{(2,2)} W_2 \oplus m^7_{(2,2)} W_7 \oplus 
m^6_{(2,2)} W_6 \\
&\cong& 
V_{q,(4)}(2) \otimes  V_{q,(2,2)}(2) \oplus 
V_{q,(2,2)}(2) \otimes  V_{q,(4)}(2) \\
&& \quad \oplus
V_{q,(3,1)}(2) \otimes  V_{q,(3,1)}(2)  \oplus 
V_{q,(2,2)}(2) \otimes  V_{q,(2,2)}(2), \\ \\
V_{q,(2,1,1)}(4) & \cong & m^3_{(2,1,1)} W_3 \oplus m^5_{(2,1,1)} W_5 \\
&\cong& 
 V_{q,(3,1)}(2) \otimes  V_{q,(3,1)}(2) \oplus
V_{q,(2,2)}(2) \otimes  V_{q,(3,1)}(2) \\
&& \quad \oplus 
V_{q,(3,1)}(2) \otimes  V_{q,(2,2)}(2), \\ \\
V_{q,(1,1,1,1)}(4) &\cong& m^1_{(1,1,1,1)} W_1 \\
&\cong& V_{q,(2,2)}(2) \otimes  V_{q,(2,2)}(2),
\end{array}
\]
in accordance with the reciprocity conjecture.

We are unable to verify the refined reciprocity conjecture on computer
since  the necessary symbolic computations turn out to be
beyond the reach of the  desktop MATLAB/Maple.

\begin{figure}[p]
\[
\begin {array}{|c|ccccccccc|} \hline 
\sigma& \hat  A_\sigma \\ \hline
1&17738&17738&17550&17362&16994&16626&16114&15602&14933\\\noalign{\medskip}&14264&13550&12836&12008&11180&10392&9604&8790&7976\\\noalign{\medskip}&7226&6476&5806&5136&4518&3900&3418&2936&2504\\\noalign{\medskip}&2072&1762&1452&1202&952&776&600&482&364\\\noalign{\medskip}&280&196&152&108&77&46&34&22&14\\\noalign{\medskip}&6&4&2&1&&&&&\\ \hline\noalign{\medskip}121&20322&20322&20083&19844&19354&18864&18211&17558&16668\\\noalign{\medskip}&15778&14890&14002&12934&11866&10916&9966&8962&7958\\\noalign{\medskip}&7092&6226&5470&4714&4047&3380&2895&2410&1982\\\noalign{\medskip}&1554&1287&1020&804&588&463&338&253&168\\\noalign{\medskip}&122&76&53&30&19&8&5&2&1\\ \hline\noalign{\medskip}12121&9078&9078&8973&8868&8623&8378&8051&7724&7245\\\noalign{\medskip}&6766&6335&5904&5363&4822&4382&3942&3443&2944\\\noalign{\medskip}&2562&2180&1851&1522&1267&1012&831&650&506\\\noalign{\medskip}&362&284&206&151&96&70&44&28&12\\\noalign{\medskip}&7&2&1&&&&&&\\ \hline\noalign{\medskip}1212121&1918&1918&1913&1908&1866&1824&1742&1660&1523\\\noalign{\medskip}&1386&1277&1168&1042&916&821&726&603&480\\\noalign{\medskip}&395&310&246&182&145&108&83&58&40\\\noalign{\medskip}&22&14&6&3&&&&&\\ \hline\noalign{\medskip}121212121&25032&24784&24124&23136&21978&20808&19710&18768&17934\\\noalign{\medskip}&17160&16358&15440&14384&13168&11849&10484&9139&7880\\\noalign{\medskip}&6725&5692&4765&3928&3170&2480&1868&1344&914\\\noalign{\medskip}&584&345&188&93&40&15&4&1& \\ \hline\end {array} 
\]
\caption{Positivity and unimodality 
of $\hat A_\sigma$'s}
\label{fposq}
\end{figure}

\ignore{
\begin{figure}[p]
\[
\begin {array}{|c|ccccccccc|} \hline 
\sigma& \hat B_\sigma \\ \hline
1&17738&17738&17550&17362&16994&16626&16114&15602&14933\\\noalign{\medskip}&14264&13550&12836&12008&11180&10392&9604&8790&7976\\\noalign{\medskip}&7226&6476&5806&5136&4518&3900&3418&2936&2504\\\noalign{\medskip}&2072&1762&1452&1202&952&776&600&482&364\\\noalign{\medskip}&280&196&152&108&77&46&34&22&14\\\noalign{\medskip}&6&4&2&1&&&&&\\ \hline\noalign{\medskip}121&20322&20322&20083&19844&19354&18864&18211&17558&16668\\\noalign{\medskip}&15778&14890&14002&12934&11866&10916&9966&8962&7958\\\noalign{\medskip}&7092&6226&5470&4714&4047&3380&2895&2410&1982\\\noalign{\medskip}&1554&1287&1020&804&588&463&338&253&168\\\noalign{\medskip}&122&76&53&30&19&8&5&2&1\\ \hline\noalign{\medskip}12121&9078&9078&8973&8868&8623&8378&8051&7724&7245\\\noalign{\medskip}&6766&6335&5904&5363&4822&4382&3942&3443&2944\\\noalign{\medskip}&2562&2180&1851&1522&1267&1012&831&650&506\\\noalign{\medskip}&362&284&206&151&96&70&44&28&12\\\noalign{\medskip}&7&2&1&&&&&&\\ \hline\noalign{\medskip}1212121&1918&1918&1913&1908&1866&1824&1742&1660&1523\\\noalign{\medskip}&1386&1277&1168&1042&916&821&726&603&480\\\noalign{\medskip}&395&310&246&182&145&108&83&58&40\\\noalign{\medskip}&22&14&6&3&&&&&\\ \hline\noalign{\medskip}121212121&172&172&181&190&195&200&191&182&159\\\noalign{\medskip}&136&118&100&86&72&64&56&43&30\\\noalign{\medskip}&21&12&7&2&1&&&&\\ \hline\end {array}
\]
\caption{Positivity of the coefficients of the first basic relation among ${\cal Q}_i$'s}
\end{figure}
}

\begin{figure} [p]
\[
\begin {array}{|c|ccccccccc|} \hline 
\sigma& \hat B_\sigma \\ \hline
\emptyset&390464&389128&385120&378581&369652&358471&345176&330055&313396\\\noalign{\medskip}&295506&276692&257207&237304&217316&197576&178283&159636&141795\\\noalign{\medskip}&124920&109152&94632&81349&69292&58469&48888&40497&33244\\\noalign{\medskip}&27011&21680&17198&13512&10505&8060&6095&4528&3314\\\noalign{\medskip}&2408&1731&1204&812&540&357&232&147&84\\\noalign{\medskip}&43&24&16&8&2&&&&\\ \hline\noalign{\medskip}2&102390&101996&100847&98976&96425&93236&89466&85172&80462\\\noalign{\medskip}&75444&70190&64772&59280&53804&48429&43240&38271&33556\\\noalign{\medskip}&29145&25088&21393&18068&15099&12472&10185&8236&6586\\\noalign{\medskip}&5196&4040&3092&2333&1744&1286&920&640&440\\\noalign{\medskip}&300&200&129&76&41&24&16&8&2\\ \hline \noalign{\medskip}1&50420&50420&49799&49178&48066&46954&45325&43696&41665\\\noalign{\medskip}&39634&37420&35206&32782&30358&27969&25580&23303&21026\\\noalign{\medskip}&18902&16778&14947&13116&11553&9990&8713&7436&6455\\\noalign{\medskip}&5474&4724&3974&3416&2858&2490&2122&1831&1540\\\noalign{\medskip}&1350&1160&1031&902&779&656&582&508&441\\\noalign{\medskip}&374&313&252&213&174&144&114&86&58\\\noalign{\medskip}&47&36&27&18&11&4&4&4&2\\
\hline\noalign{\medskip}12&13180&13086&12992&12744&12496&12124&11752&11225&10698\\\noalign{\medskip}&10112&9526&8890&8254&7584&6914&6294&5674&5083\\\noalign{\medskip}&4492&3979&3466&3036&2606&2256&1906&1638&1370\\\noalign{\medskip}&1178&986&840&694&603&512&450&388&335\\\noalign{\medskip}&282&259&236&206&176&153&130&116&102\\\noalign{\medskip}&85&68&54&40&34&28&21&14&9\\\noalign{\medskip}&4&4&4&2&&&&& \\ \hline\end {array} 
\]
\caption{The vectors  $\hat B_\sigma$}
\label{fposp1}
\end{figure}

\begin{figure}[p]
\[
\begin {array}{|c|ccccccccc|} \hline 
\sigma& \hat B_\sigma \\ \hline
212&3432&3432&3379&3326&3242&3158&3033&2908&2744\\\noalign{\medskip}&2580&2417&2254&2069&1884&1709&1534&1371&1208\\\noalign{\medskip}&1062&916&797&678&581&484&411&338&287\\\noalign{\medskip}&236&202&168&143&118&108&98&84&70\\\noalign{\medskip}&65&60&56&52&43&34&30&26&23\\\noalign{\medskip}&20&15&10&7&4&4&4&2&\\ \hline\noalign{\medskip}121&51252&51252&50661&50070&48941&47812&46219&44626&42589\\\noalign{\medskip}&40552&38328&36104&33645&31186&28756&26326&23948&21570\\\noalign{\medskip}&19376&17182&15217&13252&11581&9910&8522&7134&6030\\\noalign{\medskip}&4926&4106&3286&2664&2042&1632&1222&941&660\\\noalign{\medskip}&497&334&233&132&89&46&25&4&-4\\\noalign{\medskip}&-12&-10&-8&-6&-4&-4&-4&-2&\\ \hline\noalign{\medskip}1212&13352&13285&13218&12957&12696&12341&11986&11462&10938\\\noalign{\medskip}&10358&9778&9131&8484&7812&7140&6498&5856&5240\\\noalign{\medskip}&4624&4088&3552&3079&2606&2236&1866&1552&1238\\\noalign{\medskip}&1026&814&646&478&373&268&198&128&93\\\noalign{\medskip}&58&35&12&4&-4&-4&-4&-4&-4\\\noalign{\medskip}&-4&-4&-2&&&&&&\\ \hline\noalign{\medskip}21212&3472&3472&3427&3382&3293&3204&3093&2982&2810\\\noalign{\medskip}&2638&2483&2328&2132&1936&1772&1608&1423&1238\\\noalign{\medskip}&1098&958&820&682&587&492&398&304&249\\\noalign{\medskip}&194&153&112&85&58&37&16&10&4\\\noalign{\medskip}&2&0&-2&-4&-4&-4&-2&&\\ \hline\noalign{\medskip}12121&20922&20922&20625&20328&19815&19302&18558&17814&16848\\\noalign{\medskip}&15882&14868&13854&12740&11626&10537&9448&8430&7412\\\noalign{\medskip}&6509&5606&4830&4054&3439&2824&2349&1874&1526\\\noalign{\medskip}&1178&945&712&550&388&301&214&155&96\\\noalign{\medskip}&70&44&30&16&11&6&3&& \\ \hline\end {array} 
\]
\caption{The vectors $\hat B_\sigma$ (cont.) }
\label{fposp2}
\end{figure}

\begin{figure} [p!]
\[
\begin {array}{|c|ccccccccc|} \hline 
\sigma& \hat B_\sigma \\ \hline
121212&5496&5453&5410&5286&5162&5008&4854&4600&4346\\\noalign{\medskip}&4068&3790&3497&3204&2894&2584&2295&2006&1749\\\noalign{\medskip}&1492&1280&1068&889&710&586&462&366&270\\\noalign{\medskip}&215&160&117&74&56&38&27&16&11\\\noalign{\medskip}&6&3&&&&&&&\\ \hline \noalign{\medskip}2121212&1434&1434&1406&1378&1346&1314&1267&1220&1128\\\noalign{\medskip}&1036&961&886&799&712&633&554&470&386\\\noalign{\medskip}&334&282&231&180&146&112&82&52&42\\\noalign{\medskip}&32&24&16&11&6&3&&&\\ \hline\noalign{\medskip}1212121&3800&3800&3735&3670&3573&3476&3326&3176&2974\\\noalign{\medskip}&2772&2567&2362&2138&1914&1692&1470&1277&1084\\\noalign{\medskip}&921&758&624&490&396&302&238&174&131\\\noalign{\medskip}&88&65&42&29&16&11&6&3&\\ \hline\noalign{\medskip}12121212&1004&992&980&957&934&908&882&829&776\\\noalign{\medskip}&716&656&597&538&472&406&346&286&240\\\noalign{\medskip}&194&158&122&94&66&51&36&26&16\\\noalign{\medskip}&11&6&3&&&&&&\\ \hline\noalign{\medskip}212121212&258&258&252&246&245&244&237&230&208\\\noalign{\medskip}&186&168&150&132&114&95&76&60&44\\\noalign{\medskip}&37&30&23&16&11&6&3&&\\ \hline\noalign{\medskip}1212121212&68&67&66&65&64&63&62&58&54\\\noalign{\medskip}&49&44&39&34&28&22&17&12&9\\\noalign{\medskip}&6&4&2&1&&&&& \\ \hline\end {array} 
\]
\caption{The vectors $\hat B_\sigma$ (cont)}
\label{fposp3}
\end{figure}

\begin{figure}[h!] 
\[ \begin{array}{|c|c|c|}\hline
Number & Coefficient & \sigma \\ \hline 
1 & \frac{ 20.q^{0}+104.q^{1}+256.q^{2}-113.q^{3}-49.q^{4}-113.q^{5}+256.q^{6}+104.q^{7}+20.q^{8} }{ 2.q^{3}+12.q^{4}+2.q^{5}} &  1\\ \hline
2 & \frac{ -16.q^{0}-64.q^{1}-128.q^{2}-192.q^{3}-224.q^{4}-192.q^{5}-128.q^{6}-64.q^{7}-16.q^{8} }{ 2.q^{3}+12.q^{4}+2.q^{5}} &  2\\ \hline
3 & \frac{ -4.q^{0}-16.q^{1}-28.q^{2}-32.q^{3}-28.q^{4}-16.q^{5}-4.q^{6} }{ 2.q^{3}} &  3\\ \hline
4 & \frac{ 1.q^{0}-4.q^{2}+6.q^{4}-4.q^{6}+1.q^{8} }{ 2.q^{3}+2.q^{5}} &  12\\ \hline
5 & \frac{ -1.q^{0}-18.q^{1}-65.q^{2}-128.q^{3}-190.q^{4}-220.q^{5}-190.q^{6}-128.q^{7}-65.q^{8}-18.q^{9}-1.q^{10} }{ 2.q^{4}+12.q^{5}+2.q^{6}} &  13\\ \hline
6 & \frac{ 1.q^{0}+5.q^{1}+17.q^{2}+36.q^{3}+46.q^{4}+46.q^{5}+46.q^{6}+36.q^{7}+17.q^{8}+5.q^{9}+1.q^{10} }{ 2.q^{4}+2.q^{6}} &  21\\ \hline
7 & \frac{ 7.q^{0}+26.q^{1}+75.q^{2}+152.q^{3}+174.q^{4}+156.q^{5}+174.q^{6}+152.q^{7}+75.q^{8}+26.q^{9}+7.q^{10} }{ 2.q^{3}+12.q^{4}+4.q^{5}+12.q^{6}+2.q^{7}} &  23\\ \hline
8 & \frac{ -1.q^{0}-8.q^{1}-20.q^{2}-24.q^{3}-22.q^{4}-24.q^{5}-20.q^{6}-8.q^{7}-1.q^{8} }{ 2.q^{3}+2.q^{5}} &  32\\ \hline
9 & \frac{ -22.q^{0}-92.q^{1}-170.q^{2}-200.q^{3}-170.q^{4}-92.q^{5}-22.q^{6} }{ 2.q^{2}+12.q^{3}+2.q^{4}} &  121\\ \hline
10 & \frac{ 2.q^{0}+2.q^{1}+12.q^{2}+14.q^{3}+4.q^{4}+14.q^{5}+12.q^{6}+2.q^{7}+2.q^{8} }{ 2.q^{3}+2.q^{5}} &  132\\ \hline
11 & \frac{ -2.q^{0}-12.q^{1}-40.q^{2}-52.q^{3}-44.q^{4}-52.q^{5}-40.q^{6}-12.q^{7}-2.q^{8} }{ 2.q^{3}+12.q^{4}+2.q^{5}} &  212\\ \hline
12 & \frac{ -1.q^{0}-2.q^{1}-12.q^{2}-14.q^{3}-6.q^{4}-14.q^{5}-12.q^{6}-2.q^{7}-1.q^{8} }{ 2.q^{3}+2.q^{5}} &  213\\ \hline
13 & \frac{ 1.q^{0}+22.q^{1}+88.q^{2}+170.q^{3}+206.q^{4}+170.q^{5}+88.q^{6}+22.q^{7}+1.q^{8} }{ 2.q^{3}+12.q^{4}+2.q^{5}} &  232\\ \hline
14 & \frac{ 6.q^{0}+8.q^{1}+4.q^{2}+8.q^{3}+6.q^{4} }{ 2.q^{2}} &  323\\ \hline
15 & \frac{ 3.q^{0}+6.q^{1}+5.q^{2}+4.q^{3}+5.q^{4}+6.q^{5}+3.q^{6} }{ 2.q^{2}+2.q^{4}} &  1212\\ \hline
16 & \frac{ 12.q^{0}+32.q^{1}+40.q^{2}+32.q^{3}+12.q^{4} }{ 2.q^{1}+12.q^{2}+2.q^{3}} &  1213\\ \hline
17 & \frac{ -3.q^{0}-2.q^{1}-5.q^{2}-12.q^{3}-5.q^{4}-2.q^{5}-3.q^{6} }{ 2.q^{2}+2.q^{4}} &  1232\\ \hline
18 & \frac{ 1.q^{0}+4.q^{1}+11.q^{2}+16.q^{3}+11.q^{4}+4.q^{5}+1.q^{6} }{ 2.q^{3}} &  1321\\ \hline
19 & \frac{ 8.q^{0}+12.q^{1}+24.q^{2}+40.q^{3}+24.q^{4}+12.q^{5}+8.q^{6} }{ 2.q^{2}+12.q^{3}+2.q^{4}} &  1323\\ \hline
20 & \frac{ -6.q^{0}-8.q^{1}-4.q^{2}-8.q^{3}-6.q^{4} }{ 2.q^{1}+2.q^{3}} &  2121\\ \hline
21 & \frac{ -5.q^{0}-4.q^{1}-44.q^{2}-60.q^{3}-30.q^{4}-60.q^{5}-44.q^{6}-4.q^{7}-5.q^{8} }{ 2.q^{2}+12.q^{3}+4.q^{4}+12.q^{5}+2.q^{6}} &  2123\\ \hline
22 & \frac{ -1.q^{0}-5.q^{1}-11.q^{2}-14.q^{3}-11.q^{4}-5.q^{5}-1.q^{6} }{ 2.q^{3}} &  2321\\ \hline
23 & \frac{ -3.q^{0}-6.q^{1}-5.q^{2}-4.q^{3}-5.q^{4}-6.q^{5}-3.q^{6} }{ 2.q^{2}+2.q^{4}} &  2323\\ \hline
24 & \frac{ 2.q^{0}+4.q^{1}+4.q^{2}+4.q^{3}+2.q^{4} }{ 2.q^{2}} &  3212\\ \hline
25 & \frac{ -1.q^{0}-4.q^{1}-6.q^{2}-4.q^{3}-1.q^{4} }{ 2.q^{2}} &  3213\\ \hline
26 & \frac{ 6.q^{0}+8.q^{1}+4.q^{2}+8.q^{3}+6.q^{4} }{ 2.q^{1}+2.q^{3}} &  3232\\ \hline
27 & \frac{ 16.q^{0}+32.q^{1}+16.q^{2} }{ 2.q^{0}+12.q^{1}+2.q^{2}} &  12121\\ \hline
28 & \frac{ 4.q^{0}+8.q^{1}+40.q^{2}+8.q^{3}+4.q^{4} }{ 2.q^{1}+12.q^{2}+2.q^{3}} &  12123\\ \hline
29 & \frac{ -3.q^{0}-8.q^{1}-4.q^{2}-8.q^{3}+46.q^{4}-8.q^{5}-4.q^{6}-8.q^{7}-3.q^{8} }{ 2.q^{2}+12.q^{3}+4.q^{4}+12.q^{5}+2.q^{6}} &  12132\\ \hline
30 & \frac{ -8.q^{0} }{ 2.q^{0}} &  12321\\ \hline
\end{array} \]
\caption{A relation in ${\cal B}^H_4$ from GCT4}
\label{fb41}
\end{figure}
\begin{figure}[h!]
\[ \begin{array}{|c|c|c|}\hline
Number & Coefficient & \sigma \\ \hline 
31 & \frac{ -4.q^{0}-8.q^{1}-40.q^{2}-8.q^{3}-4.q^{4} }{ 2.q^{1}+12.q^{2}+2.q^{3}} &  12323\\ \hline
32 & \frac{ -3.q^{0}-4.q^{1}-2.q^{2}-4.q^{3}-3.q^{4} }{ 2.q^{1}+2.q^{3}} &  13212\\ \hline
33 & \frac{ -9.q^{0}-6.q^{1}-55.q^{2}+12.q^{3}-55.q^{4}-6.q^{5}-9.q^{6} }{ 2.q^{1}+12.q^{2}+4.q^{3}+12.q^{4}+2.q^{5}} &  13232\\ \hline
34 & \frac{ 9.q^{0}+6.q^{1}+55.q^{2}-12.q^{3}+55.q^{4}+6.q^{5}+9.q^{6} }{ 2.q^{1}+12.q^{2}+4.q^{3}+12.q^{4}+2.q^{5}} &  21213\\ \hline
35 & \frac{ 1.q^{0}-1.q^{1}+3.q^{2}-6.q^{3}+3.q^{4}-1.q^{5}+1.q^{6} }{ 2.q^{2}+2.q^{4}} &  21232\\ \hline
36 & \frac{ -1.q^{0}+2.q^{2}-1.q^{4} }{ 2.q^{2}} &  21321\\ \hline
37 & \frac{ 2.q^{0}+3.q^{1}+6.q^{2}-3.q^{3}-16.q^{4}-3.q^{5}+6.q^{6}+3.q^{7}+2.q^{8} }{ 2.q^{2}+12.q^{3}+4.q^{4}+12.q^{5}+2.q^{6}} &  21323\\ \hline
38 & \frac{ 3.q^{0}+4.q^{1}+2.q^{2}+4.q^{3}+3.q^{4} }{ 2.q^{1}+2.q^{3}} &  23213\\ \hline
39 & \frac{ -16.q^{0}-32.q^{1}-16.q^{2} }{ 2.q^{0}+12.q^{1}+2.q^{2}} &  23232\\ \hline
40 & \frac{ 3.q^{0}+4.q^{1}+2.q^{2}+4.q^{3}+3.q^{4} }{ 2.q^{1}+2.q^{3}} &  32121\\ \hline
41 & \frac{ 8.q^{0} }{ 2.q^{0}} &  32123\\ \hline
42 & \frac{ 1.q^{0}-2.q^{2}+1.q^{4} }{ 2.q^{2}} &  32132\\ \hline
43 & \frac{ -3.q^{0}-4.q^{1}-2.q^{2}-4.q^{3}-3.q^{4} }{ 2.q^{1}+2.q^{3}} &  32321\\ \hline
44 & \frac{ -8.q^{0}-16.q^{1}-8.q^{2} }{ 2.q^{0}+12.q^{1}+2.q^{2}} &  121213\\ \hline
45 & \frac{ -1.q^{0}-14.q^{1}-15.q^{2}-4.q^{3}-15.q^{4}-14.q^{5}-1.q^{6} }{ 2.q^{1}+12.q^{2}+4.q^{3}+12.q^{4}+2.q^{5}} &  121232\\ \hline
46 & \frac{ -2.q^{0}-4.q^{1}-2.q^{2} }{ 2.q^{1}} &  121321\\ \hline
47 & \frac{ -2.q^{0}+4.q^{2}-2.q^{4} }{ 2.q^{1}+12.q^{2}+2.q^{3}} &  123213\\ \hline
48 & \frac{ 8.q^{0}+16.q^{1}+8.q^{2} }{ 2.q^{0}+12.q^{1}+2.q^{2}} &  123232\\ \hline
49 & \frac{ -1.q^{0}-2.q^{1}-1.q^{2} }{ 2.q^{1}} &  132121\\ \hline
50 & \frac{ 2.q^{0}-4.q^{2}+2.q^{4} }{ 2.q^{1}+12.q^{2}+2.q^{3}} &  132123\\ \hline
51 & \frac{ 2.q^{0}+8.q^{1}+12.q^{2}+8.q^{3}+2.q^{4} }{ 2.q^{1}+12.q^{2}+2.q^{3}} &  212132\\ \hline
52 & \frac{ 2.q^{0}+4.q^{1}+2.q^{2} }{ 2.q^{1}} &  212321\\ \hline
53 & \frac{ 1.q^{0}+14.q^{1}+15.q^{2}+4.q^{3}+15.q^{4}+14.q^{5}+1.q^{6} }{ 2.q^{1}+12.q^{2}+4.q^{3}+12.q^{4}+2.q^{5}} &  212323\\ \hline
54 & \frac{ 3.q^{0}+8.q^{1}+10.q^{2}+8.q^{3}+3.q^{4} }{ 2.q^{1}+12.q^{2}+2.q^{3}} &  213212\\ \hline
55 & \frac{ -2.q^{0}-8.q^{1}-12.q^{2}-8.q^{3}-2.q^{4} }{ 2.q^{1}+12.q^{2}+2.q^{3}} &  213232\\ \hline
56 & \frac{ -1.q^{0}-2.q^{1}-1.q^{2} }{ 2.q^{1}} &  232121\\ \hline
57 & \frac{ -3.q^{0}-8.q^{1}-10.q^{2}-8.q^{3}-3.q^{4} }{ 2.q^{1}+12.q^{2}+2.q^{3}} &  232132\\ \hline
58 & \frac{ 1.q^{0}+2.q^{1}+1.q^{2} }{ 2.q^{1}} &  232321\\ \hline
59 & \frac{ -2.q^{0}-4.q^{1}-2.q^{2} }{ 2.q^{1}} &  321232\\ \hline
60 & \frac{ 2.q^{0}+4.q^{1}+2.q^{2} }{ 2.q^{1}} &  321323\\ \hline
\end{array} \]
\caption{A relation in ${\cal B}^H_4$ from GCT4 continued.}
\label{fb42}
\end{figure}
\begin{figure}[h!] 
\[ \begin{array}{|c|c|c|}\hline
Number & Coefficient & \sigma \\ \hline 
61 & \frac{ 1.q^{0}+2.q^{1}+1.q^{2} }{ 2.q^{1}} &  323212\\ \hline
62 & \frac{ 1.q^{0}-2.q^{1}+1.q^{2} }{ 2.q^{0}+2.q^{2}} &  1212132\\ \hline
63 & \frac{ 2.q^{0} }{ 2.q^{0}} &  1213213\\ \hline
64 & \frac{ 1.q^{0}-2.q^{1}+1.q^{2} }{ 2.q^{0}+2.q^{2}} &  1213232\\ \hline
65 & \frac{ 2.q^{0} }{ 2.q^{0}} &  1232121\\ \hline
66 & \frac{ 2.q^{0}-4.q^{1}+2.q^{2} }{ 2.q^{0}+12.q^{1}+2.q^{2}} &  1232132\\ \hline
67 & \frac{ 16.q^{1} }{ 2.q^{0}+12.q^{1}+2.q^{2}} &  1321232\\ \hline
68 & \frac{ -2.q^{0} }{ 2.q^{0}} &  1321323\\ \hline
69 & \frac{ -1.q^{0}+2.q^{1}-1.q^{2} }{ 2.q^{0}+2.q^{2}} &  2121323\\ \hline
70 & \frac{ -4.q^{0}-8.q^{1}-4.q^{2} }{ 2.q^{0}+12.q^{1}+2.q^{2}} &  2123212\\ \hline
71 & \frac{ -16.q^{1} }{ 2.q^{0}+12.q^{1}+2.q^{2}} &  2123213\\ \hline
72 & \frac{ -1.q^{0}+2.q^{1}-1.q^{2} }{ 2.q^{0}+2.q^{2}} &  2123232\\ \hline
73 & \frac{ -2.q^{0}+4.q^{1}-2.q^{2} }{ 2.q^{0}+12.q^{1}+2.q^{2}} &  2132123\\ \hline
74 & \frac{ 4.q^{0}+8.q^{1}+4.q^{2} }{ 2.q^{0}+12.q^{1}+2.q^{2}} &  2321232\\ \hline
\end{array} \]
\caption{A relation in ${\cal B}^H_4$ from GCT4 continued.}
\label{fb43}
\end{figure}

\end{document}